\documentclass[english]{JHEP3}
\usepackage[T1]{fontenc}
\usepackage[latin9]{inputenc}
\usepackage{float}
\usepackage{amsmath}
\usepackage{graphicx}
\usepackage{amssymb}
\usepackage{esint}

\makeatletter

\newcommand{\noun}[1]{\textsc{#1}}
\newcommand{\lyxdot}{.}

\newenvironment{lyxlist}[1]
{\begin{list}{}
{\settowidth{\labelwidth}{#1}
 \setlength{\leftmargin}{\labelwidth}
 \addtolength{\leftmargin}{\labelsep}
 }}
{\end{list}}

\usepackage{cite}
\usepackage[mathscr]{euscript}

\title{A positive-weight next-to-leading order simulation of weak boson pair production.}

\author{Keith Hamilton \\ INFN, Sezione di Milano-Bicocca, Piazza della Scienza 3, 20126 Milan, Italy. \\ Email: \email{keith.hamilton@mib.infn.it}}

\preprint{ MCnet/10/18 }

\abstract{
In this article we describe simulations of $ZZ$, $W^{\pm}Z$ and $W^{+}W^{-}$ production based on the positive weight next-to-leading-order\,(NLO) matching scheme, \noun{Powheg}, in the \noun{Herwig++} event generator. Building on earlier efforts within the \noun{Herwig++} framework, the simulation includes a full description of truncated showering effects, required to correctly model soft, wide angle, emissions in angular-ordered parton showers. We utilize simple relations among each of the diboson cross sections, holding to $\cal{O}(\alpha_{S})$, in constructing the simulation. Spin correlation effects are also included in the decays of the vector bosons at the tree order. A large part of this work is concerned with a full and thorough validation of the simulations through comparisons with alternative methods and calculations.}

\keywords{QCD, Monte Carlo, NLO Computations, Resummation, Collider Physics.}

\newcommand{\splusminus}{{\mathchoice%
{\vplusminus\displaystyle}%
{\vplusminus\scriptstyle}%
{\vplusminus\scriptscriptstyle}%
{\vplusminus\scriptscriptstyle}%
}}

\newdimen\hbigcirc
\newdimen\wbigcirc
\newcommand{\vplusminus}[1]{%
\settoheight{\hbigcirc}{$#1\bigcirc$}%
\settowidth{\wbigcirc}{$#1\bigcirc$}%
\makebox[\wbigcirc]{%
\makebox[0pt]{\rule[0.4\hbigcirc]{0.5\wbigcirc}{0.05\hbigcirc}}%
\makebox[0pt]{\rule[0.1\hbigcirc]{0.5\wbigcirc}{0.05\hbigcirc}}%
\makebox[0pt]{\rule[0.1\hbigcirc]{0.05\wbigcirc}{0.6\hbigcirc}}%
\makebox[0pt]{$#1\bigcirc$}}%
}

\makeatother

\usepackage{babel}

\begin{document}

\section{Introduction\label{sec:Introduction}}

The production of electroweak gauge boson pairs is a subject of significant
interest in collider physics. Weak boson pair production was first
studied in detail at LEP2, at centre-of-mass energies of up to $209\,\mathrm{GeV}$
\cite{EWWG:2008,Barate:2000gi,Abreu:2000kf,Acciarri:2000pc,Abbiendi:2000eg},
with the aim of precisely measuring the $W$ boson mass and with a
view to directly probing and testing the non-Abelian character of
the Standard Model gauge structure. In the period 1996-2000 approximately
40000 $W^{+}W^{-}$ and 1000 $ZZ$ events were recorded by the four
LEP experiments \cite{Natale:2004aa}. At the Tevatron $W^{+}W^{-}$
and $W^{\pm}Z$ production have also been the focus of extensive analysis
by the CDF and D0 collaborations \cite{Abazov:2004kc,Abazov:2005ys,Acosta:2005mu,Abazov:2007rab,Abulencia:2007tu,Aaltonen:2009fd,Aaltonen:2009us,Abazov:2009tr,Aaltonen:2010rq},
albeit with an estimated factor of twenty less diboson events per
experiment, based on their most recent publications. Recently the
observation of $ZZ$ production has also been reported by CDF and
D0.

Although all of these LEP2 and Tevatron measurements are in agreement
with Standard Model predictions, it is generally considered that should
non-Standard Model physics play a role in this process, its effects
will be manifest greater at greater energy scales. In particular,
the effects of anomalous $WW\gamma$ and $WWZ$ triple gauge boson
couplings, are predicted to grow with the invariant mass of the gauge
boson pair \cite{Ellison:1998uy}. Thus, even though fewer events
will be recorded, the high energy data still being accumulated by
CDF and D0 combined, has the potential to surpass the sensitivity
of the LEP2 data to such effects. At the LHC the inclusive cross section
for vector boson pair production is around a factor of \emph{ten }higher
than at the Tevatron, moreover, the luminosity is also substantially
higher. It is then expected that in the near future very large quantities
of high energy weak boson pairs will be recorded by the LHC experiments
and that the non-Abelian nature of the weak bosons will be tested
with a new level of vigor. To facilitate such precision tests theoretical
predictions, ideally in the form of fully exclusive Monte Carlo simulations,
should be made as accurate as possible.

Aside from being an interesting process in its own right, weak boson
pair production deserves to be thoroughly understood given its role
as an important background in many searches for currently undiscovered
particles, most notably the Higgs boson. Notwithstanding the fact
that the latest Tevatron analysis excludes, at the 95\% confidence
level, the Standard Model Higgs boson from having a mass in the range
$158<m_{\mathrm{H}}<175$ GeV \cite{TEVNPH:2010ar}, if the mass is
above $\sim135$ GeV the Higgs boson will primarily decay into $W^{+}W^{-}$
or $ZZ$ pairs \cite{Rizzo:1980gz,Keung:1984hn,Eichten:1984eu,Djouadi:1995gv,Carena:2000yx}.
Moreover, alternative models with extended Higgs sectors can yield
large branching fractions for charged Higgs bosons decaying into $W^{\pm}Z$
pairs \cite{Gunion:1989we}. $W^{\pm}Z$ pair production is also a
problematic background in studies of strong $WW$ scattering \cite{Chanowitz:1994ap}
and in many supersymmetry search channels, specifically, those associated
with a trilepton final state and missing transverse energy \cite{Matchev:1999nb,Baer:1999bq,ATLASTDR:1999fr}.
As with the dedicated studies of weak boson pair production discussed
above, the discovery potential of these ongoing and imminent analyses
depends heavily on the quality of the theoretical inputs.

In general, the main source of uncertainty associated with leading
order theoretical predictions for hadronic collisions comes in the
form of next-to-leading order QCD corrections. Next-to-leading order
(NLO) QCD corrections to $ZZ$, $W^{\pm}Z$ and $W^{+}W^{-}$ production
were first calculated and studied in the early nineties by two separate
groups\emph{ }\cite{Mele:1990bq,Ohnemus:1990za,Ohnemus:1991kk,Ohnemus:1991gb,Frixione:1992pj,Frixione:1993yp}.
Some years later these results were improved upon by Dixon \emph{et
al}, who performed the calculations at the level of helicity amplitudes,
thus providing full knowledge of the $\mathcal{O}\left(\alpha_{\mathrm{S}}\right)$
corrections to these processes \emph{including} the leptonic decays
of the massive vector bosons \cite{Dixon:1998py,Dixon:1999di}. Further
improvements were made by Campbell and Ellis who extended the results
of the latter work beyond the narrow width approximation, including
contributions from singly resonant Feynman diagrams and interference
effects between intermediate $Z$ bosons and photons\cite{Campbell:1999ah}.

Following these results, in the early part of the last decade, a number
of ground breaking developments took place in the field of Monte Carlo
event generator research, most significantly, the invention of the
CKKW(-L) and MLM algorithms, combining parton shower simulations together
with those based on tree level matrix elements \cite{Catani:2001cc,Krauss:2002up,Schalicke:2005nv,Lonnblad:2001iq,Mangano:2001xp,Mrenna:2003if}
and, separately, the \noun{Mc}{\footnotesize @}\noun{nlo }\cite{Frixione:2002ik}
and \noun{Powheg} \cite{Nason:2004rx,Frixione:2007vw} formalisms
for consistently including fully differential NLO corrections in parton
shower simulations. In fact the \noun{Mc}{\footnotesize @}\noun{nlo
}formalism was first successfully demonstrated through the simulation
of $W^{+}W^{-}$ pair production and the first public release of the
program comprised of $W^{+}W^{-}$, $W^{\pm}Z$ and $ZZ$ production,
two years after the work of Campbell and Ellis. An impressive number
of important Standard Model processes can now be simulated with the
\noun{Mc}{\footnotesize @}\noun{nlo }program \cite{Frixione:2003ei,Frixione:2005vw,Frixione:2006gn,Frixione:2008yi},
in addition, the modeling of unstable particle production has been
enhanced through the inclusion of full spin correlations in the leading
order and real contributions by the method developed in Ref.\,\cite{Frixione:2007zp}.
Nowadays a similar number of processes may also be simulated with
publicly available programs based on the alternative \noun{Powheg}
formalism \cite{Frixione:2007nw,Frixione:2007nu,Alioli:2008gx,Hamilton:2008pd,Alioli:2008tz,Hamilton:2009za,Alioli:2009je,Nason:2009ai,Alioli:2010xd},
first realized in the case of $ZZ$ hadroproduction in 2006 \cite{Nason:2006hfa}.

Lately significant steps have been taken in the direction of automating
the \noun{Mc}{\footnotesize @}\noun{nlo} and \noun{Powheg} methods
\cite{Frederix:2009yq,Alioli:2010xd,Hoeche:2010pf}. The most recent
of these efforts features \noun{Powheg} simulations of $W^{+}W^{-}$
and $ZZ$ production, integrating the calculation of Campbell and
Ellis within the framework of the \noun{Sherpa} event generator
\cite{Hoeche:2010pf}. This simulation therefore includes singly
resonant contributions, in particular $Z/\gamma^{*}$ interference
effects in $W^{\pm}Z$ and $Z$ pair hadroproduction, which are not
included in the program discussed here. 
For reasons of technical simplicity, greatly increased computational
efficiency, and minded to best utilize the existing \noun{Herwig++}
infrastructure, \emph{e.g.} the facility to include higher order QCD
and QED corrections to the \emph{decays} of vector bosons, we have
based our work on the original calculations of Frixione \emph{et al}.
\cite{Mele:1990bq,Frixione:1992pj,Frixione:1993yp}, as in
\noun{Mc}{\footnotesize @}\noun{nlo}, valid in the double pole
approximation.

We also note that many important collider physics analyses, such as
the study of the di-vector boson signal
\cite{Abazov:2005ys,Abulencia:2007tu,Abazov:2007rab,Aaltonen:2008mv,Abazov:2008gya,Abazov:2009eu,Aaltonen:2010ws,Brigljevic:2007zz} and anomalous triple
gauge boson couplings \cite{Abazov:2007rab,Aaltonen:2007sd,Abazov:2009hk},
employ invariant mass cuts on the vector boson decay products -- almost
always to $Z$ bosons -- which generally reduces the singly resonant
contributions to negligible levels. The same is true of Higgs boson searches
involving final-states comprising of $Z$ bosons \cite{Aaltonen:2010pa}.
In the case of the $W$ boson the same invariant mass cuts around the boson
decay products are not necessary since, in the case of the $Z$, the singly
resonant contributions are only able to become sizable through $Z/\gamma$
interference. Thus, in $W$ pair production the singly resonant terms are
subject to a simple Breit-Wigner suppression. Accordingly, the program
presented here can be applied to the simulation of $WW$ pair production
for study as a signal process and as a background to Higgs boson production,
as an alternative to the \noun{Mc}{\footnotesize @}\noun{nlo} program 
\cite{Aaltonen:2009us,Aaltonen:2010yv,Aaltonen:2010cm,Abazov:2010ct}.
Furthermore, as is usually the case in higher order calculations involving
unstable particles, an \emph{ad hoc} scheme is adopted in
Ref.\,\cite{Campbell:1999ah} to include the width of the vector bosons which,
in this case, mistreats the singly resonant parts, particularly in the doubly
resonant region. Nevertheless, it is incumbent upon us to point out that
should analyses not include some form of cut on $Z$ boson decay products,
limiting, in particular, $Z/\gamma$ interference contributions, the matrix
elements of Campbell and Ellis will continue to provide a good
description of the production and decay processes, whereas those used here,
like those of the underlying LO \noun{Herwig++} simulation, will fail.
Analyses not employing such cuts are typical of general searches for physics
beyond the Standard Model \emph{e.g.} the search for SUSY in the trilepton
channel, for which singly resonant contributions are expected to add around
15\% to the $W^{\pm}Z$ doubly resonant background
\footnote{A study of the impact of singly resonant diagrams
on such an analysis can be found in Ref.~\cite{Campbell:1999ah}.}.

In what follows we present an application of the \noun{Powheg} method
to Monte Carlo simulations of weak boson pair production within the
\noun{Herwig++} event generator \cite{Bahr:2008tx,Bahr:2008pv}, including
a full validation. In Sect. \ref{sec:Next-to-leading-order-cross}
we briefly recall the basic \noun{Powheg} algorithm and give details
concerning how the NLO cross section is organized in a form amenable
to its implementation; the section closes with a discussion of the
associated NLO matrix elements \cite{Mele:1990bq,Frixione:1992pj,Frixione:1993yp}\emph{,
}noting exact relations between the $W^{\pm}Z$, $W^{+}W^{-}$, and
$ZZ$ next-to-leading order cross sections\emph{. }In Section~\ref{sec:Implementation}
we describe the key points in the simulation process, including the
implementation of truncated and vetoed parton showers, occurring after
the hardest emission has been generated. In Section~\ref{sec:Results}
we show results from our implementation, comparing it to two independent
NLO programs, \noun{Mcfm} and \noun{Mc}{\footnotesize @}\noun{nlo},
before summarizing our findings in Sect. \ref{sec:Conclusion}.

\section{Hardest emission cross sections\label{sec:Next-to-leading-order-cross}}

The starting point for all \noun{Powheg} simulations is the so-called
\emph{hardest} \emph{emission} \emph{cross} \emph{section}, specifically,
a matching between the constituents of the exact NLO differential
cross section and the corresponding leading-log resummed cross section
implicit in parton shower simulations \cite{Nason:2004rx}. For simple
processes, such as the one we are considering, it can be written simply
as \begin{equation}
\mathrm{d}\sigma=\overline{B}\left(\Phi_{B}\right)\,\mathrm{d}\Phi_{B}\,\left[\Delta_{\hat{R}}\left(k_{T,min}\right)+\frac{\widehat{R}\left(\Phi_{B},\Phi_{R}\right)}{B\left(\Phi_{B}\right)}\,\Delta_{\hat{R}}\left(k_{T}\left(\Phi_{B},\Phi_{R}\right)\right)\,\mathrm{d}\Phi_{R}\right],\label{eq:powheg_1}\end{equation}
where $\Phi_{B}$ are \emph{Born variables}, which fully determine
the kinematics of leading order configurations, and $\Phi_{R}$ are
\emph{radiative variables}, parametrizing the kinematics of the hardest
emission with respect to $\Phi_{B}$. $B\left(\Phi_{B}\right)$ and
$\overline{B}\left(\Phi_{B}\right)$ are the leading and next-to-leading
order differential cross sections respectively, while $\widehat{R}\left(\Phi_{B},\Phi_{R}\right)$
is simply the \emph{bare}, tree level, real emission cross section.
The $\overline{B}\left(\Phi_{B}\right)$ function may be expressed
as \begin{equation}
\overline{B}\left(\Phi_{B}\right)=B\left(\Phi_{B}\right)+V\left(\Phi_{B}\right)+\int\,\mathrm{d}\Phi_{R}\, R\left(\Phi_{B},\Phi_{R}\right)\,,\label{eq:powheg_2}\end{equation}
$V\left(\Phi_{B}\right)$ being the finite combination of, unresolvable,
soft emission and virtual loop corrections, while $R\left(\Phi_{B},\Phi_{R}\right)$
corresponds to the remaining, regularized, real emission corrections.
The \noun{Powheg} Sudakov form factor for the hardest emission, $\Delta_{\hat{R}}\left(p_{T}\right)$,
is defined as\begin{equation}
\Delta_{\hat{R}}\left(p_{T}\right)=\exp\left[-\int\mathrm{d}\Phi_{R}\,\frac{\widehat{R}\left(\Phi_{B},\Phi_{R}\right)}{B\left(\Phi_{B}\right)}\,\theta\left(k_{T}\left(\Phi_{B},\Phi_{R}\right)-p_{T}\right)\right],\label{eq:powheg_3}\end{equation}
where $k_{T}\left(\Phi_{B},\Phi_{R}\right)$ tends to the transverse
momentum of the emitted parton in the soft and collinear limits. Emissions
for which $k_{T}\le k_{T,min}$ are considered as being \emph{unresolvable}.
In the ensuing subsections we give details concerning the components
of the hardest emission cross section for the case of weak boson pair
production (Eq.\,\ref{eq:powheg_1}) , focusing on the $\overline{B}\left(\Phi_{B}\right)$
function and its regularization.

\subsection{Kinematics and phase space\label{sub:Kinematics-and-phase}}

In this subsection we begin by giving details regarding our parametrization
of the kinematics for the hadroproduction of a pair of weak vector
bosons, with and without the emission of an additional radiated parton.
For both classes of event we denote the momenta of the partons incident
from the $\pm z$ directions by $p_{\splusminus}$, while those of
the weak bosons are labelled $p_{1}$ and $p_{2}$. In discussing
three-body, real emission, contributions the momentum of the additional,
radiated, parton is denoted $k$. Also, since we shall frequently
refer to the sum of the weak boson momenta, we define $p\equiv p_{1}+p_{2}$. 

The parametrization of our two- and three-body kinematics is taken
to be identical to that in Refs.\,\cite{Mele:1990bq,Frixione:1992pj,Frixione:1993yp}.
To this end we first define precisely what we mean by the rest frame
of the vector boson pair. In the context of three-body events, for
$q\bar{q}$ and $qg$ collisions, we shall use this term to refer
to the frame in which the vector bosons are balanced in their three-momenta,
with the incoming quark defining the $+z$ axis and the transverse
momentum of the other initial-state parton defining the $+y$ axis;
for $g\bar{q}$ collisions the gluon replaces the quark in defining
the $+z$ axis. For genuine two-body events the latter criterion,
concerning the definition of the $+y$ axis, is omitted since for
such events $p_{\oplus}$ and $p_{\ominus}$ are naturally acolinear.

We now introduce a set of \emph{Born variables} $\Phi_{B}=\left\{ p^{2},\,\mathrm{y},\,\theta\right\} $
and a set of \emph{radiative} \emph{variables} $\Phi_{R}=\left\{ x,\, y,\,\phi\right\} $,
clearly defined as follows:
\begin{lyxlist}{00.00.0000}
\item [{~~~~~~~~~~$p^{2}$}] -~~~the invariant mass squared
of the vector boson pair 
\item [{~~~~~~~~~~$\mathrm{y}$}] -~~~the rapidity of the
weak boson pair in the lab frame
\item [{~~~~~~~~~~$\theta$}] -~~~the polar angle of $p_{1}$
in the rest frame of the vector boson pair
\item [{~~~~~~~~~~$\phi$}] -~~~the azimuthal angle of $p_{1}$
in the rest frame of the vector boson pair
\item [{~~~~~~~~~~$x$}] -~~~the ratio $p^{2}/\hat{s}$ where,
as usual, $\hat{s}=\left(p_{\oplus}+p_{\ominus}\right)^{2}$ 
\item [{~~~~~~~~~~$y$}] -~~~the cosine of the polar angle
of momentum $k$ in the partonic\vspace*{-3mm}

\item [{~}] ~~~~centre-of-mass frame
\end{lyxlist}
Both $\theta$ and $\phi$ range between $0$ and $\pi$. Given a
set of Born variables one can readily construct the momenta in the
lab frame for two-body $p_{\oplus}+p_{\ominus}\rightarrow p_{1}+p_{2}$
reactions, while augmenting these with a set of radiative variables
enables one to fully reconstruct radiative processes $p_{\oplus}+p_{\ominus}\rightarrow p_{1}+p_{2}+k$
in the lab frame. For the latter case explicit expressions for the
momenta in the rest frame of the vector boson pair are given in Ref.\,\cite{Frixione:1992pj}.
Since these are unwieldy, yet straightforward, we do not reproduce
them here. When reconstructing the three-body events using the aforesaid
formulae, all particles are then returned to the lab frame by applying
the following Lorentz transformation, $\mathbb{T}$, to each one:\begin{equation}
\mathbb{T}=\mathbb{B}_{\parallel}\,\mathbb{B}_{\perp}\,\mathbb{R}\,.\label{eq:nlo_1_1}\end{equation}
The first component of this transformation, $\mathbb{R}$, is a rotation
of angle $\arctan\, p_{T}/\sqrt{p^{2}}$ in the $y-z$ plane, where
$p_{T}$ is the transverse momentum of the vector boson system in
the lab and partonic centre-of-mass frames:\begin{equation}
p_{T}^{2}=\frac{1}{4}\,\frac{p^{2}}{x}\left(1-y^{2}\right)\left(1-x\right)^{2}\,.\label{eq:nlo_1_2}\end{equation}
Following $\mathbb{R}$ a transverse Lorentz boost, $\mathbb{B}_{\perp}$,
is carried out such that the momentum of the stationary vector boson
system, $p$, becomes $\left(E_{T},0,-p_{T},0\right)$, where we have
defined $E_{T}=\sqrt{p_{T}^{2}+p^{2}}$. Lastly a longitudinal boost,
$\mathbb{B}_{\parallel}$, gives the vector boson system rapidity
$\mathrm{y}$, returning all particles to the lab frame and completing
the momentum reconstruction. Note that the ultimate step in the generation
of all \textsc{Powheg} events involves randomizing the azimuthal orientation. 

The final kinematic quantities we wish to declare are the momentum
fractions of the incoming partons with respect to the beam particles.
These are not independent degrees of freedom but functions of $\Phi_{B}$
and, in the case of three-body final-states, $\Phi_{R}$. Taking $P_{\splusminus}$
to represent the momenta of the parent beam hadrons, the momentum
fractions $x_{\splusminus}$ are defined according to the relation
$p_{\splusminus}=x_{\splusminus}P_{\splusminus}$, whereupon it follows
that\begin{eqnarray}
x_{\splusminus}=\frac{\bar{x}_{\splusminus}}{\sqrt{x}}\sqrt{\frac{2-\left(1-x\right)\left(1\mp y\right)}{2-\left(1-x\right)\left(1\pm y\right)}} & \mathrm{\,\,\,\,\,\,\,\,\,\,\,\, and\,\,\,\,\,\,\,\,\,\,\,\,} & \bar{x}_{\splusminus}=\sqrt{\frac{p^{2}}{s}}\mathrm{e}^{\,\pm\mathrm{y}}\,,\label{eq:nlo_1_3}\end{eqnarray}
with $s$ being the hadronic centre-of-mass energy. Note that for
genuine two-body final states, namely those corresponding to leading
order and virtual contributions, the limit $x\rightarrow1$ is clearly
implied in the evaluation of the \emph{all} kinematics quantities,
including the momentum fractions \emph{i.e.} for such final states
$x_{\splusminus}=\bar{x}_{\splusminus}$.

Having described the parameterization of the kinematics we move to
specify the integration measures in the two- and three-body phase
spaces, for non-radiative and radiative events respectively. With
these definitions in hand the phase space for the leading-order process
can then be written as \begin{equation}
\mathrm{d}\Phi_{B}\mbox{ }=\mbox{ }\mathrm{d}x_{\oplus}\,\mathrm{d}x_{\ominus}\,\mathrm{d}\hat{\Phi}_{B}\mbox{ }=\mbox{ }\frac{1}{s}\,\mathrm{d}p^{2}\,\mathrm{d}\mathrm{y}\,\mathrm{d}\hat{\Phi}_{B}\,,\label{eq:nlo_1_4}\end{equation}
where $\mathrm{d}\hat{\Phi}_{B}$ the two-body phase space measure
for the vector boson system. In the centre-of-mass frame of $p$,
using conventional dimensional regularization, with $n=4-2\epsilon$
dimensions \begin{eqnarray}
\mathrm{d}\hat{\Phi}_{B} & = & \frac{\left(4\pi\right)^{\epsilon}}{\Gamma\left(1-\epsilon\right)}\,\left(p^{*}\sin\theta\right)^{-2\epsilon}\,\frac{1}{8\pi}\,\frac{p^{*}}{\sqrt{p^{2}}}\,\mathrm{d}\cos\theta\,,\label{eq:nlo_1_5}\end{eqnarray}
where $p^{*}$ is the magnitude of the three momentum of either weak
boson in their rest frame. 

To parametrize the three-body phase space we factorize it into a product
of two two-body phase spaces for the reactions $p_{\oplus}+p_{\ominus}\rightarrow p+k$
and $p\rightarrow p_{1}+p_{2}$, by inserting the identity in the
form of an integral over a delta function defining $p$ as $p_{1}+p_{2}$.
In this way we may readily write the phase space measure as\begin{eqnarray}
\mathrm{d}\Phi & = & \mathrm{d}\Phi_{B}\,\mathrm{d}\Phi_{R}\,\frac{p^{2}}{\left(4\pi\right)^{2}x^{2}}\,\left(\frac{1}{p^{2}}\right)^{\epsilon}\, c_{\Gamma}\,\mathcal{J}\left(x,y\right)\,\mathcal{J}\left(\phi\right)\,,\label{eq:nlo_1_6}\end{eqnarray}
where%
\footnote{An irrelevant overall factor $\frac{1}{c_{\Gamma}}\,\frac{\left(4\pi\right)^{\epsilon}}{\Gamma\left(1-\epsilon\right)}$
has been dropped in writing $\mathcal{J}\left(x,y\right)$ since it
is equal to $1+\mathcal{O}\left(\epsilon^{3}\right)$.%
}\begin{align}
\mathcal{J}\left(x,y\right) & =2^{2\epsilon}\, x^{\epsilon}\left(1-x\right)^{1-2\epsilon}\left(1-y^{2}\right)^{-\epsilon}\,, & \mathcal{J}\left(\phi\right) & =\sqrt{\pi}\,\frac{\Gamma\left(1-\epsilon\right)}{\Gamma\left(\frac{1}{2}-\epsilon\right)}\,\sin^{-2\epsilon}\phi\,,\label{eq:nlo_1_7}\end{align}
and

\begin{eqnarray}
\mathrm{d}\Phi_{R} & = & \frac{1}{2\pi}\,\mathrm{d}y\,\mathrm{d}x\,\mathrm{d}\phi\,.\label{eq:nlo_1_8}\end{eqnarray}
The constant $c_{\Gamma}$ appears due to the use of dimensional regularization,
it is given by\begin{eqnarray}
c_{\Gamma} & = & \left(4\pi\right)^{\epsilon}\frac{\Gamma\left(1-\epsilon\right)^{2}\Gamma\left(1+\epsilon\right)}{\Gamma\left(1-2\epsilon\right)}\,.\label{eq:nlo_1_9}\end{eqnarray}
We emphasize that \emph{both} $\theta$ and $\phi$ range between
$0$ and $\pi$, hence $\int\mathrm{d}\Phi_{R}=1$. It is also worth
noting that $\frac{1}{\pi}\int\mathrm{d}\phi\,\mathcal{J}\left(\phi\right)=1$
and $\lim_{\epsilon\rightarrow0}\mathcal{J}\left(\phi\right)=1$.

Since we restrict ourselves to processes for which the NLO corrections
contain at most soft and initial-state collinear singularities, the
product of $p_{T}^{2}$ with the squared real emission matrix elements
will be finite throughout the radiative phase space. With this in
mind, following Refs.\,\cite{Mele:1990bq,Mangano:1991jk}, we extract
a factor of $p_{T}^{2}$ from $\mathcal{J}\left(x,y\right)$ and then
expand it in powers of $\epsilon$, to give \begin{equation}
\mathcal{J}\left(x,y\right)=\left[\mathcal{S}\,\delta\left(1-x\right)+\mathcal{C}\left(x\right)\,\left(2\delta\left(1+y\right)+2\delta\left(1-y\right)\right)+\mathcal{H}\left(x,y\right)\right]\,\frac{p_{T}^{2}}{\hat{s}}\,,\label{eq:nlo_1_10}\end{equation}
in which \begin{eqnarray}
\mathcal{S} & = & \frac{1}{\epsilon^{2}}-\frac{\pi^{2}}{6}-\frac{4}{\epsilon}\ln\eta+8\ln^{2}\eta\,,\label{eq:nlo_1_11_a}\\
\mathcal{C}\left(x\right) & = & -\frac{1}{\epsilon}\frac{1}{\left(1-x\right)_{\rho}}-\frac{1}{\left(1-x\right)_{\rho}}\ln x+2\left(\frac{\ln\left(1-x\right)}{1-x}\right)_{\rho}\,,\label{eq:nlo_1_11_b}\\
\mathcal{H}\left(x,y\right) & = & \frac{2}{\left(1-x\right)_{\rho}}\left[\left(\frac{1}{1+y}\right)_{+}+\left(\frac{1}{1-y}\right)_{+}\right]\,,\label{eq:nlo_1_11_c}\end{eqnarray}
where $\eta=\sqrt{1-\rho}$ and $\rho=\left(m_{1}+m_{2}\right)^{2}/\hat{s}$,
$m_{1}$ and $m_{2}$ being the masses of the weak vector bosons.
The $\rho$-distributions appearing in Eqs.\,\ref{eq:nlo_1_11_b}
and \ref{eq:nlo_1_11_c} are defined according to the relation\begin{equation}
\int_{\rho}^{1}\mathrm{d}x\, h\left(x\right)\left(\frac{\ln^{n}\left(1-x\right)}{1-x}\right)_{\rho}=\int_{\rho}^{1}\mathrm{d}x\,\left(h\left(x\right)-h\left(1\right)\right)\frac{\ln^{n}\left(1-x\right)}{1-x}\,,\label{eq:nlo_1_12}\end{equation}
for any sufficiently regular test function $h\left(x\right)$: in
this case, the product of $p_{T}^{2}$ with the real emission matrix
elements.

For completeness, we note that the kinematic boundaries are \begin{equation}
\begin{array}{rlclllrcl}
\left(m_{1}+m_{2}\right)^{2} & \le & p^{2} & \le & s\,, & \,\,\,\,\,\,\,\,\,\,\,\,\,\,\,\,\,\, & \left|\mathrm{y}\right| & \le & -\frac{1}{2}\ln\left(\frac{p^{2}}{s}\right)\,,\\
\bar{x}\left(y\right) & \le & x & \le & 1\,, & \,\,\,\,\,\,\,\,\,\,\,\,\,\,\,\,\,\, & \left|y\right| & \le & 1\,,\end{array}\label{eq:nlo_1_13}\end{equation}
 with $\bar{x}\left(y\right)$ given by \begin{equation}
\begin{array}{rcl}
\bar{x}\left(y\right) & = & \mathrm{max}\left(\frac{2\left(1+y\right)\bar{x}_{\oplus}^{2}}{\sqrt{\left(1+\bar{x}_{\oplus}^{2}\right)^{2}\left(1-y\right)^{2}+16y\bar{x}_{\oplus}^{2}}+\left(1-y\right)\left(1-\bar{x}_{\oplus}^{2}\right)}\,,\,\left\{ \begin{array}{c}
1+y\leftrightarrow1-y\\
\bar{x}_{\oplus}\leftrightarrow\bar{x}_{\ominus}\end{array}\right\} \right)\end{array}\,.\label{eq:nlo_1_14}\end{equation}
In order to ease the numerical implementation of the $\rho$ distributions
we map the $x$ variable into $\tilde{x}$, defined according to \begin{align}
x\left(\tilde{x},y\right) & =\bar{x}\left(y\right)+\bar{\eta}\left(y\right)^{2}\,\tilde{x}\,, & \bar{\eta}\left(y\right) & =\sqrt{1-\bar{x}\left(y\right)}.\label{eq:nlo_1_15}\end{align}
Whereas the $x$ integration domain was dependent on $y$, the $\tilde{x}$
integral simply ranges from $0$ to $1$. 

Finally we wish to clarify that the expressions for $\mathcal{J}\left(x,y\right)$
in Eqs.\,\ref{eq:nlo_1_7} and \ref{eq:nlo_1_10} are equivalent
only up to terms of $\mathcal{O}\left(\epsilon\right)$; these terms
do not contribute to the differential cross section in the limit $\epsilon\rightarrow0$.

\subsection{Differential cross section\label{sub:Differential-cross-section}}

In this section we enumerate the various contributions to the NLO
differential cross section. These are obtained by simply considering
the product of the matrix elements with the phase space measures as
written in Eq.\,\ref{eq:nlo_1_4} and Eqs.\,\ref{eq:nlo_1_6}-\ref{eq:nlo_1_12},
exploiting simplifications arising in the soft $\left(x\rightarrow1\right)$
and collinear limits $\left(y\rightarrow\pm1\right)$ to integrate
out radiative variables. The discussion here is rather similar to
that in our earlier work \cite{Hamilton:2009za}, including some minor
changes and clarifications, indeed the formulae in that article were
derived with the current application in mind, so we shall be brief. 

In the following we shall refer to the parton types incident from
the $+$ and $-z$ directions as $a$ and $b$ respectively. We denote
the parton distribution function (PDF) for a parton of type $i$ inside
a beam hadron traveling in the $\pm z$ direction by $f_{i}^{\splusminus}\left(x_{\splusminus},\mu_{F}^{2}\right)$,
where $\mu_{F}$ is the factorization scale. For brevity we then introduce
the luminosity\emph{ }function, $\mathcal{L}_{ab}$,\emph{ }as the
product of the PDFs associated to $a$ and $b$:\emph{ }\begin{equation}
\mathcal{L}_{ab}\left(x_{\oplus},x_{\ominus}\right)=f_{a}^{\oplus}\left(x_{\oplus},\mu_{F}^{2}\right)\, f_{b}^{\ominus}\left(x_{\ominus},\mu_{F}^{2}\right)\,.\label{eq:nlo_2_1}\end{equation}

The leading order contribution to the differential cross section is
given by the product of the leading order spin and colour averaged
squared amplitude, $\mathcal{M}_{ab}^{B}$, together with the luminosity
function and flux factor: \begin{align}
\mathrm{d}\sigma^{ab} & =B\left(\Phi_{B}\right)\,\mathrm{d}\Phi_{B}\,, & B\left(\Phi_{B}\right) & =\frac{1}{2p^{2}}\,\mathcal{M}_{ab}^{B}\left(\Phi_{B}\right)\mathcal{L}_{ab}\left(\bar{x}_{\oplus},\,\bar{x}_{\ominus}\right)\,.\label{eq:nlo_2_2}\end{align}
In all cases, due to the universal nature of the infrared divergences
in the one-loop amplitudes, the virtual corrections to $\mathcal{M}_{ab}^{B}$
can be generically written in the form\begin{equation}
\mathcal{M}_{ab}^{V_{0}}\left(\Phi_{B}\right)=\mathcal{V}_{0}\,\mathcal{M}_{ab}^{B}\left(\Phi_{B}\right)+\mathcal{M}_{ab}^{V_{\mathrm{reg}}}\left(\Phi_{B}\right)\,,\label{eq:nlo_2_3}\end{equation}
where the first term is universal and divergent, with $\mathcal{V}_{0}$
given by \begin{equation}
\mathcal{V}_{0}=\frac{\alpha_{S}c_{\Gamma}}{2\pi}\,\left(\frac{\mu^{2}}{p^{2}}\right)^{\epsilon}\, C_{F}\,\left[-\frac{2}{\epsilon^{2}}-\frac{2}{\epsilon}\, p_{a\,\widetilde{ag}}-\frac{\pi^{2}}{3}\right]\,,\label{eq:nlo_2_4}\end{equation}
while the second is process dependent and finite as $\epsilon\rightarrow0$.
For completeness, we define the \emph{bare} virtual cross section
and $V_{0}\left(\Phi_{B}\right)$ function by analogy to the corresponding
leading order quantities:\begin{align}
\mathrm{d}\sigma_{ab}^{V_{0}} & =V_{0}\left(\Phi_{B}\right)\,\mathrm{d}\Phi_{B}\,, & V_{0}\left(\Phi_{B}\right) & =\frac{1}{2p^{2}}\,\mathcal{M}_{ab}^{V_{0}}\left(\Phi_{B}\right)\mathcal{L}_{ab}\left(\bar{x}_{\oplus},\,\bar{x}_{\ominus}\right)\,.\label{eq:nlo_2_5}\end{align}

The differential cross sections for the real emission processes, $a+b\rightarrow n+c$,
take the following general form\begin{eqnarray}
\mathrm{d}\sigma_{ab}^{R} & = & \frac{1}{2\hat{s}}\,\mathcal{M}_{ab}^{R}\left(\Phi_{B},\Phi_{R}\right)\,\mathcal{L}_{ab}\left(x_{\oplus},x_{\ominus}\right)\,\mathrm{d}\Phi\,.\label{eq:nlo_2_6}\end{eqnarray}
For each flavour combination we consider these corrections to comprise
of three components, corresponding to the three terms $\mathcal{S}$,
$\mathcal{C}\left(x\right)$ and $\mathcal{H}\left(x,y\right)$ in
the phase space Jacobian, Eqs.\,\ref{eq:nlo_1_11_a}-\ref{eq:nlo_1_11_c}.
We shall refer to these components as the \emph{soft}, \emph{collinear},
and \emph{hard} / \emph{resolved} contributions to the cross section.

The squared matrix elements for the real emission processes, in which
a gluon is emitted from an initial-state quark or antiquark, factorize
in the limit that the gluon is soft $\left(x\rightarrow1\right)$
according to\begin{equation}
\lim_{x\rightarrow1}\mathcal{M}_{ab}^{R}\left(\Phi_{B},\Phi_{R}\right)=8\pi\alpha_{S}\mu^{2\epsilon}\,\frac{1}{x\, p_{T}^{2}}\,2\, C_{F}\,\mathcal{M}_{ab}^{B}\left(\Phi_{B}\right)\,.\label{eq:nlo_2_7}\end{equation}
Thus, for $x=1$ the integrand in Eq.\,\ref{eq:nlo_2_6} is entirely
independent of the radiative phase space and it becomes trivial to
integrate over $\Phi_{R}$. In doing this one finds the following
expression for the soft contribution to the differential cross section:
\begin{equation}
\mathrm{d}\sigma_{ab}^{S_{0}}=\frac{\alpha_{S}c_{\Gamma}}{2\pi}\,\left(\frac{\mu^{2}}{p^{2}}\right)^{\epsilon}\, C_{F}\,\left(\frac{2}{\epsilon^{2}}-\frac{\pi^{2}}{3}-\frac{8}{\epsilon}\ln\eta+16\ln^{2}\eta\right)\, B\left(\Phi_{B}\right)\,\mathrm{d}\Phi_{B}\,.\label{eq:nlo_2_8}\end{equation}
All other sources of real corrections involve the emission of a quark
or antiquark from external initial-state gluons, as such they do not
contribute to the cross section in the soft limit; by contrast to
the case of gluon emission, the matrix elements for such processes
are regular in the limit $x\rightarrow1$, hence when they multiply
the factor of $p_{T}^{2}$ in $\mathcal{J}\left(x,y\right)$ (Eq.\,\ref{eq:nlo_1_10})
the result is zero.

We shall now turn our attention to the collinear limits. Since the
real emission corrections to the process we are considering do not
involve internal gluon lines, in the limit $y\rightarrow\pm1$ the
spin averaged squared matrix elements factorize trivially according
to%
\footnote{In Eqs.\,\ref{eq:nlo_2_7} and \ref{eq:nlo_2_9} we show only the
leading term in $\frac{1}{p_{T}^{2}}$ since all others ultimately
vanish due to the factor of $p_{T}^{2}$ in Eq.\,\ref{eq:nlo_1_10}. %
} \begin{equation}
\lim_{y\rightarrow\pm1}\mathcal{M}_{ab}^{R}\left(\Phi_{B},\Phi_{R}\right)=8\pi\alpha_{S}\mu^{2\epsilon}\,\frac{1}{x\, p_{T}^{2}}\,\left(1-x\right)\,\hat{P}_{i\,\widetilde{ic}}\left(x\,;\epsilon\right)\,\mathcal{M}_{ab}^{B}\left(\Phi_{B}\right)\,,\label{eq:nlo_2_9}\end{equation}
where $i=a$ or $b$ for $y=\pm1$ respectively and $c$ denotes the
type of the emitted parton. Considering only those terms in the phase
space Jacobian proportional to $\mathcal{C}\left(x\right)$ we then
obtain the following expression for the collinear contribution to
the cross section \begin{eqnarray}
\mathrm{d}\sigma_{ab}^{C_{0}\splusminus} & = & \mathrm{d}\sigma_{ab}^{S\, C\splusminus}+\mathrm{d}\sigma_{ab}^{C\splusminus}-\mathrm{d}\sigma_{ab}^{CT\splusminus},\nonumber \\
\mathrm{d}\sigma_{ab}^{S\, C\splusminus} & = & \frac{\alpha_{S}c_{\Gamma}}{2\pi}\,\left(\frac{\mu^{2}}{p^{2}}\right)^{\epsilon}\, C_{i\,\widetilde{ic}}\left(p_{i\,\widetilde{ic}}+4\ln\eta\right)\left(\frac{1}{\epsilon}+\ln\left(\frac{p^{2}}{\mu^{2}}\right)\right)\, B\left(\Phi_{B}\right)\,\mathrm{d}\Phi_{B},\nonumber \\
\mathrm{d}\sigma_{ab}^{C\splusminus} & = & \frac{\alpha_{S}}{2\pi}\,\frac{1}{x}\,\mathcal{C}_{i\,\widetilde{ic}}^{\splusminus}\left(x\right)\,\widehat{\mathcal{L}}_{ab}^{\splusminus}\left(x_{\oplus},\, x_{\ominus}\right)\, B\left(\Phi_{B}\right)\,\mathrm{d}\Phi_{B}\,\mathrm{d}x,\label{eq:nlo_2_10}\\
\mathcal{C}_{i\,\widetilde{ic}}^{\splusminus}\left(x\right) & = & \left[\frac{1}{\left(1-x\right)_{\rho}}\ln\left(\frac{p^{2}}{\mu^{2}x}\right)+2\left(\frac{\ln\left(1-x\right)}{1-x}\right)_{\rho}\right]\,\left(1-x\right)\hat{P}_{i\,\widetilde{ic}}\left(x\right)-\hat{P}_{i\,\widetilde{ic}}^{\epsilon}\left(x\right),\nonumber \\
\mathrm{d}\sigma_{ab}^{CT\splusminus} & = & \frac{1}{\bar{\epsilon}}\,\frac{\alpha_{S}}{2\pi}\,\frac{1}{x}\, P_{i\,\widetilde{ic}}\left(x\right)\,\widehat{\mathcal{L}}_{ab}^{\splusminus}\left(x_{\oplus},\, x_{\ominus}\right)\, B\left(\Phi_{B}\right)\,\mathrm{d}\Phi_{B}\,\mathrm{d}x,\nonumber \end{eqnarray}
where $i=a$ in the case that parton $a$ splits to produce parton
$c$, and $i=b$ for the case that parton $b$ branches to produce
$c$. The functions $\hat{P}_{i\,\widetilde{ic}}\left(x;\epsilon\right)$
are the bare Altarelli-Parisi splitting kernels in $n=4-2\epsilon$
dimensions, explicit expressions for which can be found throughout
the literature \emph{e.g.} Refs.\,\cite{Catani:1996vz,Hamilton:2009za},
while $P_{i\,\widetilde{ic}}\left(x\right)$ denotes their regularized
counterparts (Appendix\,\ref{sec:Splitting-functions}); the soft-collinear
cross section, $\mathrm{d}\sigma_{ab}^{S\, C\splusminus}$, is entirely
due to the $\delta\left(1-x\right)$ term present in the latter. Lastly
we declare $\widehat{\mathcal{L}}_{ab}^{\splusminus}$ to be the ratio
of the luminosity function evaluated at $y=\pm1$ with respect to
that found in the leading order cross section \emph{viz} \begin{align}
\widehat{\mathcal{L}}_{ab}^{\splusminus}\left(x_{\oplus},\, x_{\ominus}\right) & =\left.\widehat{\mathcal{L}}_{ab}\left(x_{\oplus},\, x_{\ominus}\right)\right|_{y=\pm1}\,, & \widehat{\mathcal{L}}_{ab}\left(x_{\oplus},\, x_{\ominus}\right) & =\frac{\mathcal{L}_{ab}\left(x_{\oplus},\, x_{\ominus}\right)}{\mathcal{L}_{ab}\left(\bar{x}_{\oplus},\,\bar{x}_{\ominus}\right)}\,.\label{eq:nlo_2_11}\end{align}
The singular parts of the collinear contribution proportional to the
regularized Altarelli-Parisi functions, $\mathrm{d}\sigma_{ab}^{CT\splusminus}$,
are exactly canceled by collinear counterterms in the $\overline{\mathrm{MS}}$
scheme, thus they play no further role in our discussion.

Finally, considering the third part of the $\mathcal{J}\left(x,y\right)$
phase space Jacobian, we have that the hard / resolved contribution
to the real emission cross section is given by 

\begin{eqnarray}
\mathrm{d}\sigma_{ab}^{H} & = & \frac{\alpha_{\mathrm{S}}}{2\pi}\,\frac{1}{x}\,\mathcal{H}_{ab}\,\widehat{\mathcal{L}}_{ab}\left(x_{\oplus},\, x_{\ominus}\right)\, B\left(\Phi_{B}\right)\,\mathrm{d}\Phi_{B}\,\mathrm{d}\Phi_{R}\,,\label{eq:nlo_2_12}\\
\mathcal{H}_{ab} & = & x\, p_{T}^{2}\,\frac{1}{8\pi\alpha_{S}}\,\frac{\mathcal{M}_{ab}^{R}\left(\Phi_{B},\Phi_{R}\right)}{\mathcal{M}_{ab}^{B}\left(\Phi_{B}\right)}\,\mathcal{H}\left(x,y\right)\,.\end{eqnarray}

Combining all of the various pieces together we are able to write
the next-to-leading order differential cross section as \begin{equation}
\mathrm{d}\sigma=B\left(\Phi_{B}\right)\,\mathrm{d}\Phi_{B}+V\left(\Phi_{B}\right)\,\mathrm{d}\Phi_{B}+R\left(\Phi_{B},\Phi_{R}\right)\,\mathrm{d}\Phi_{B}\,\mathrm{d}\Phi_{R},\label{eq:nlo_2_13}\end{equation}
where the second term is given by the finite sum of the bare virtual
corrections, $\mathrm{d}\sigma^{V_{0}}$, the soft real emission contributions,
$\mathrm{d}\sigma^{S_{0}}$, and the soft-collinear contributions,
$\mathrm{d}\sigma^{S\, C\splusminus}$. The $R\left(\Phi_{B},\Phi_{R}\right)$
function is comprised of the remaining collinear and hard / resolved
real emission corrections, \begin{eqnarray}
R\left(\Phi_{B},\Phi_{R}\right) & = & \frac{\alpha_{S}}{2\pi}\,\frac{1}{x}\,\mathcal{R}_{ab}\,\widehat{\mathcal{L}}_{ab}\left(x_{\oplus},x_{\ominus}\right)\, B\left(\Phi_{B}\right)\,,\label{eq:nlo_2_14}\\
\mathcal{R}_{ab} & = & 2\,\mathcal{C}_{a\,\widetilde{ac}}^{\oplus}\left(x\right)\delta\left(1-y\right)+2\,\mathcal{C}_{b\,\widetilde{bc}}^{\ominus}\left(x\right)\delta\left(1+y\right)+\mathcal{H}_{ab}\,.\nonumber \end{eqnarray}
Although it is not explicitly stated above, a summation over all contributing
channels and flavour combinations is implied. Furthermore, in the
case of the $qg$ and $g\bar{q}$ initiated real emission corrections,
one of the terms proportional to $\mathcal{C}_{i\,\widetilde{ic}}^{\splusminus}\left(x\right)$
in the $\mathcal{R}_{ab}$ function should be omitted, either $2\mathcal{C}_{a\,\widetilde{ac}}^{\oplus}\left(x\right)\delta\left(1-y\right)$
if $b$ is a gluon, or $2\,\mathcal{C}_{b\,\widetilde{bc}}^{\ominus}\left(x\right)\delta\left(1+y\right)$
if $a$ is a gluon, since collinear branchings of the initial-state
quarks do not occur in those processes.

\subsection{Matrix elements \label{sub:vv_matrix_elements}}

As noted in the introduction the simulation described in this article
uses the NLO QCD calculations for $ZZ$, $W^{\pm}Z$ and $W^{+}W^{-}$
production of Frixione \emph{et al}. \cite{Mele:1990bq,Frixione:1992pj,Frixione:1993yp}.
We have observed that these computations share a common underlying
structure which we have exploited here. In particular we find that
\emph{all} $W^{\pm}Z$ production matrix elements, leading-order,
real and virtual, are simply related to the corresponding $ZZ$ production
matrix elements by the following replacements:

\begin{equation}
\begin{array}{lclclcl}
F_{ij} & \rightarrow & 1\,, & \,\,\,\,\,\,\,\,\,\,\,\,\,\,\,\,\,\,\,\,\,\,\,\,\,\,\, & g_{\mathrm{u,L}} & \rightarrow & \frac{1}{2}\sqrt{g_{\mathrm{V}}^{4}+g_{\mathrm{A}}^{4}+6g_{\mathrm{V}}^{2}g_{\mathrm{A}}^{2}}\,,\\
e_{\mathrm{Z}}^{2} & \rightarrow & 0\,, & \,\,\,\,\,\,\,\,\,\,\,\,\,\,\,\,\,\,\,\,\,\,\,\,\,\,\, & g_{\mathrm{d,L}} & \rightarrow & \frac{1}{2}\sqrt{g_{\mathrm{V}}^{4}+g_{\mathrm{A}}^{4}+6g_{\mathrm{V}}^{2}g_{\mathrm{A}}^{2}}\,,\\
e_{\mathrm{Z}} & \rightarrow & 0\,, & \,\,\,\,\,\,\,\,\,\,\,\,\,\,\,\,\,\,\,\,\,\,\,\,\,\,\, & m_{\mathrm{W}} & \rightarrow & m_{\mathrm{Z}}\,,\end{array}\label{eq:WZ_to_ZZ_replacements}\end{equation}

\noindent where the quantities on the left are written using the notation
of the $W^{\pm}Z$ production publication \cite{Frixione:1992pj}
and those on the right use that of the $ZZ$ article \cite{Mele:1990bq}.
In the latter work $g_{\mathrm{d},\mathrm{L}}$ and $g_{\mathrm{u,L}}$
denote the left handed couplings of the $Z$ boson to up- and down-type
quarks, $F_{ij}$ is the $W$ boson coupling to quark flavours $i$
and $j$, while $e_{\mathrm{Z}}$ is the trilinear gauge coupling
$\left(e_{\mathrm{Z}}=g_{\mathrm{u,L}}-g_{\mathrm{d,L}}\right)$.
In Ref.\,\cite{Mele:1990bq} $g_{\mathrm{V}}$ and $g_{\mathrm{A}}$
are the vector and axial-vector couplings of the $Z$ bosons to the
colliding quarks. 

We also find that the $W^{\pm}Z$ matrix elements are simply related
to those of $W^{+}W^{-}$ production by

\begin{equation}
\begin{array}{lclclcl}
K_{ij} & \rightarrow & 1\,, & \,\,\,\,\,\,\,\,\,\,\,\,\,\,\,\,\,\,\,\,\,\,\,\,\,\,\, & e_{\mathrm{Z}}^{2} & \rightarrow & \frac{4}{g_{\mathrm{W}}^{2}}\,\left(p^{2}-m_{\mathrm{W}}^{2}\right)^{2}\, c_{i}^{ss}\left(p^{2}\right)\,,\\
g_{\mathrm{i,L}} & \rightarrow & 0\,, & \,\,\,\,\,\,\,\,\,\,\,\,\,\,\,\,\,\,\,\,\,\,\,\,\,\,\, & e_{\mathrm{Z}} & \rightarrow & -\frac{\sqrt{2}}{g_{\mathrm{W}}}\,\frac{4}{g_{\mathrm{W}}^{2}}\,\left(p^{2}-m_{\mathrm{W}}^{2}\right)\, c_{i}^{ts}\left(p^{2}\right)\,,\\
g_{\not\mathrm{i},\mathrm{L}} & \rightarrow & \frac{g_{\mathrm{W}}}{\sqrt{2}}\,, & \,\,\,\,\,\,\,\,\,\,\,\,\,\,\,\,\,\,\,\,\,\,\,\,\,\,\, & m_{\mathrm{Z}} & \rightarrow & m_{\mathrm{W}}\,,\end{array}\label{eq:WZ_to_WW_replacements}\end{equation}
where on the left $K_{ij}$ denotes the relevant Cabibbo-Kobayashi-Maskawa
matrix element in $W^{\pm}Z$ production, while on the right $i$
is used to refer to the flavour of the colliding quarks (up- or down-type).
The functions $c_{i}^{ss}\left(p^{2}\right)$ and $c_{i}^{ts}\left(p^{2}\right)$
are the coefficients of those parts of the $W^{+}W^{-}$ matrix elements
corresponding to s-channel trilinear gauge-boson graphs interfering
with themselves and with t-channel graphs respectively.

The validity of these relations was examined analytically using Mathematica%
\footnote{A Mathematica file detailing these checks for all cross section formulae
in Refs.\,\cite{Mele:1990bq,Frixione:1992pj,Frixione:1993yp} is
available from the author on request. These checks also reveal that,
in both cases, all matrix elements satisfy additional symmetries involving
the momenta of the final state vector bosons \emph{before} the last
transformation, $m_{\mathrm{W}}\rightarrow m_{\mathrm{Z}}$ / $m_{\mathrm{Z}}\rightarrow m_{\mathrm{W}}$,
is carried out; further exploration of this point is beyond the scope
of this work.%
}. In our simulation we have employed the matrix elements of Ref.\,\cite{Frixione:1992pj}
for $W^{\pm}Z$ production and applied the transformations in Eqs.\,\ref{eq:WZ_to_ZZ_replacements}
and \ref{eq:WZ_to_WW_replacements} to these when generating $ $$ZZ$
and $W^{+}W^{-}$ production events respectively. The correctness
of relations Eqs.\,\ref{eq:WZ_to_ZZ_replacements} and \ref{eq:WZ_to_WW_replacements}
at NLO is tested again, numerically, by comparing to alternative calculations
in \noun{Mcfm} and \noun{Mc}{\footnotesize @}\noun{nlo}. 

We attribute the fact that these relations hold at the NLO level to
the deceptively simple Dirac structure in the real and virtual corrections.
In the beginnings of Refs.\,\cite{Mele:1990bq,Frixione:1992pj,Frixione:1993yp}
it is observed that, for both sets of radiative corrections, all of
the Dirac traces can be expressed in the form $\mathrm{Tr}\left[\mbox{\ensuremath{\left(a+b\gamma_{5}\right)\Gamma}}\right]$,
where $a$ and $b$ are constants and $\Gamma$ is an arbitrary string
of $\gamma$ matrices, excluding $\gamma_{5}$. In the case of the
virtual corrections the $\gamma_{5}$ term never contributes to the
final trace: it leads to a term proportional to the antisymmetric
Levi-Civita tensor, for which there are not enough linearly independent
momenta available to give rise to a non-vanishing contribution. Similarly,
in the case of the real corrections this term gives a purely imaginary
contribution, where a real part must be taken. In fact, it is remarked
in Refs.\,\cite{Frixione:1992pj,Frixione:1993yp} that the NLO $W^{\pm}Z$
and $W^{+}W^{-}$ cross sections could be considered to have arisen
from a fictious theory, with suitably redefined vector couplings and
no axial ones. 

Although the results in \cite{Mele:1990bq,Frixione:1992pj,Frixione:1993yp}
were updated in Ref.\,\cite{Dixon:1998py,Dixon:1999di}, at the end
of the event generation process we simulate the decays of the vector
bosons according to either the full $2\rightarrow4$ or $2\rightarrow5$
tree order matrix elements, using basically the same procedure as
is adopted in several other \noun{Mc}{\footnotesize @}\noun{nlo }and
\noun{Powheg} simulations \cite{Frixione:2005vw,Frixione:2006gn,Frixione:2007zp,Frixione:2007nw,Alioli:2009je,Re:2010bp}.
The $2\rightarrow4$ and $2\rightarrow5$ particle helicity amplitudes
were constructed using the C++ libraries present in ThePEG event generator
tool kit \cite{Lonnblad:2006pt}, based on the \noun{Helas} formalism
\cite{Murayama:1992gi}; in practice we have implemented these matrix
elements as a convolution of the $2\rightarrow2$ and $2\rightarrow3$
particle production spin density matrices with $1\rightarrow2$ particle
decay matrices (Sect.~\ref{sub:decays}). We have checked that the
corresponding $2\rightarrow2$ and $2\rightarrow3$ `undecayed' matrix
elements reproduce the results obtained with those in Refs.\,\cite{Mele:1990bq,Frixione:1992pj,Frixione:1993yp},
however, since the latter are considerably faster to evaluate we refrain
from using them until the last step of the hardest emission event
generation \emph{i.e.} the vector boson decays. Further details of
the decay simulation procedure are deferred to Section~\ref{sec:Implementation}.

\section{Event generation\label{sec:Implementation}}

In this section we describe the steps by which the event simulation
is carried out in practice: the generation of the hardest emission
about a $2\rightarrow2$ \emph{underlying} \emph{Born} configuration,
the decays of the vector boson pairs according to $2\rightarrow5$
particle matrix elements, and the subsequent parton showering.

\subsection{Hardest radiation generation\label{sub:production}}

The first step in the simulation process involves generating a set
of Born variables, $\Phi_{B}$, and hence a $2\rightarrow2$ kinematic
configuration, according to the $\overline{B}\left(\Phi_{B}\right)$
function (Eqs.\,\ref{eq:powheg_2} and \ref{eq:nlo_2_13}). The $\overline{B}\left(\Phi_{B}\right)$
function is numerically implemented directly according to the formulae
given in Section~\ref{sec:Next-to-leading-order-cross}, having applied
the transformation in Eq.\,\ref{eq:nlo_1_15} to $x$. To this end
we generate a set of Born and radiative variables $\left\{ \Phi_{B},\Phi_{R}\right\} $
by sampling Eq.\,\ref{eq:nlo_2_13} using a VEGAS based algorithm
\cite{Lonnblad:2006pt}, the $\Phi_{R}$ are then simply discarded,
leaving $\Phi_{B}$ distributed according to the integral\emph{ }with
respect to $\Phi_{R}$, in other words $\overline{B}\left(\Phi_{B}\right)$. 

With $\Phi_{B}$ in hand we proceed to generate the hardest radiation
according to the square bracketed term in Eq.\,\ref{eq:powheg_1}.
The exponent in the \noun{Powheg} Sudakov form factor Eq.\,\ref{eq:powheg_3}
consists of an integral over a sum of terms, \begin{equation}
\frac{\widehat{R}_{ab}\left(\Phi_{B},\Phi_{R}\right)}{B\left(\Phi_{B}\right)}=\frac{\alpha_{S}}{2\pi}\,\frac{1}{x}\,\widehat{\mathcal{H}}_{ab}\,\widehat{\mathcal{L}}_{ab}\left(x_{\oplus},x_{\ominus}\right)\,,\label{eq:hardest_1}\end{equation}
one for each real emission process, $\widehat{\mathcal{H}}_{ab}$
being equal to $\mathcal{H}_{ab}$ with the plus and $\rho$ regularization
prescriptions omitted. Here, in implementing the generation of $\Phi_{R}$,
we have opted not to generate $x$ and $y$ directly but rather we
re-express the $\mathrm{d}\Phi_{R}$ integration measure in terms
of $p_{T}$ and $\mathrm{y}_{k}$, where $\mathrm{y}_{k}$ is rapidity
of the hardest emission in the hadronic centre-of-mass system. Making
this change of variables removes the slightly awkward $\theta$-function
in Eq.\,\ref{eq:powheg_3}, replacing it by a lower bound on the
integration over $p_{T}$. The distribution of the radiative variables
\emph{viz. }the square bracketed term in Eq.\,\ref{eq:powheg_1},
is then sampled using \emph{the} \emph{veto algorithm} \cite{Sjostrand:2006za}.

With the resulting set of Born and radiative variables $\left\{ \Phi_{B},\Phi_{R}\right\} $
we may then fully reconstruct the kinematics of the $2\rightarrow3$
hardest emission events directly using the general formulae given
in Ref.\,\cite{Frixione:1992pj}. Alternatively in the rare event
that $p_{T}\le k_{T,min}$, where in this work we have chosen $k_{T,min}=2\,\mathrm{GeV}$,
the emission is considered to be unresolvable, in which case only
the $2\rightarrow2$ kinematics are reconstructed before proceeding
to the next step of the event generation.

Finally we note that in generating $\Phi_{R}$ for the hardest emission
we have used $p_{T}$ as the factorization scale in the PDFs and the
renormalization scale in the strong coupling constant, in accordance
with the DDT formulation of the Sudakov form factor \cite{Dokshitzer:1978hw,Nason:2006hfa}.
This completes the generation of the hardest emission kinematics according
to Equation \ref{eq:powheg_1}.

\subsection{Spin correlations and vector boson decays\label{sub:decays}}

Having generated a set of $2\rightarrow3$ kinematics we now calculate
and store the production spin density matrix, $\mathcal{M}_{\lambda_{1}\bar{\lambda}_{1}\lambda_{2}\bar{\lambda}_{2}}^{R}$,
from the associated tree order helicity amplitudes, where the pairs
of indices $\left\{ \lambda_{1},\lambda_{2}\right\} $ and $\left\{ \bar{\lambda}_{1},\bar{\lambda}_{2}\right\} $
label the helicities of each vector boson in the amplitudes and conjugate
amplitudes respectively. These helicity amplitudes were constructed
using the C++ libraries present in ThePEG event generator tool kit
\cite{Lonnblad:2006pt}, based on the \noun{Helas} formalism \cite{Murayama:1992gi}.
A set of two-body decay kinematics are then generated for each vector
boson, isotropic in their rest frames, from which corresponding decay
matrices $\mathcal{M}_{\lambda_{1}\bar{\lambda}_{1}}^{D_{1}}$ and
$\mathcal{M}_{\lambda_{2}\bar{\lambda}_{2}}^{D_{2}}$ are calculated
and contracted with the production matrix element. The decay kinematics
are then kept provided \begin{equation}
\mathcal{R}\,\,\le\,\,\frac{\mathcal{M}_{\lambda_{1}\bar{\lambda}_{1}\lambda_{2}\bar{\lambda}_{2}}^{R}\mathcal{M}_{\lambda_{1}\bar{\lambda}_{1}}^{D_{1}}\mathcal{M}_{\lambda_{2}\bar{\lambda}_{2}}^{D_{2}}}{\mathcal{M}^{R}\,\,\mathcal{M}^{D_{1}}\,\,\mathcal{M}^{D_{2}}}\,,\label{eq:decay_algorithm}\end{equation}
where $\mathcal{R}$ is a random number in the range $\left[0,1\right]$,
while $\mathcal{M}^{D_{1}}$ and $\mathcal{M}^{D_{2}}$ correspond
to the traces of the decay matrices. If the decay kinematics are rejected
the process is repeated, using newly generated sets of momenta for
the decay products, until the inequality in Eq.\,\ref{eq:decay_algorithm}
is satisfied. By generating the decays of the vector bosons in this
way, breaking the $2\rightarrow5$ process down into a $2\rightarrow3$
process followed by two-body decays, we avoid the more intensive operation
of computing the full $2\rightarrow5$ body helicity amplitudes and
summing over helicity amplitudes repeatedly. Lastly, we note that
for the tiny fraction of events in which $p_{T}\le k_{T,min}$ the
same procedure is applied to generate the decay kinematics, albeit
using the helicity amplitudes of the $2\rightarrow2$ leading order
process to compute the production spin density matrix as opposed to
those of the $2\rightarrow3$ processes.

\subsection{Truncated and vetoed parton showers\label{sub:Truncated-and-vetoed}}

In order to shower the hardest emission configurations, we first compute
a set of parton shower branching kinematics, $\Phi_{R}^{\mathrm{HW++}}$,
corresponding to the hard radiation. More specifically, we precisely
determine the \noun{Herwig++} branching variables \cite{Gieseke:2003rz}
which would have been assigned to \emph{exactly }this momentum configuration
had it been generated by initiating the parton shower from the underlying
Born configuration%
\footnote{ For technical details and formulae pertaining to this \emph{inverse
reshuffling} procedure we refer the reader to Refs.\,\cite{Hamilton:2008pd,Hamilton:2009ne}. %
}. Having ascertained how the hardest emission event may be reproduced
by the usual parton shower apparatus, we return to the underlying
Born configuration and proceed as follows:
\begin{itemize}
\item the external leg deemed to have produced the hardest emission is evolved
from the default shower starting scale to that of $\Phi_{R}^{\mathrm{HW++}}$,
with the imposition that intervening branchings conserve flavour and
have transverse momenta less than $p_{T}$: the \emph{truncated shower}
\cite{Nason:2004rx}.
\item the set of branching parameters $\Phi_{R}^{\mathrm{HW++}}$ is then
inserted in this shower.
\item the evolution continues down to the cut-off scale, vetoing any emissions
with transverse momenta greater than $p_{T}$: the \emph{vetoed shower}.
\item the non-emitting leg is evolved from the default shower starting scale
down to the cut-off scale with a further vetoed shower.
\end{itemize}
In the event that the hardest emission occurs in a region of phase
space inaccessible to the parton shower, \emph{i.e.} the wide angle
/ high $p_{T}$ \emph{dead zone} \cite{Gieseke:2003rz}, subsequent
emissions will have sufficient resolving power to \emph{see} the widely
separated emitters individually. It follows that no truncated shower
is then required, since this models coherent, large angle emission
from more collimated configurations of partons, and so we proceed
directly to the vetoed shower.

\section{Results\label{sec:Results}}

In the following we present predictions from our \noun{Powheg} simulations
in comparison with results obtained by independent calculations and
alternative methods. The main aim of this work is to provide a robust
validation of the simulations, such that they may be considered suitable
for use in real physics analyses.

All of the attendant tests have been carried out at nominal Tevatron
and LHC centre-of-mass energies, respectively, 1.96 and 14 TeV. In
\emph{all} programs we have elected to use the MRST2002 NLO PDF set
\cite{Martin:2002aw} interfaced through the LHAPDF package \cite{Whalley:2005nh}.
To analyze jet structure in the events we have used the $k_{\perp}$-jet
measure with the $R$ parameter set to $0.7$ \cite{Ellis:1993tq,Catani:1993hr},
as implemented in the \noun{FastJet} jet finder package \cite{Cacciari:2005hq},
to carry out the associated clustering.

\pagebreak
\subsection{Inclusive observables\label{sub:Inclusive_observables}}

In order to check the calculation of the \noun{Powheg} differential
cross section and $\overline{B}\left(\Phi_{B}\right)$ functions,
Eqs.\,\ref{eq:powheg_1}-\ref{eq:powheg_2}, we have compared our
predictions for total cross sections and numerous inclusive observables
against those of the NLO Monte Carlo calculator \noun{Mcfm} \cite{Campbell:1999ah,Campbell:2000bg,Campbell:2006xx}.
Since \noun{Mcfm} computes these quantities in fixed order perturbation
theory and since we wish to test the various components of the simulation
systematically, the \noun{Powheg} results shown here have been obtained
at the parton level, prior to showering. In other words, the predictions
from our simulations here solely reflect the implementation of the
hardest emission cross section, Eq.\,\ref{eq:powheg_1}.

To facilitate these investigations we have chosen to work with a fixed
value of 100 GeV for the renormalization and factorization scales
in both \noun{Mcfm} and the $\overline{B}\left(\Phi_{B}\right)$ functions
in the \noun{Powheg} simulations. Again, in order to have a meaningful
comparison with \noun{Mcfm} we have run it in a mode where the narrow
width approximation is assumed, moreover, we have arranged for each
program to use the same vector boson decay channels. 

\begin{figure}[H]
\begin{centering}
\includegraphics[scale=0.29,angle=90]{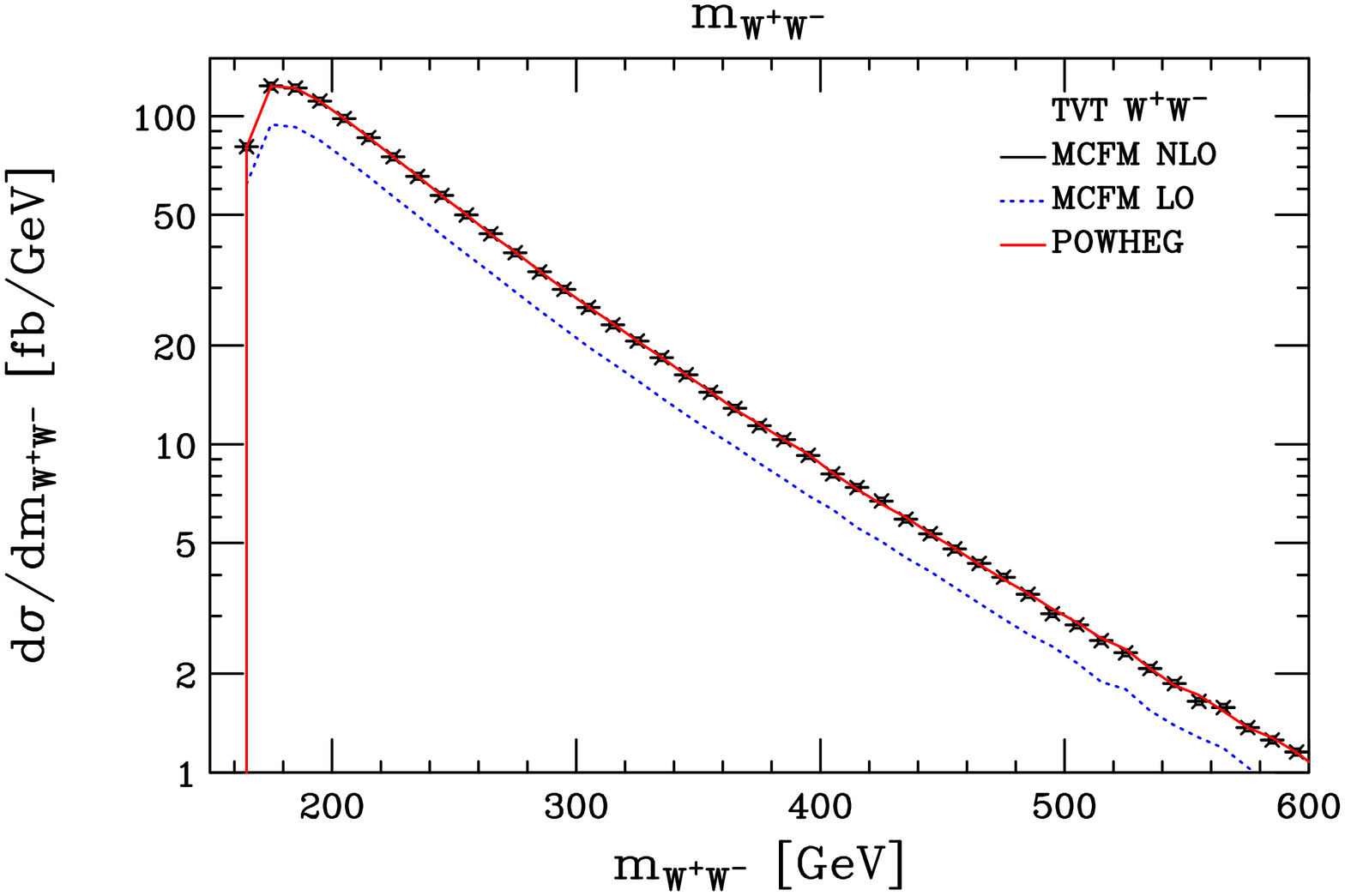}\hfill{}\includegraphics[scale=0.29,angle=90]{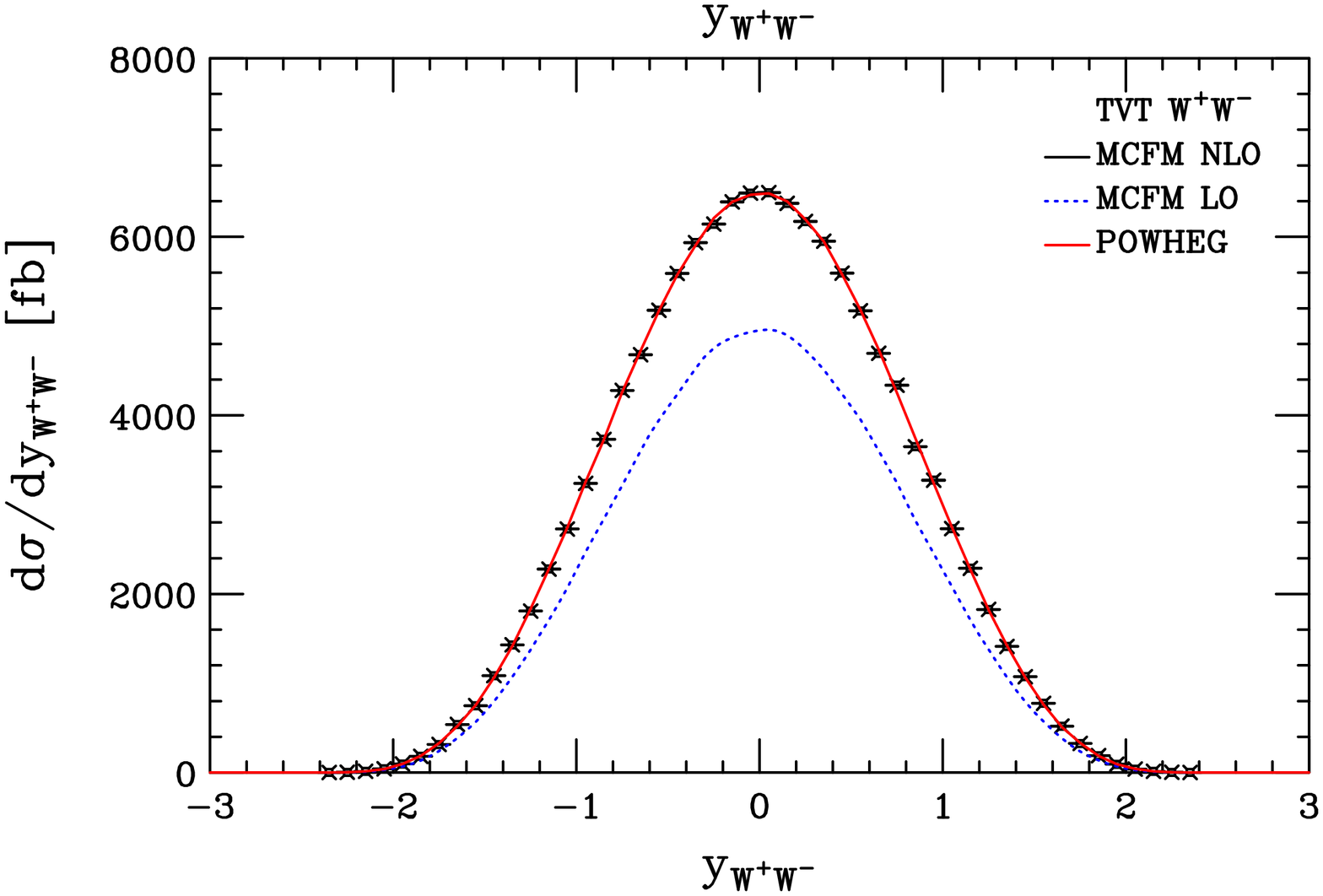}\vspace{7mm}

\par\end{centering}

\begin{centering}
\includegraphics[scale=0.29,angle=90]{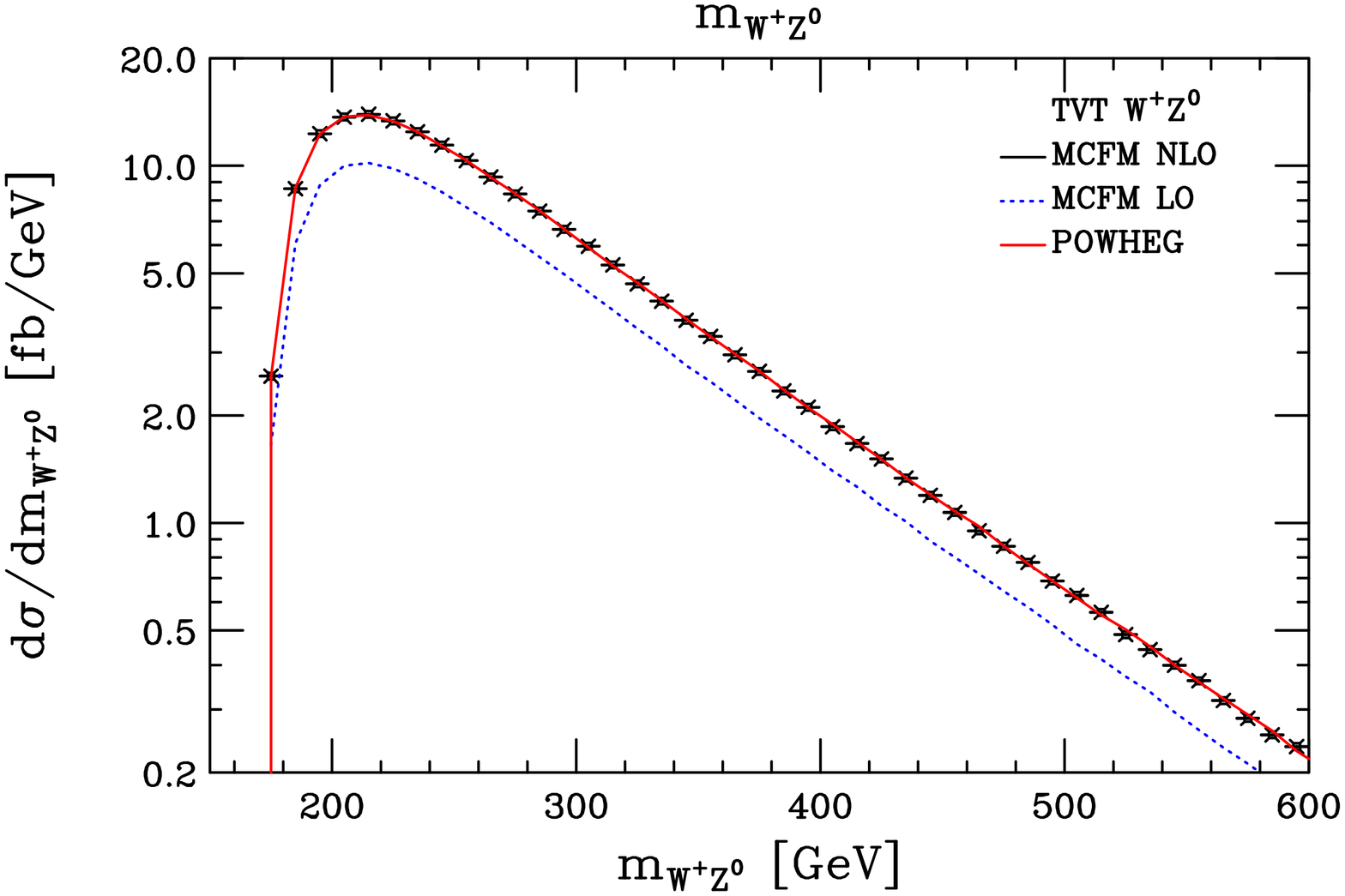}\hfill{}\includegraphics[scale=0.29,angle=90]{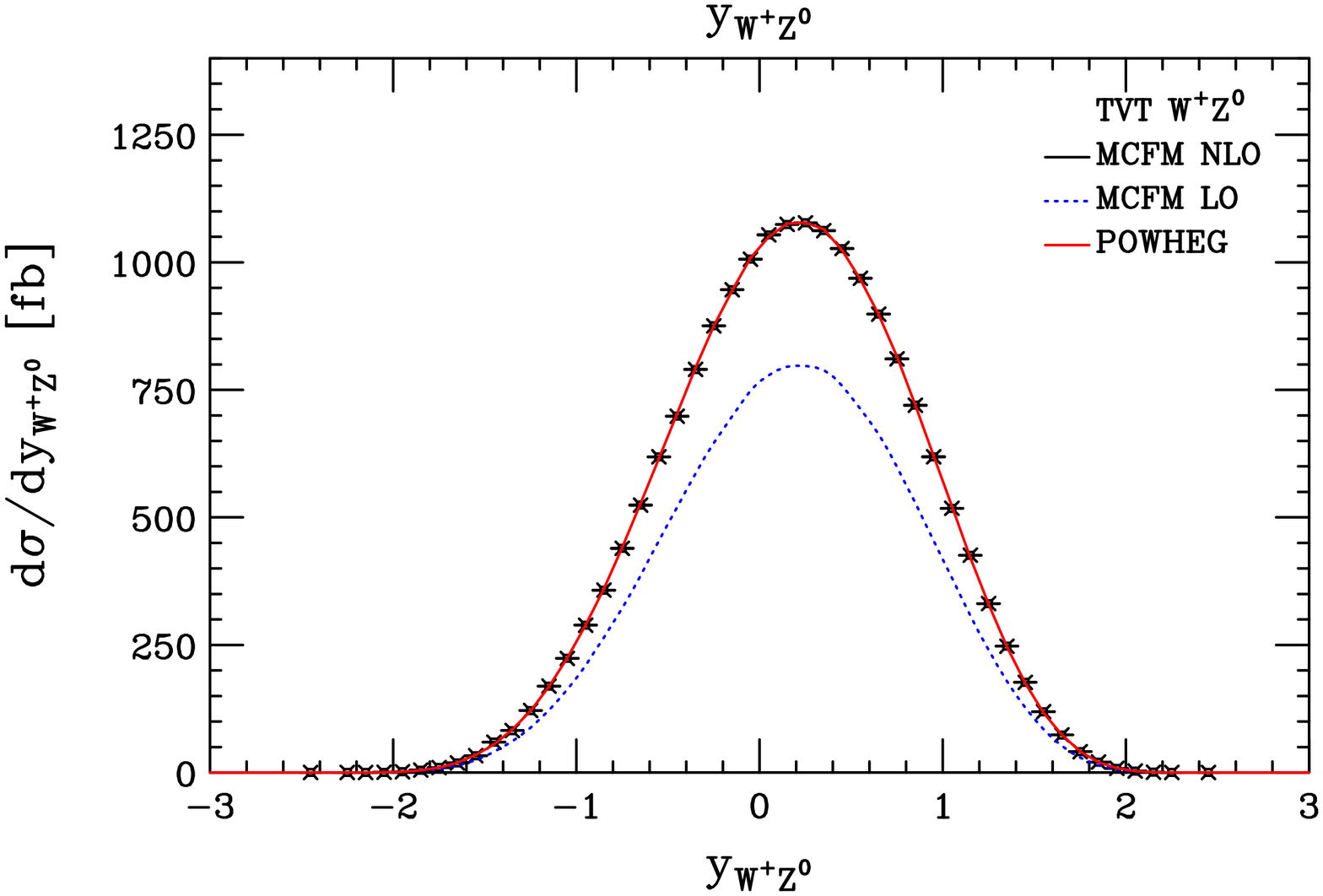}
\par\end{centering}

\begin{centering}
\vspace{7mm}
\includegraphics[scale=0.29,angle=90]{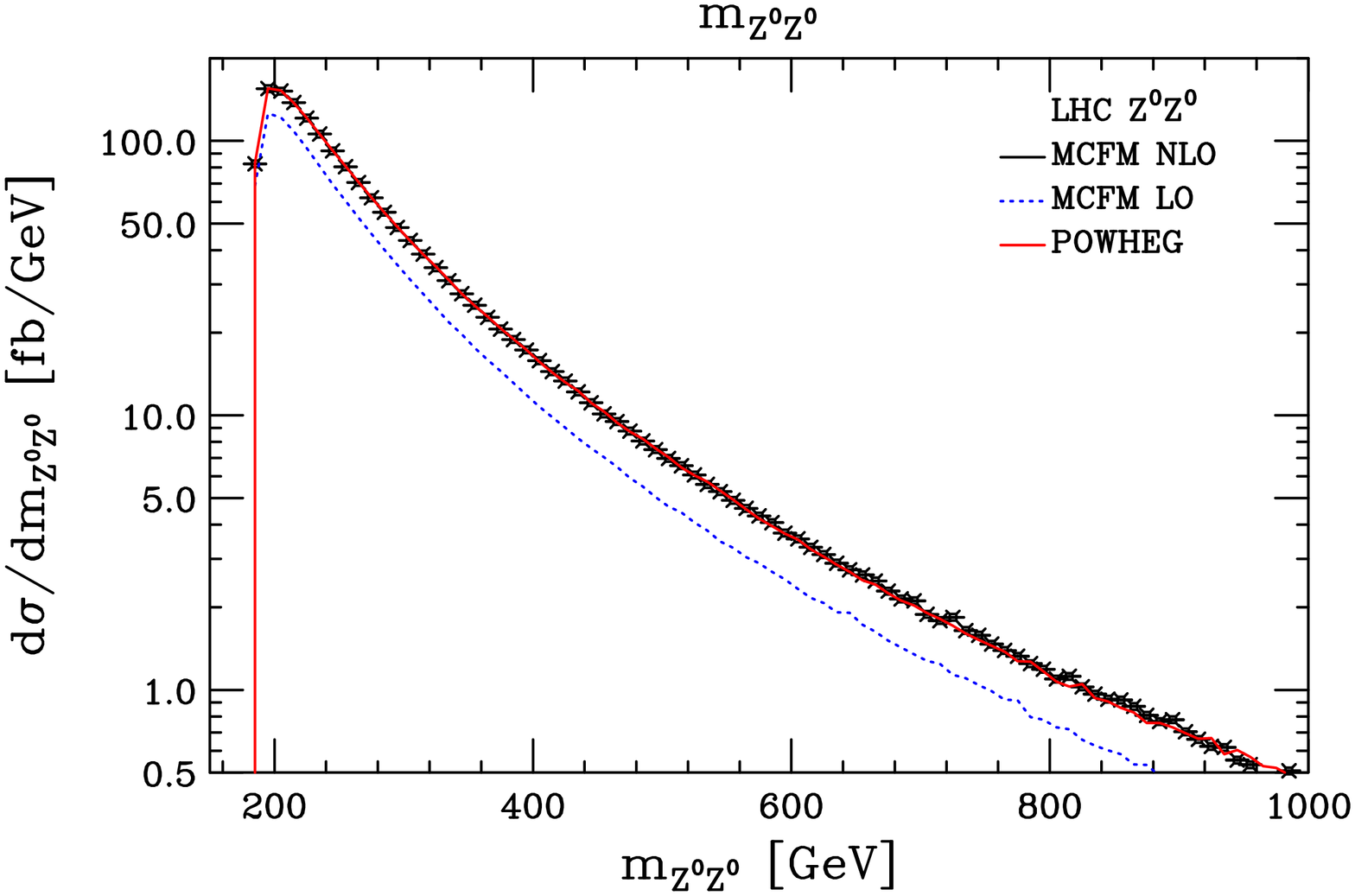}
\hfill{}\includegraphics[scale=0.29,angle=90]{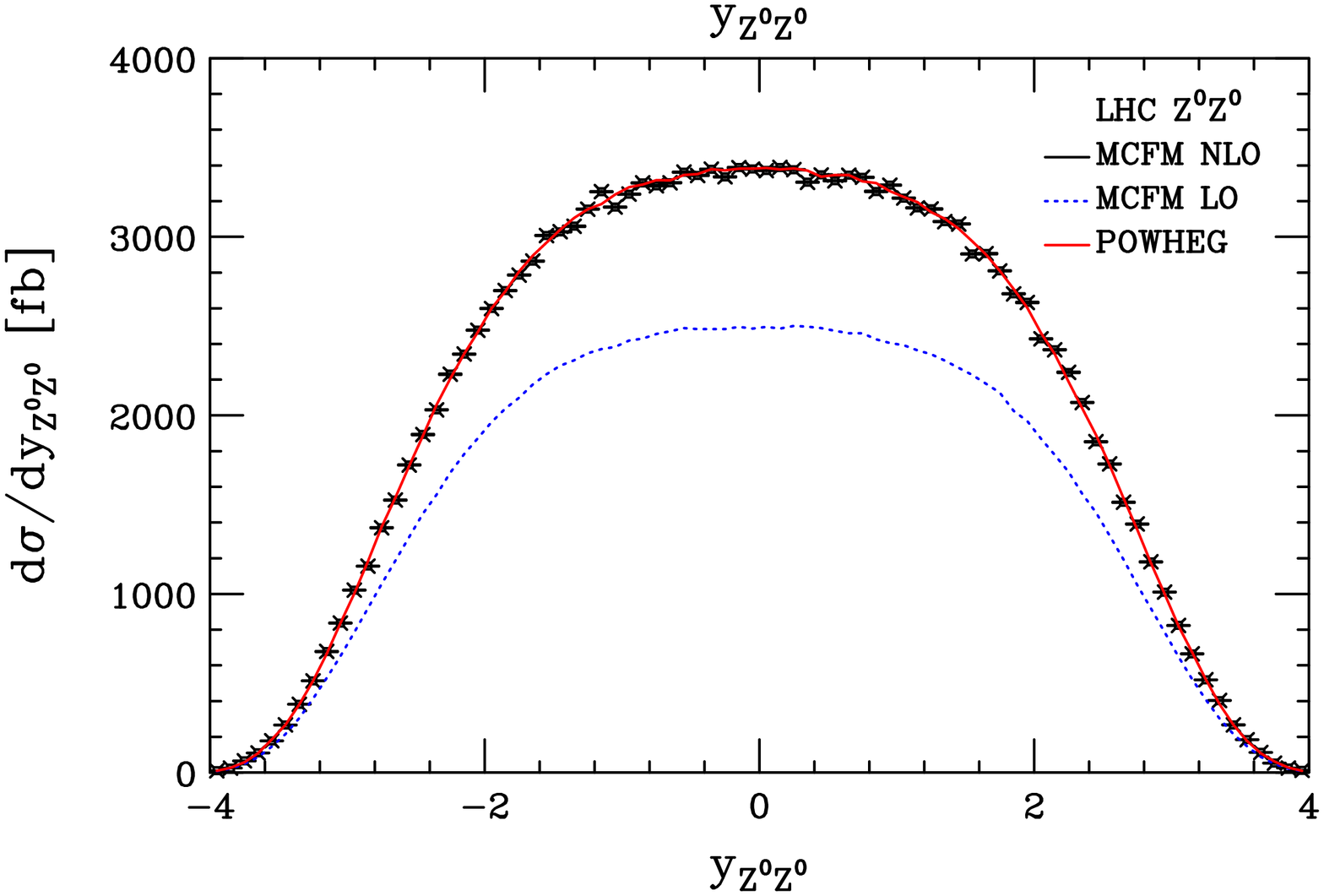}
\par\end{centering}

\begin{centering}
\vspace{7mm}
 \includegraphics[scale=0.29,angle=90]{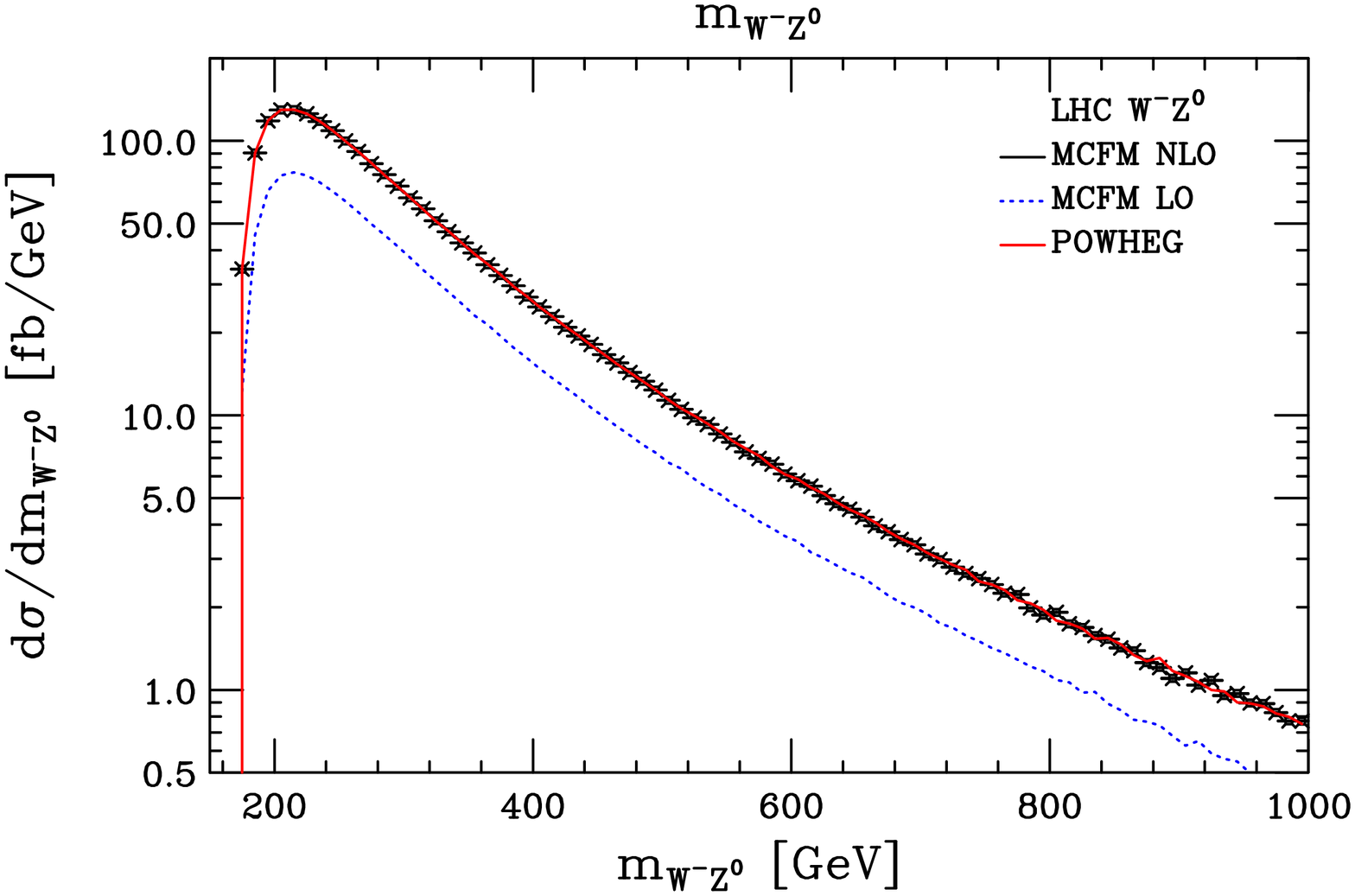}\hfill{}\includegraphics[scale=0.29,angle=90]{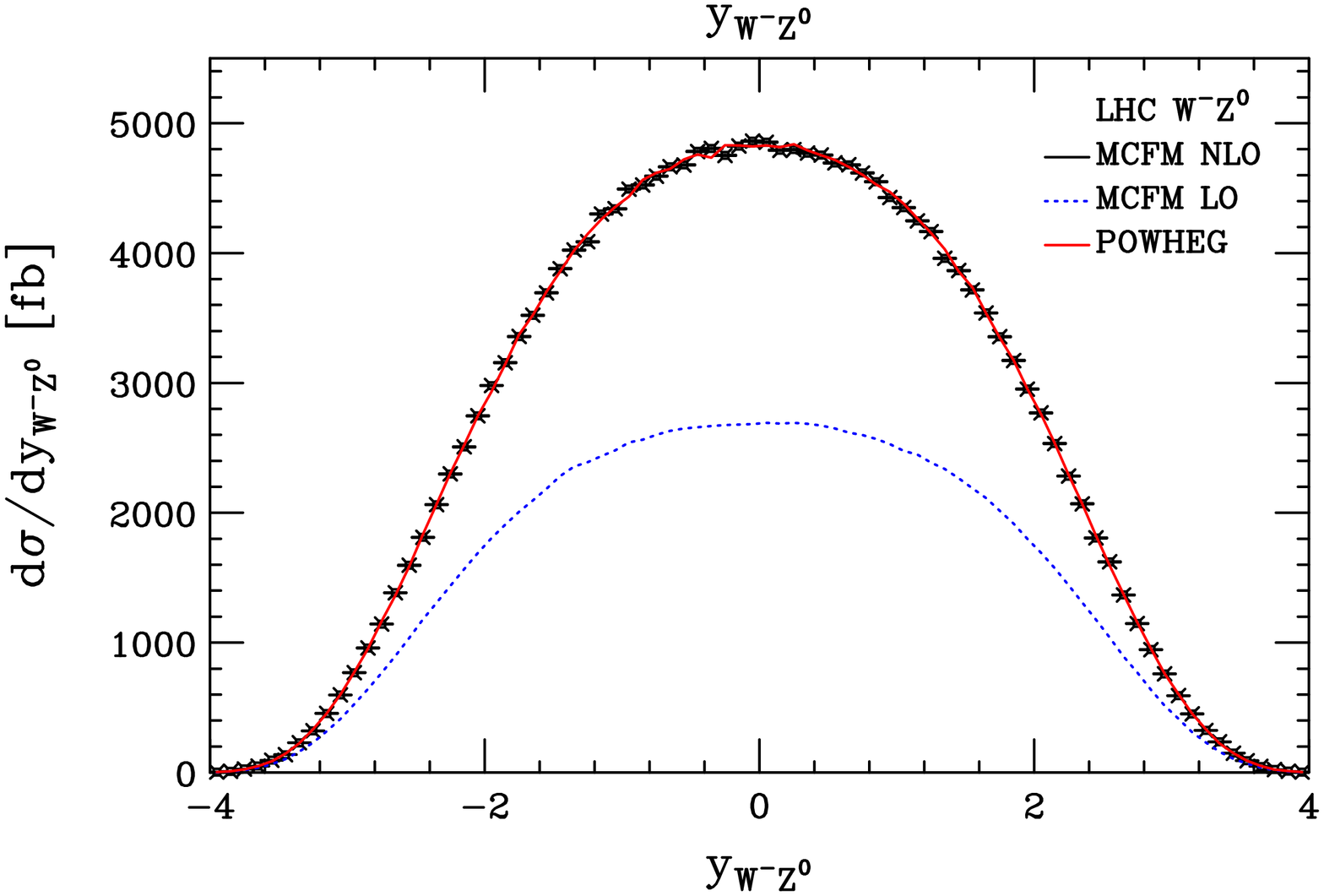}
\par\end{centering}

\caption{In this figure we show predictions for the invariant mass $\left(p^{2}\right)$
and rapidity $\left(\mathrm{y}\right)$ of the vector boson pair system,
in the left- and right-hand columns respectively; the results obtained
using the \noun{Powheg }simulation are shown in red while the blue
dotted line and the black points represent the leading and next-to-leading
order predictions from \noun{Mcfm}. Since $p^{2}$ and $\mathrm{y}$
are Born variables in the \noun{Powheg} simulation, they are distributed
purely according to the $\overline{B}\left(\Phi_{B}\right)$ function,
hence they must follow \emph{exactly }the corresponding NLO prediction
(Sect.~\ref{sub:Kinematics-and-phase}). }

\label{fig:VV_p2_and_y} 
\end{figure}

On the left of Fig.~\ref{fig:VV_p2_and_y} we show the invariant
mass of the weak boson system, $p^{2}$, one of the Born variables
for this process, for which there is excellent agreement between the
\noun{Powheg} result and that of \noun{Mcfm}. Predictions for the
$\mathrm{y}$ Born variable, on the right-hand side of Fig.~\ref{fig:VV_p2_and_y},
are also seemingly indistinguishable from the corresponding \noun{Mcfm}
results. Recall that in the \noun{Powheg }framework the Born variables
are \emph{exactly} preserved in the process of generating the hardest
emission: they are distributed \emph{purely} according to the $\overline{B}\left(\Phi_{B}\right)$
function. Consequently, if the \noun{Powheg }simulations and the underlying
NLO calculations are implemented correctly these Born variables must\emph{
}agree \emph{exactly} with fixed order NLO predictions. The results
in Figure \ref{fig:VV_p2_and_y} already therefore constitute a very
sensitive test of our implementation. 

\begin{figure}[H]
\begin{centering}
\includegraphics[scale=0.31,angle=90]{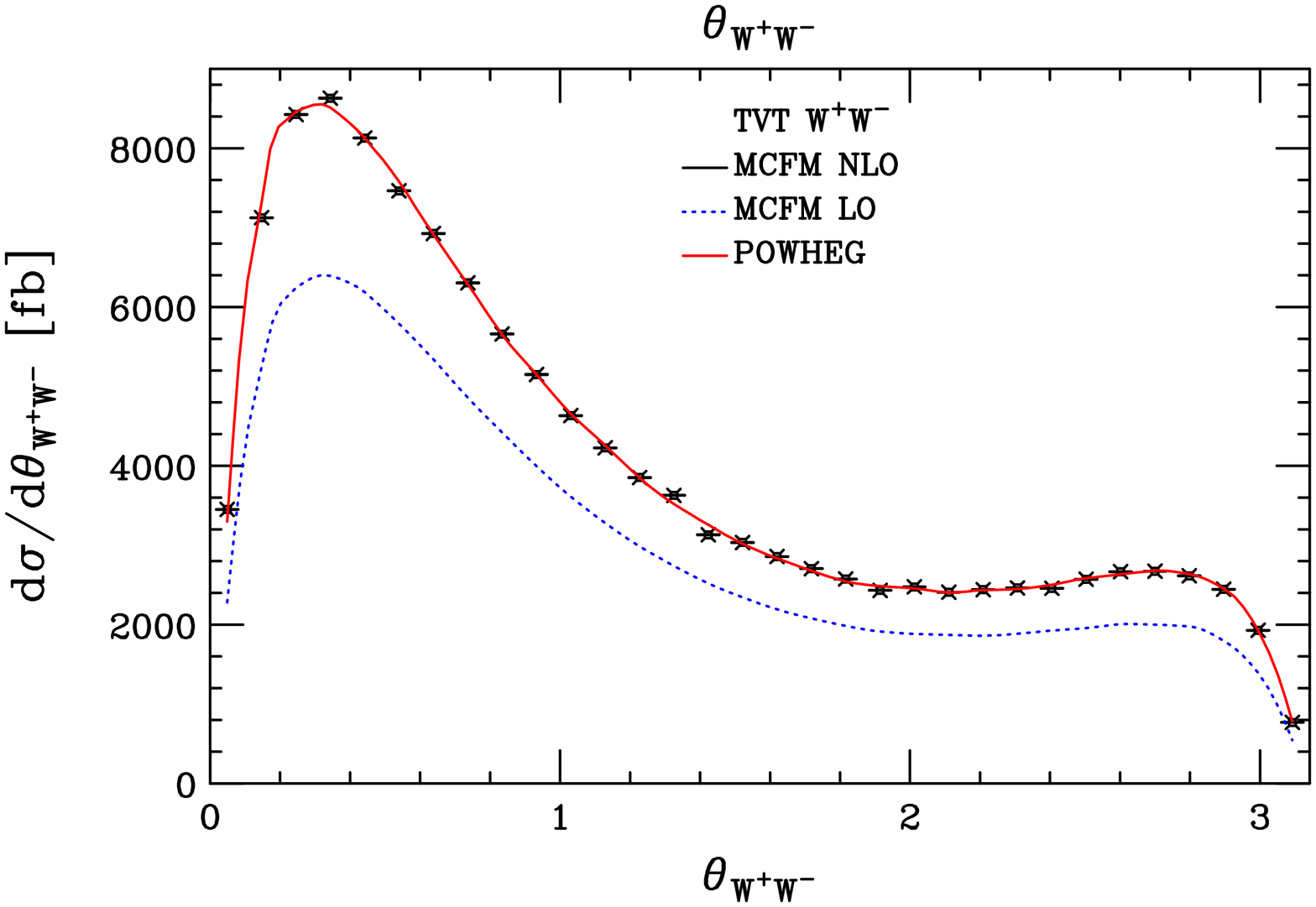}\hfill{}\includegraphics[scale=0.31,angle=90]{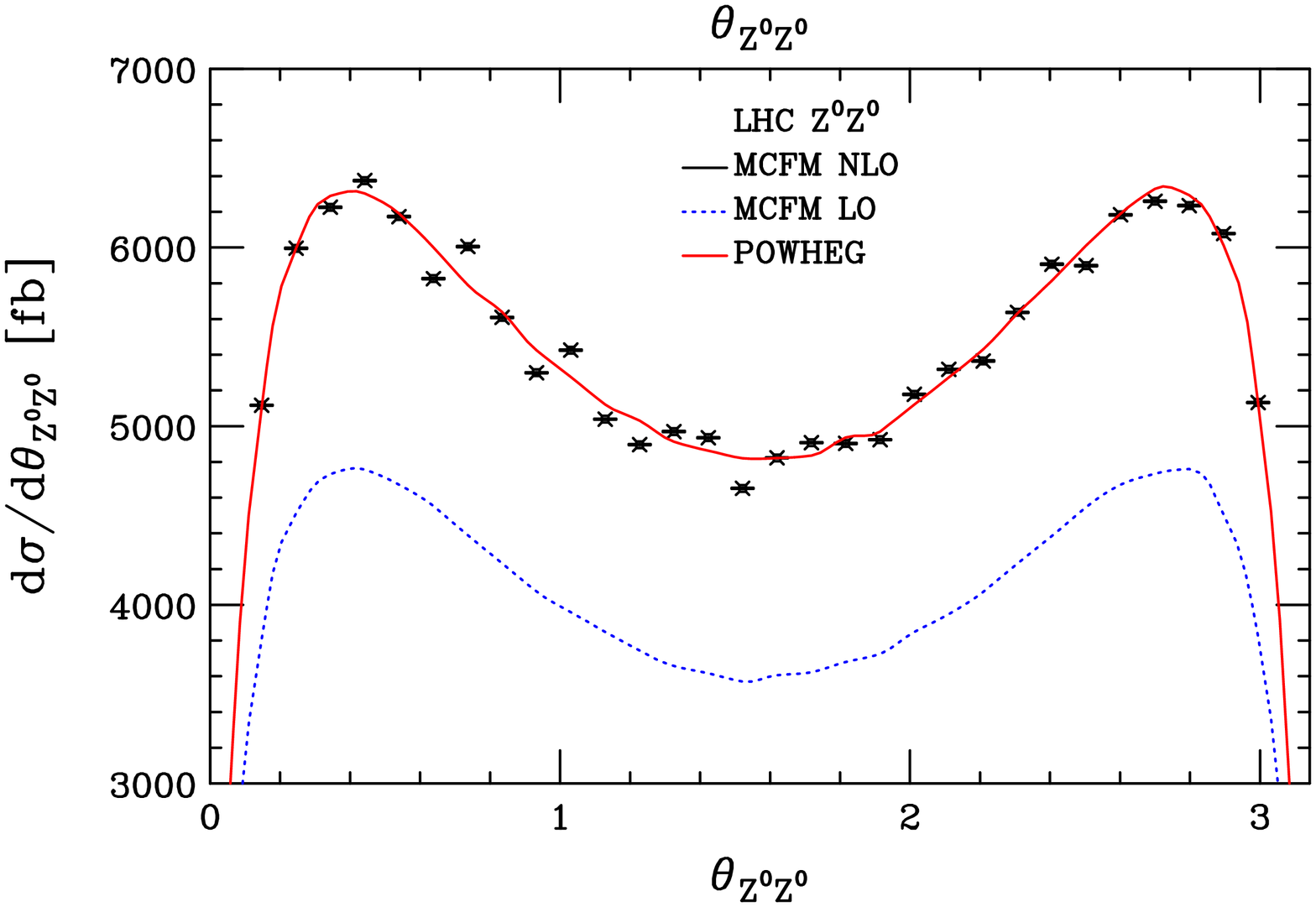}
\par\end{centering}

\begin{centering}
\vspace{5mm}
 \includegraphics[scale=0.31,angle=90]{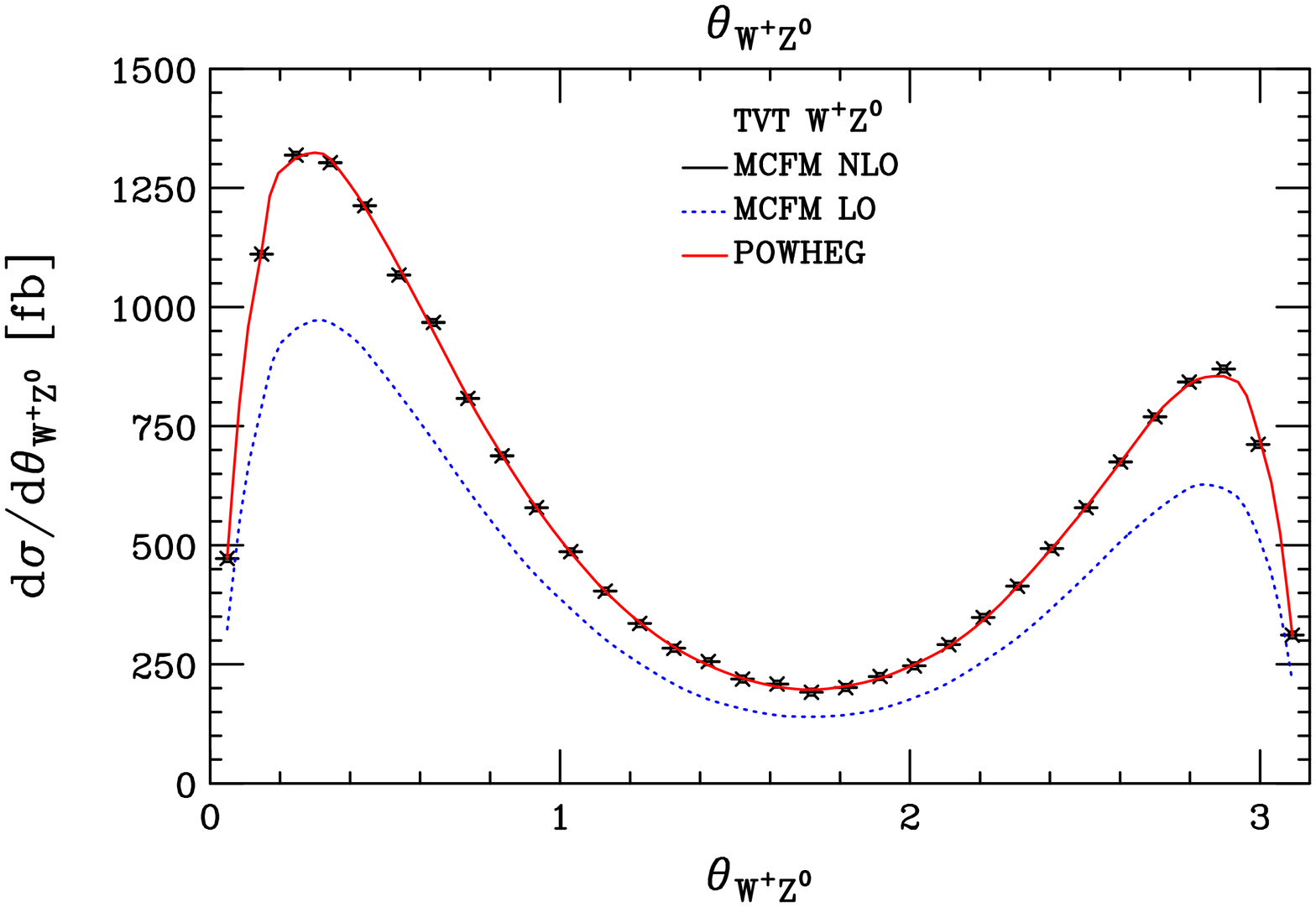}\hfill{}\includegraphics[scale=0.31,angle=90]{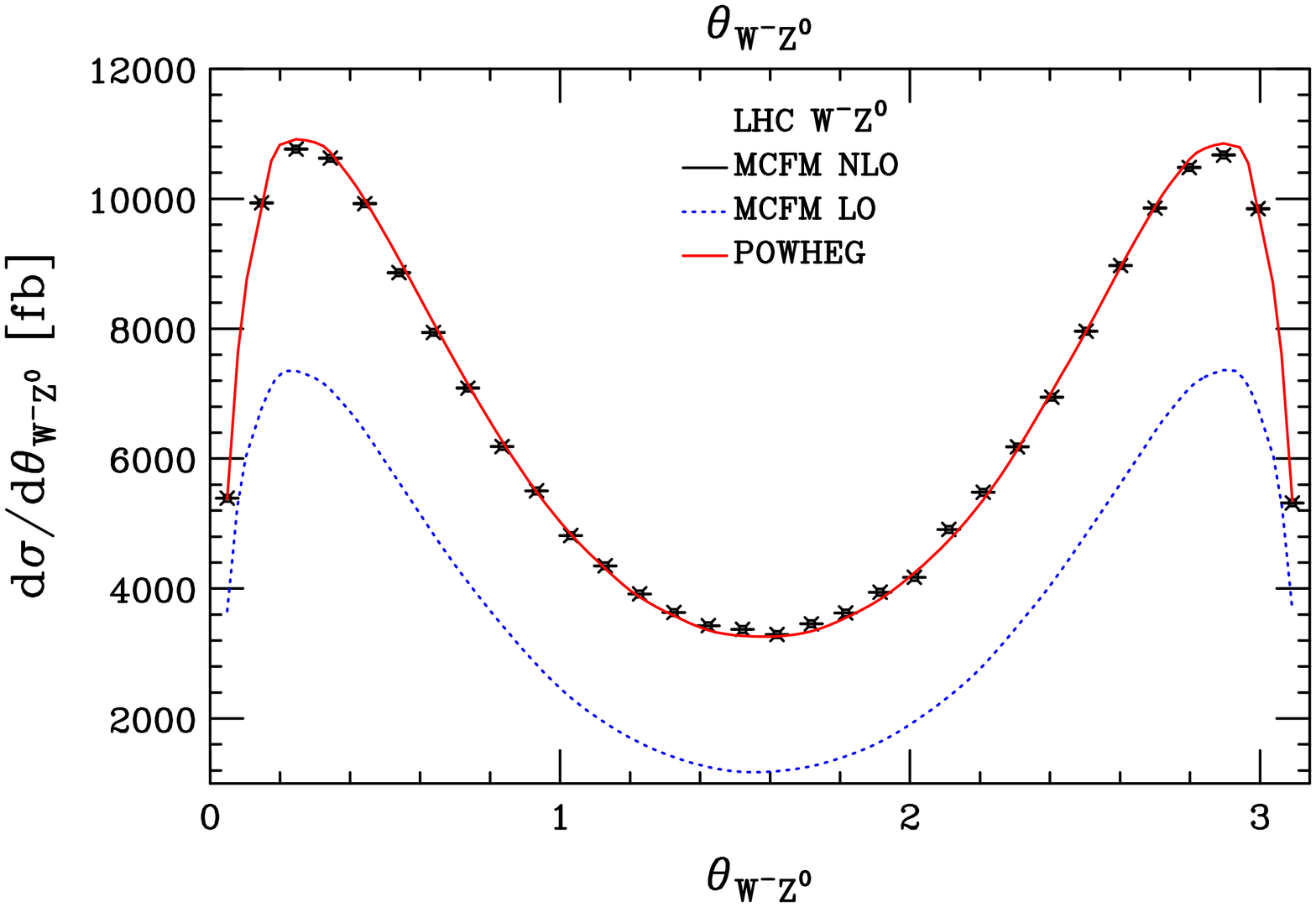}\\

\par\end{centering}

\caption{The polar angle between the incident parton traveling in the $+z$
direction and the first of the produced vector bosons in their rest
frame; in $W^{\pm}Z$ and $W^{+}W^{-}$ production these are taken
to be the $W^{\pm}$ and $W^{+}$ bosons respectively. Since this
variable is fully inclusive and a close relative of the Born variable,
$\theta$, the level of agreement shown here between the \noun{Powheg
}and \noun{Mcfm }predictions provides strong confirmation as to the
correctness of our implementation.}

\label{fig:theta_plot} 
\end{figure}

The distribution of the third Born variable $\theta$ is strongly
reflected in Figure~\ref{fig:theta_plot}, which shows the polar
angle between the incident parton traveling in the $+z$ direction
and the first of the produced vector bosons in their rest frame; in
$W^{\pm}Z$ and $W^{+}W^{-}$ production these are taken to be the
$W^{\pm}$ and $W^{+}$ bosons respectively. This quantity is different
to the $\theta$ Born variable in that the latter is defined as the
polar angle of the first vector boson with respect to the quark in
$q\bar{q}$ and $qg$ collisions, and the gluon in $g\bar{q}$ collisions,
which may or may not be traveling in the $+z$ direction in the lab.
Nevertheless, since this variable is fully inclusive and a close relative
of the Born variable, $\theta$, the level of agreement shown here
between the \noun{Powheg }and \noun{Mcfm }predictions provides further
strong assurance as to the correctness of our implementation. We add
that we have also examined rapidity and pseudorapidity distributions
of the individual vector bosons from our programs (not shown) which,
like those in Figures~\ref{fig:VV_p2_and_y} and \ref{fig:theta_plot},
exhibit no discernible deviations from those given by \noun{Mcfm}. 

\begin{figure}[H]
\begin{centering}
\includegraphics[scale=0.29,angle=90]{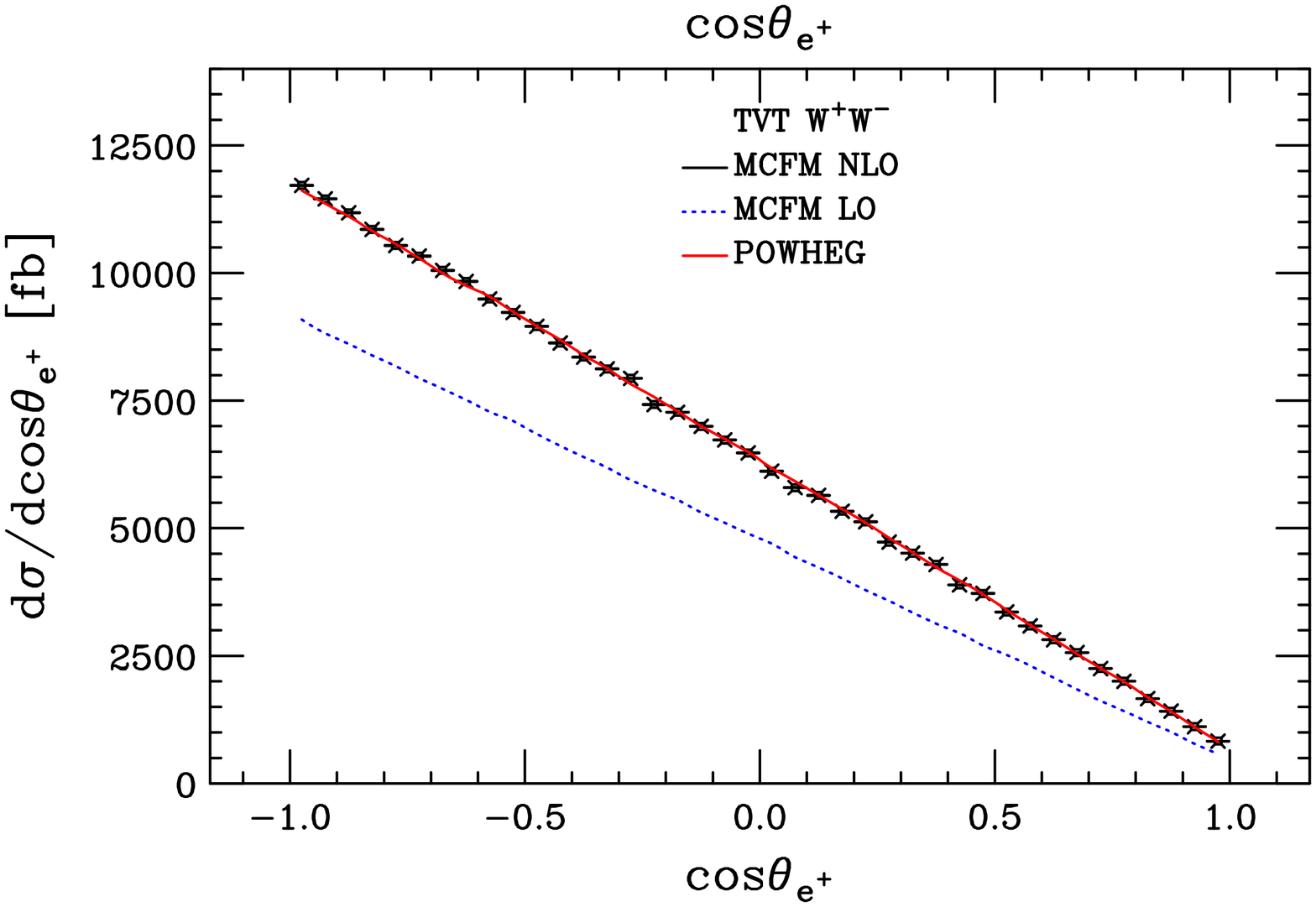}\hfill{}\includegraphics[scale=0.29,angle=90]{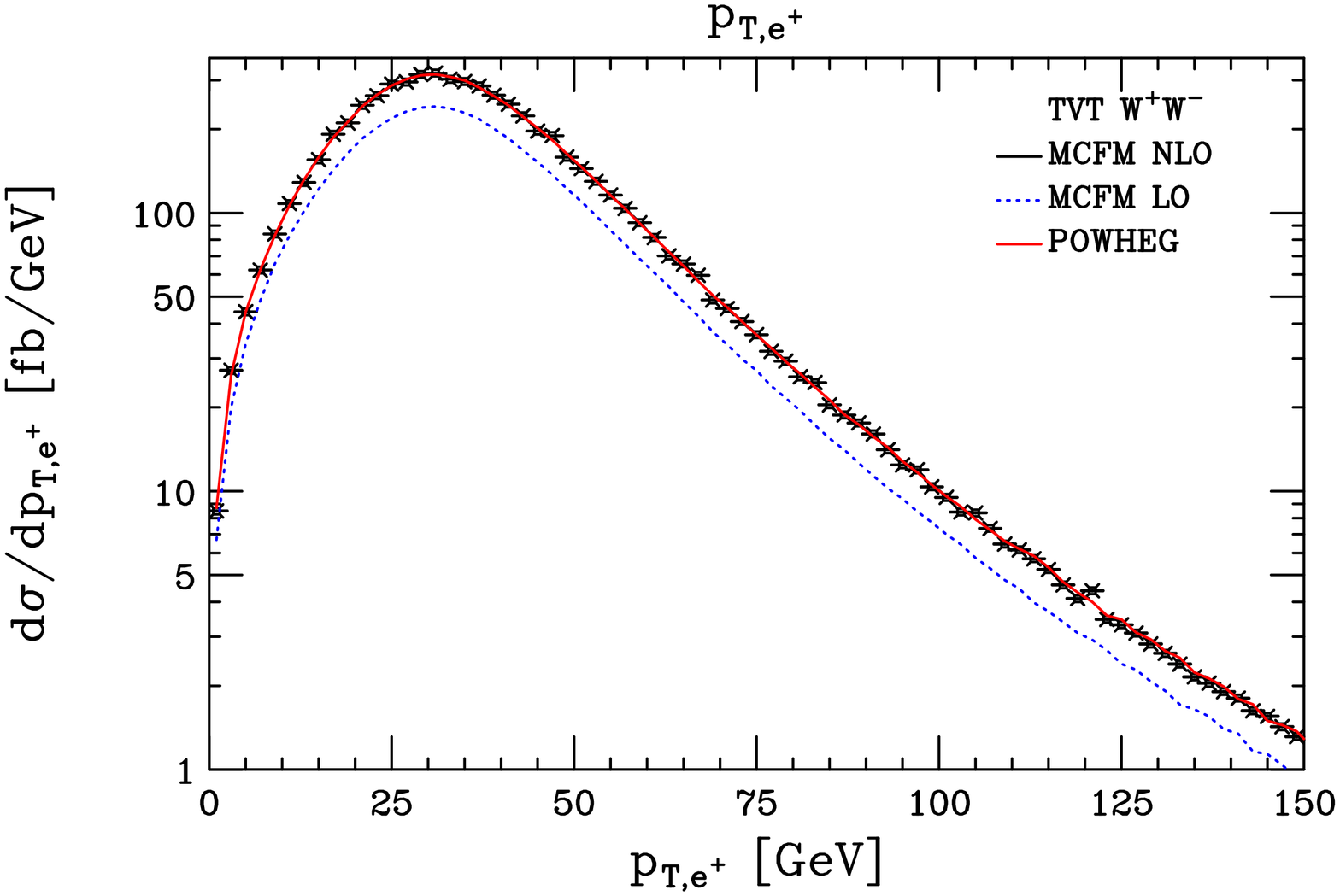}\vspace{7mm}
 \includegraphics[scale=0.29,angle=90]{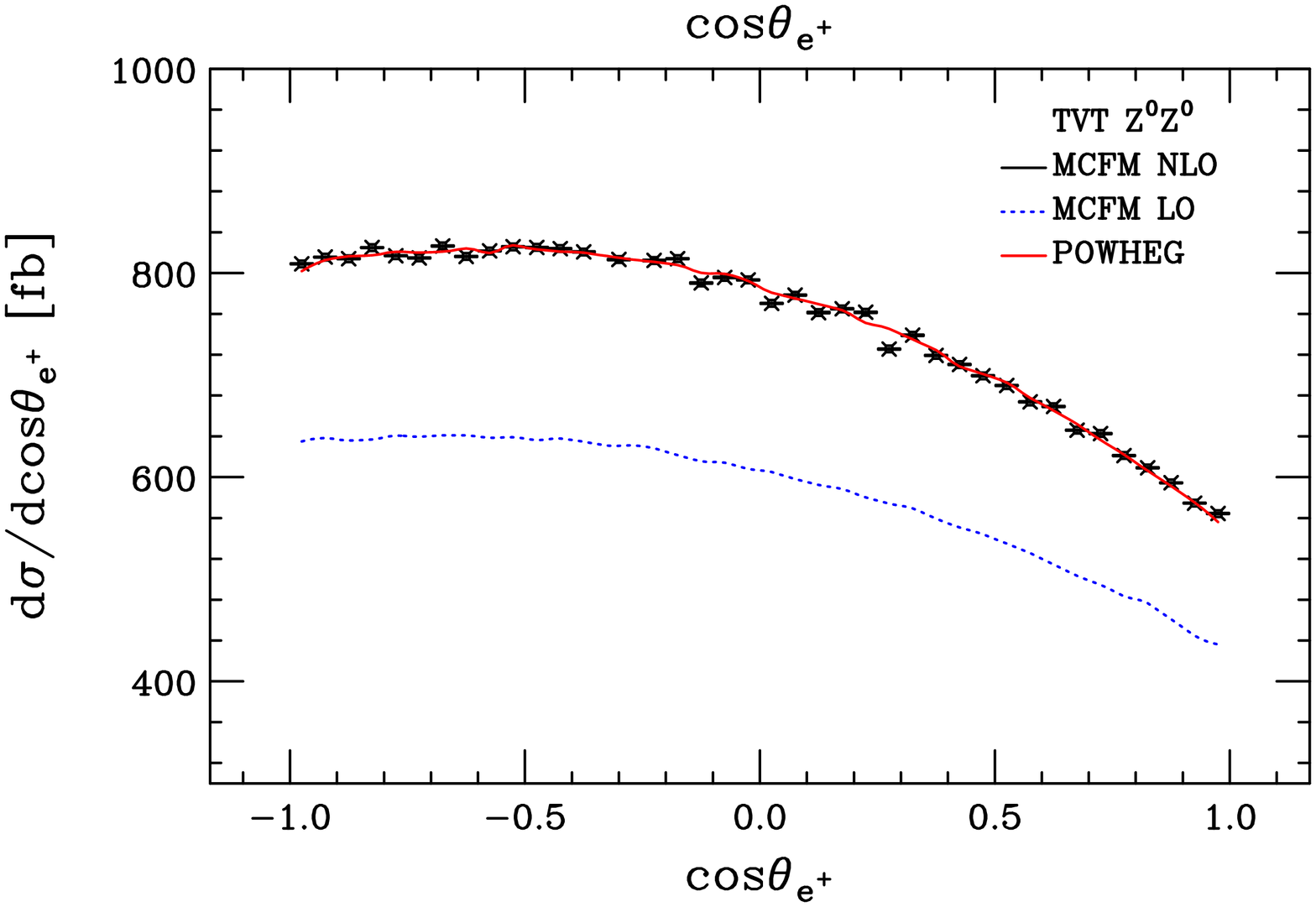}\hfill{}\includegraphics[scale=0.29,angle=90]{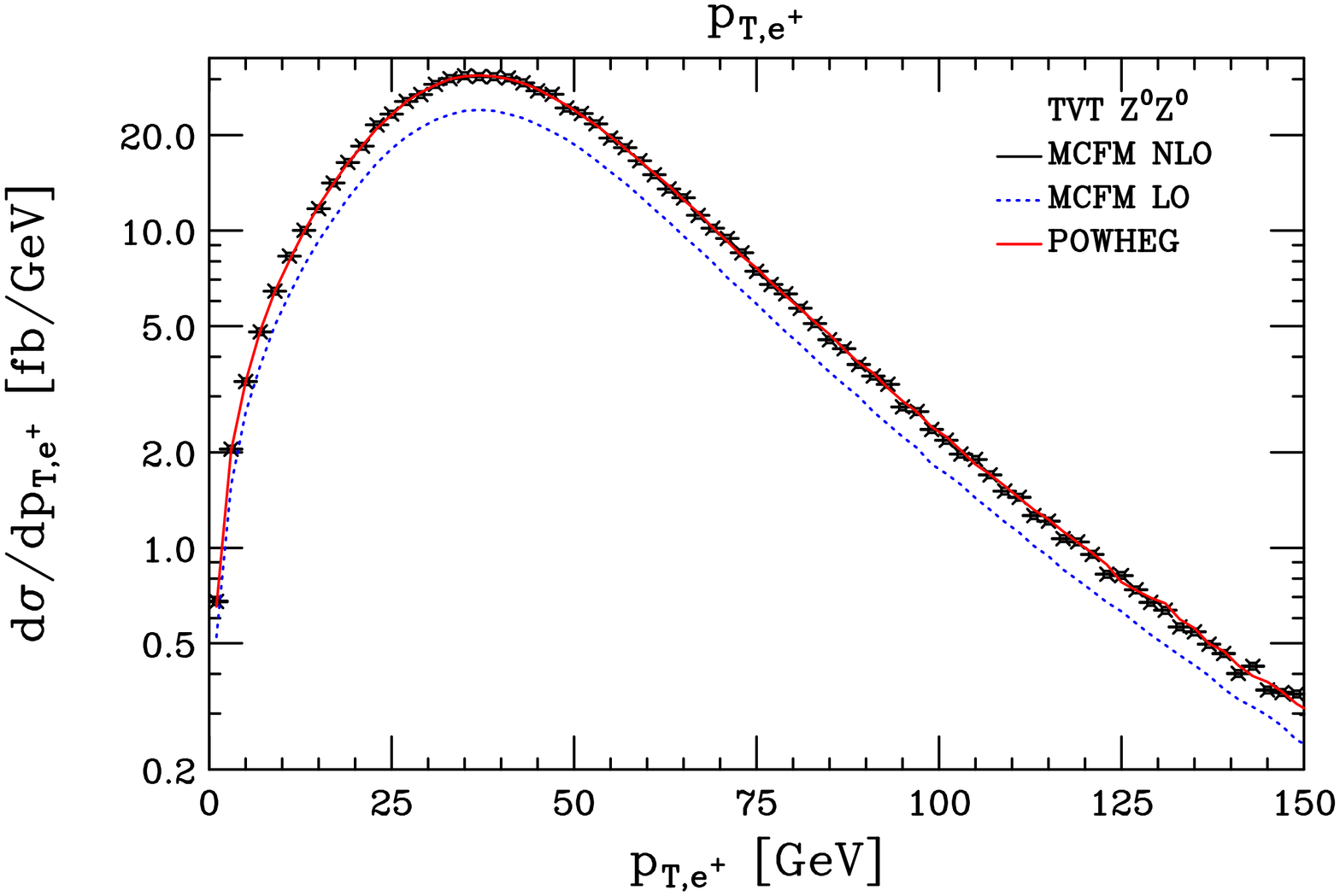}
\par\end{centering}

\begin{centering}
\vspace{7mm}
 \includegraphics[scale=0.29,angle=90]{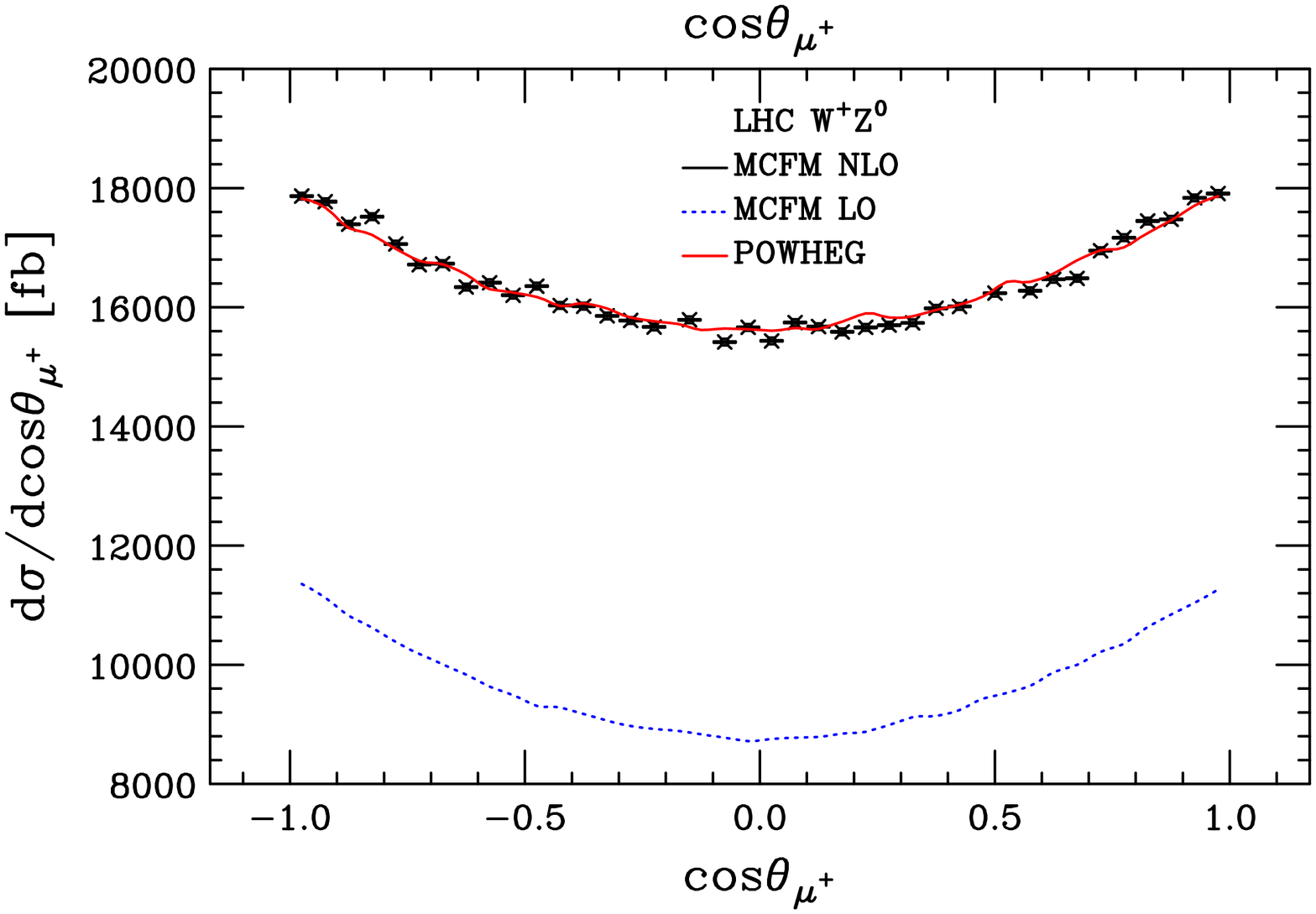}\hfill{}\includegraphics[scale=0.29,angle=90]{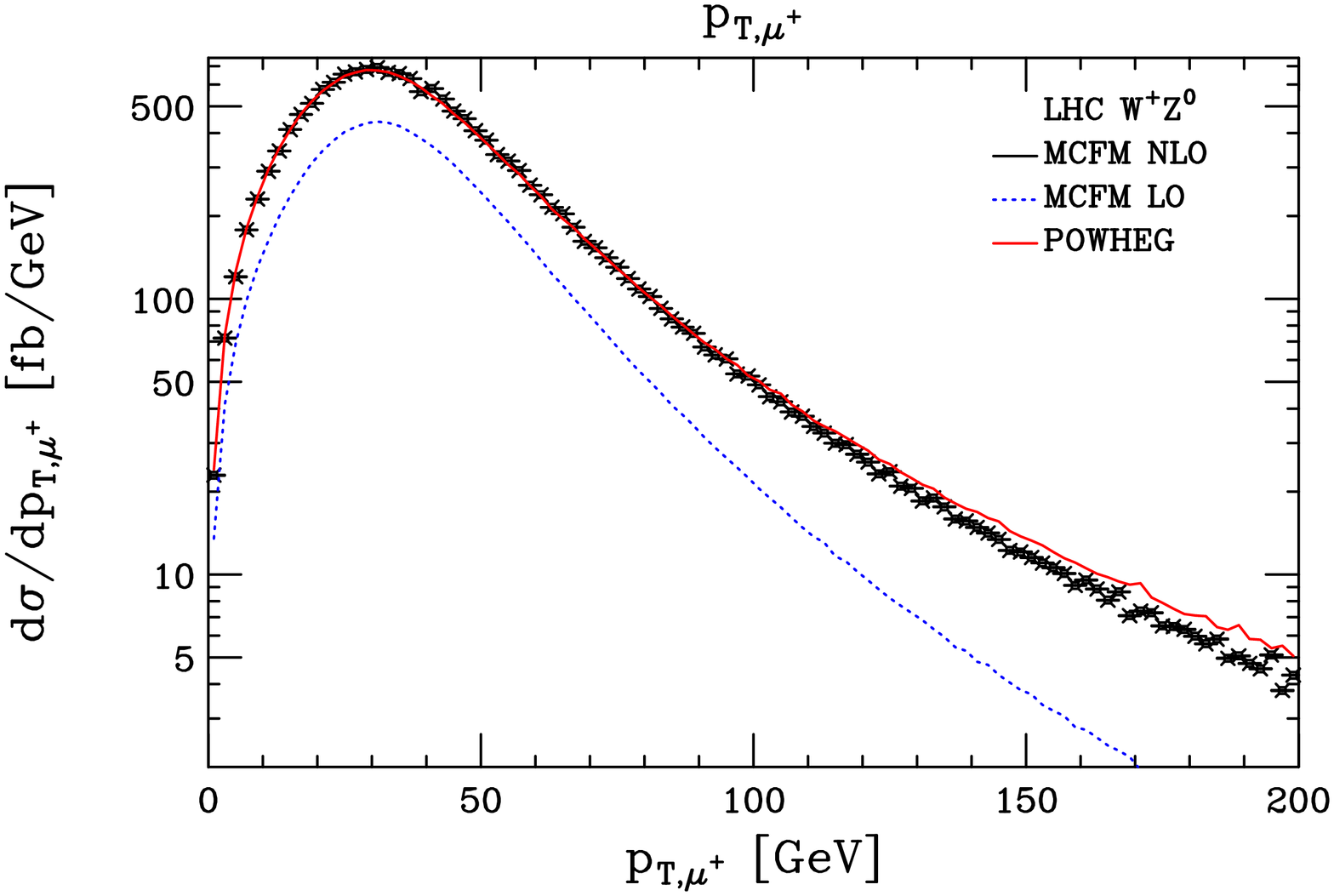}\vspace{7mm}
 \includegraphics[scale=0.29,angle=90]{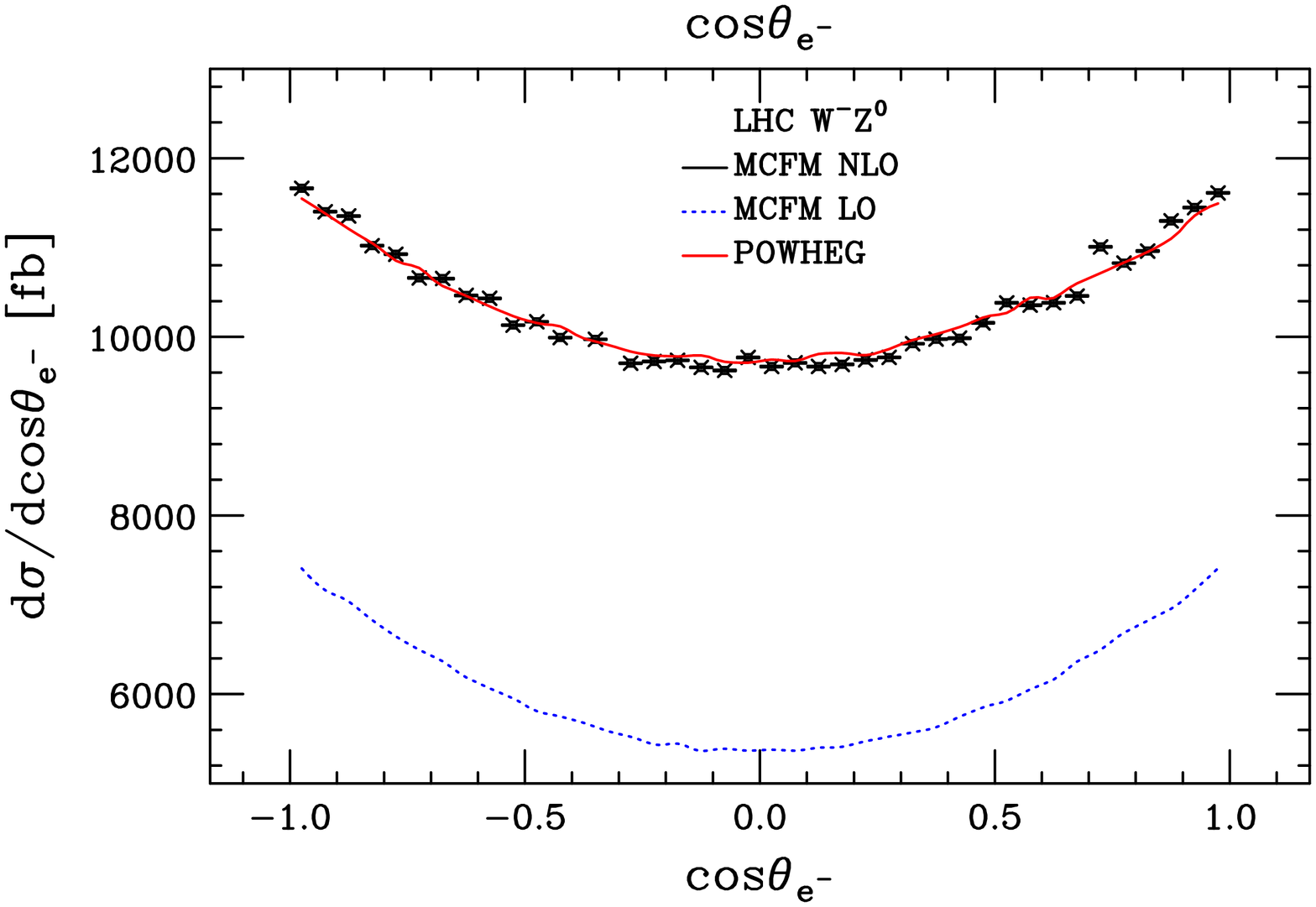}\hfill{}\includegraphics[scale=0.29,angle=90]{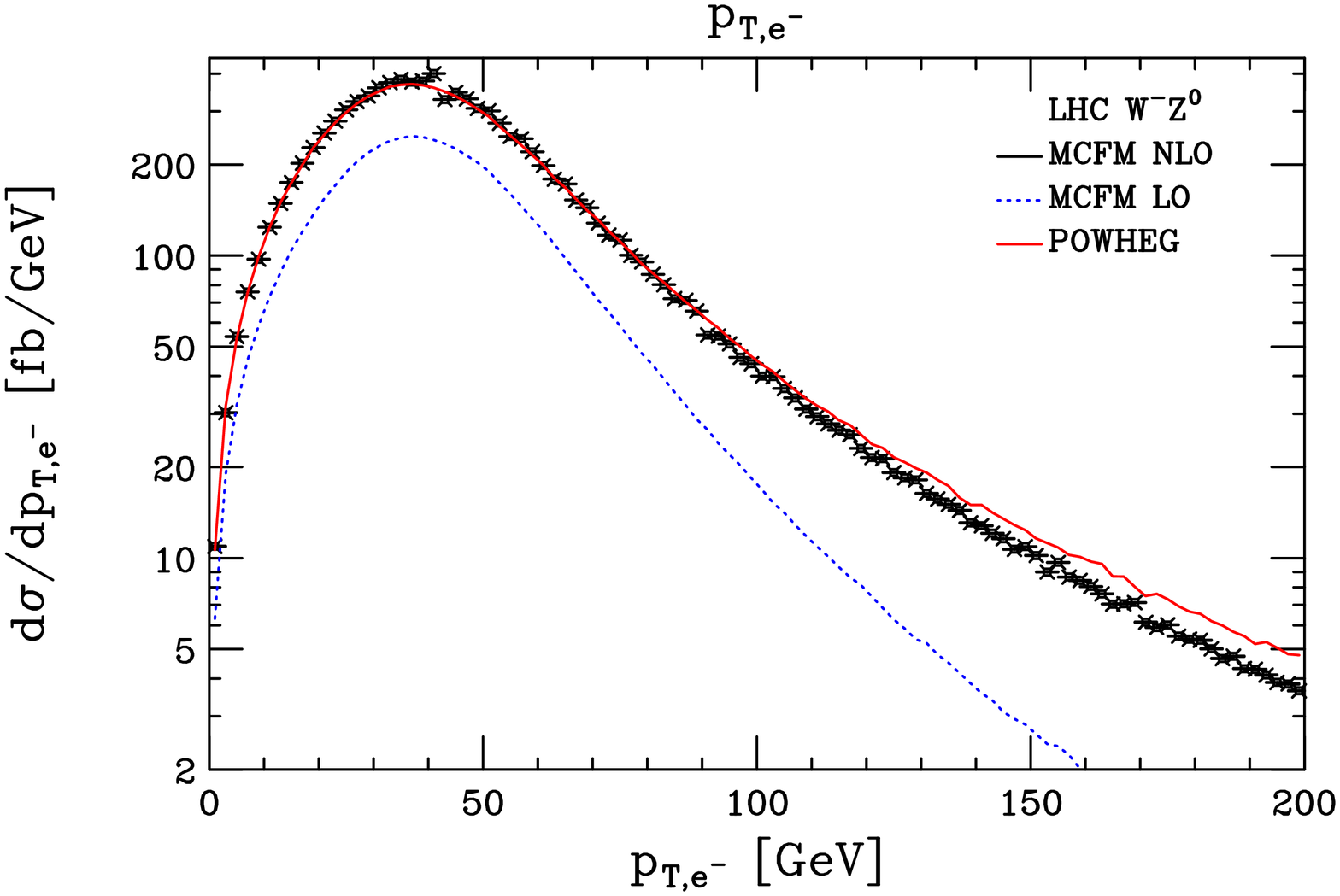}
\par\end{centering}

\caption{On the left we show the distribution of the polar angle of one of
the leptons emitted by one of the decaying weak bosons, in the rest
frame of the decay, while on the right hand-side we show the corresponding
transverse momentum spectrum. The colouring of the different predictions
is as in the previous figures. Note that in plotting these quantities
the branching fractions of the vector boson decays have been divided
out. }

\label{fig:lepton_polar_angles_and_pTs} 
\end{figure}

In Figure~\ref{fig:lepton_polar_angles_and_pTs} we have displayed
a number of distributions sensitive to the details of the decays of
the vector bosons, specifically, the polar angle of one of the leptons
produced by one of the decaying, resonant, vector bosons in its rest
frame and, separately, the corresponding transverse momentum spectra.
In all cases the agreement between our predictions and those of \noun{Mcfm}
is remarkably good. It is interesting to note that although our simulation
only includes spin correlation effects in the leading order and real
emission contributions to the cross section, it nevertheless reproduces
very well the distributions predicted by \noun{Mcfm}, which also includes
the effects of NLO virtual corrections at the level of the spin correlations.
Corroborating evidence for such behaviour can be found in the work
of Grazzini \cite{Grazzini:2005vw,Frederix:2008vb}, whose calculations
dealt with full NLO corrections to spin correlation matrices together
with soft gluon resummation effects. 

Actually, at Tevatron energies we find that the leading order production
spin density matrix, calculated using the underlying Born kinematics
$\left(\Phi_{B}\right)$, yields more-or-less identical distributions
to those shown in Figure~\ref{fig:lepton_polar_angles_and_pTs},
on the contrary, at the LHC this somewhat naive procedure produces
$\cos\theta_{l}$ distributions close in shape to the leading order
prediction, which is markedly different to the NLO one. These observations
are very much in keeping with others in the literature \cite{Frixione:2007zp},
suggesting that virtual corrections to spin correlation matrices are
typically small, whereas real emission corrections (which are generally
larger at the LHC than at the Tevatron) can lead to sizeable effects. 

Finally we wish to draw attention to the tails of the lepton $p_{T}$
distributions. In Figure~\ref{fig:lepton_polar_angles_and_pTs} we
see that for distributions forecast at Tevatron energies the \noun{Powheg
}and \noun{Mcfm} predictions agree very well over the whole spectrum,
whereas at the LHC we see that the two distributions overlap identically
in the low $p_{T}$ region but in the high $p_{T}$ region the \noun{Powheg
}result is approximately 30\% above that of \noun{Mcfm}.

In fact this level of disagreement with respect to fixed order NLO
predictions, in this region of phase space, is not unexpected. Events
in which the final-state leptons have very high transverse momenta
will, by their nature, typically contain associated high energy QCD
radiation; this fact is substantiated by the shape of the corresponding
leading order predictions shown in the blue dotted lines. Recall that,
in the\emph{ }\emph{\noun{Powheg }}hardest emission cross section,
the term responsible for the generation of the radiative variables
is not simply equal to the NLO real emission cross section, but rather
the NLO real emission cross section multiplied by a factor $\overline{B}\left(\Phi_{B}\right)/B\left(\Phi_{B}\right)$
and the Sudakov form factor. In the high $p_{T}$ limit the latter
factor tends to one, hence, \noun{Powheg }events with high transverse
momentum are generated according to the NLO real emission cross section
multiplied by $\overline{B}\left(\Phi_{B}\right)/B\left(\Phi_{B}\right)$.
Loosely speaking this factor is characteristic of the NLO total cross
section \emph{K}-factor, thus one expects the rate of events including
high $p_{T}$ emissions to be different in \noun{Powheg} with respect
to a pure NLO calculation by such an amount (different by terms beyond
NLO accuracy). This is indeed what we observe in Figure~\ref{fig:lepton_polar_angles_and_pTs}.
Finally we add that the Born variables are unique exceptions to this
reasoning since they are fully preserved, by construction, when radiation
is generated, hence, unlike other inclusive observables the Born variables
will always \emph{exactly }equal the NLO prediction, regardless of
whether the events are associated to high $p_{T}$ emissions or not.

\subsection{Exclusive observables\label{sub:Exclusive_observables}}

In this section we shift the focus of our validation onto observables
which more directly assess the generation of the hardest emission
(Sect.~\ref{sub:production}) and additional radiation arising from
the subsequent vetoed and truncated showers (Sect.~\ref{sub:Truncated-and-vetoed}).
To this end we compare our results to two different approaches, namely,
the default angular-ordered parton shower simulation in \noun{Herwig++}
and also \noun{Mc}{\footnotesize @}\noun{nlo} v3.4 \cite{Frixione:2008ym}.
Since the latter program matches the same NLO matrix elements with
the older \noun{Herwig }parton shower \cite{Corcella:2000bw,Corcella:2002jc}
it is formally of equivalent accuracy to our \noun{Powheg }simulation,
on the other hand, the former includes only LO matrix elements, hence,
it is anticipated that it will fail to adequately model high $p_{T}$
radiation.

All of the results presented in this subsection, from each of the
three approaches, were obtained at the parton level, after parton
showering. The predictions from our \noun{Powheg }simulation within
\noun{Herwig++} are displayed as red dashed lines, while those of
\noun{Mc}{\footnotesize @}\noun{nlo }and \noun{Herwig++} with the
\noun{Powheg }feature disabled are shown as black and blue dotted
lines respectively. 

\begin{figure}[t]
\noindent \begin{centering}
\includegraphics[scale=0.3,angle=90]{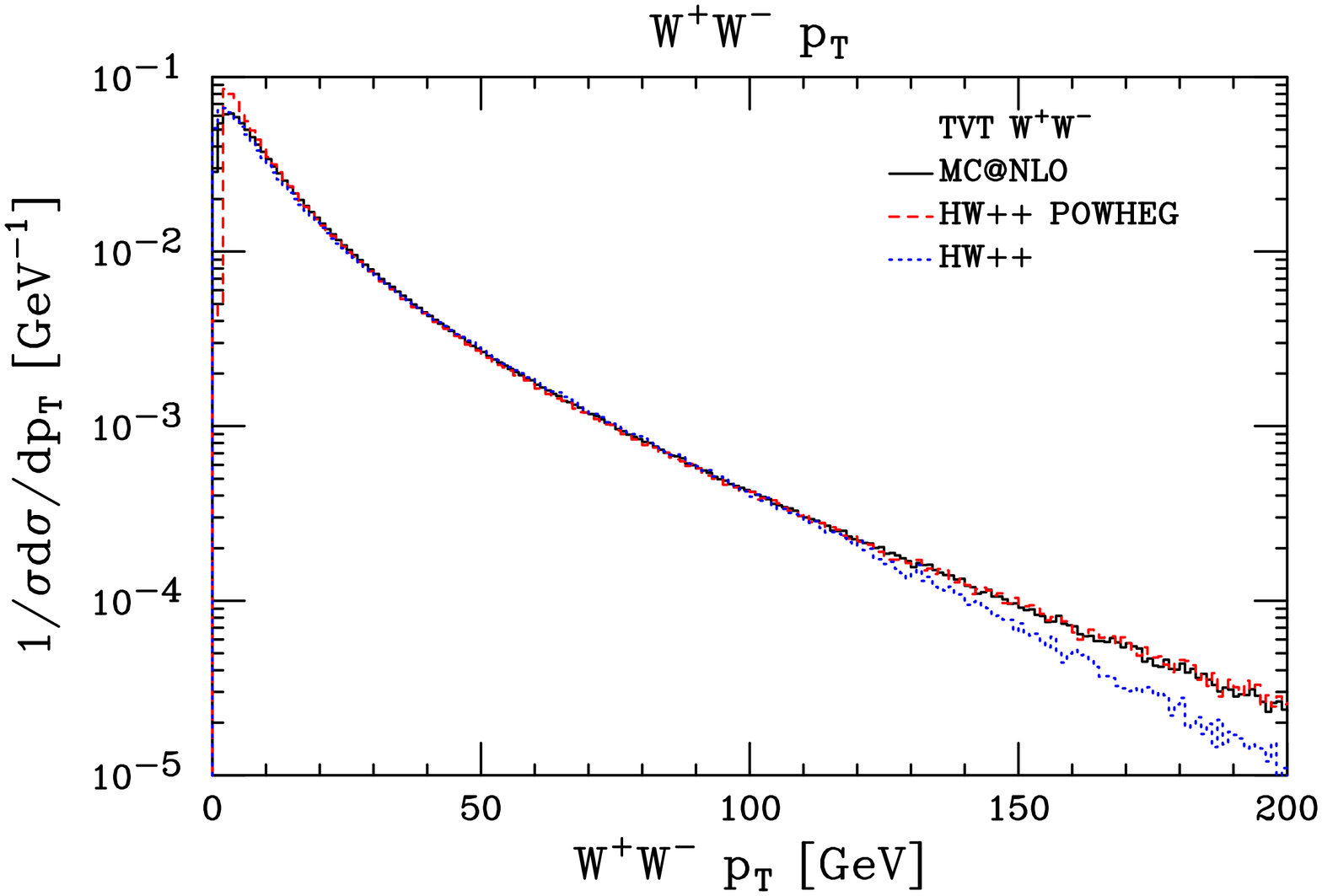}\hfill{}\includegraphics[scale=0.3,angle=90]{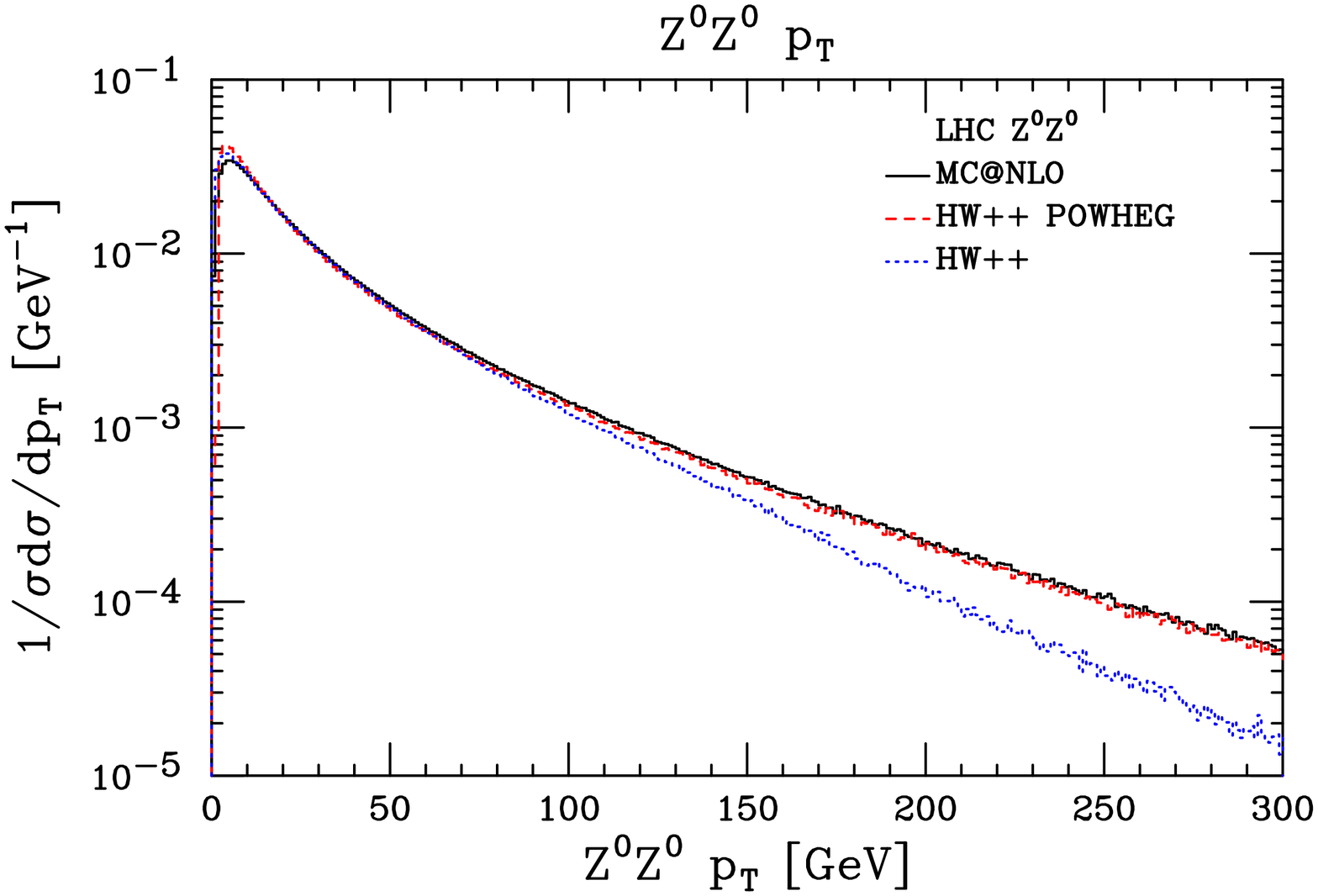}
\par\end{centering}

\begin{centering}
\vspace{7mm}

\par\end{centering}

\begin{centering}
\includegraphics[scale=0.3,angle=90]{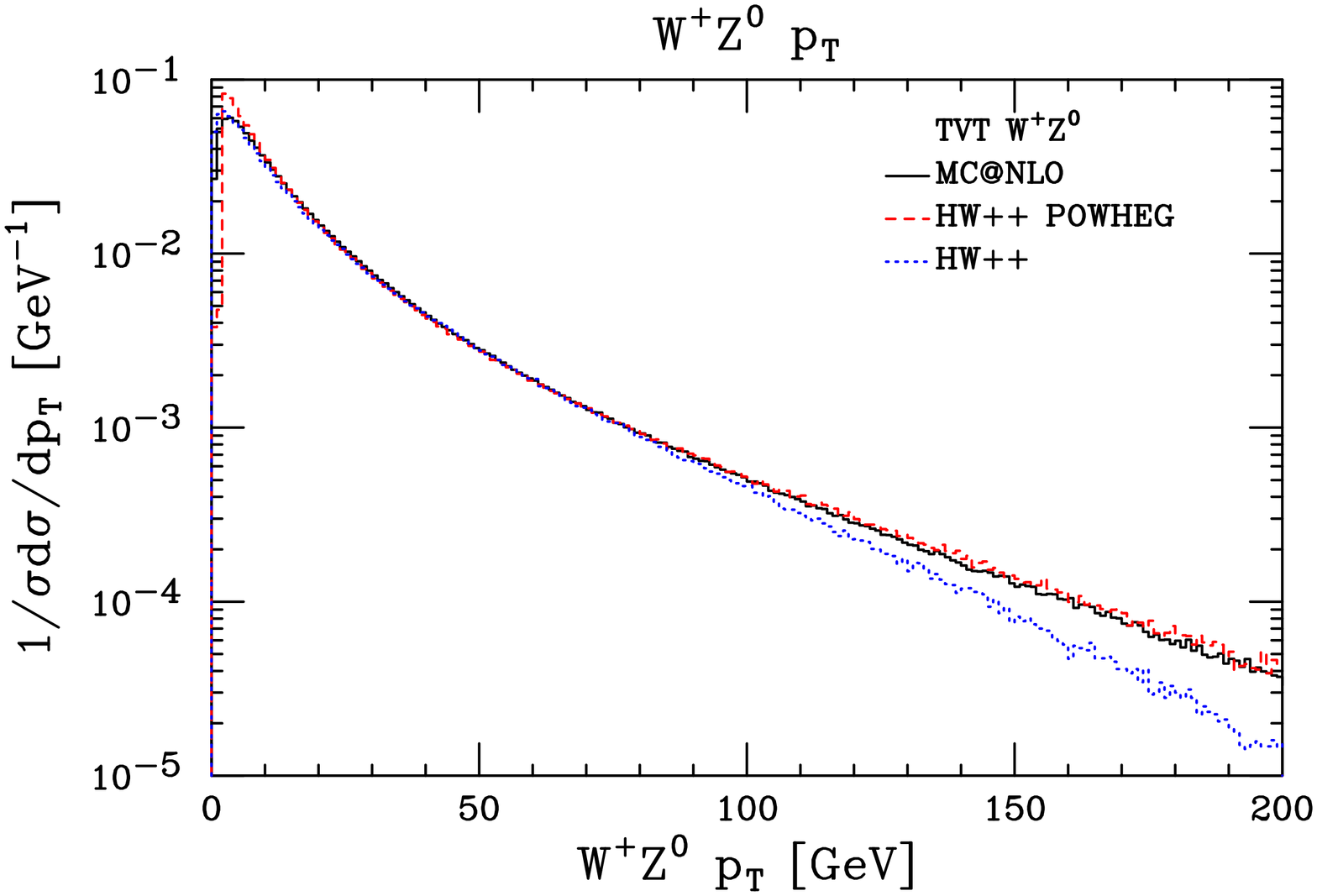}\hfill{}\includegraphics[scale=0.3,angle=90]{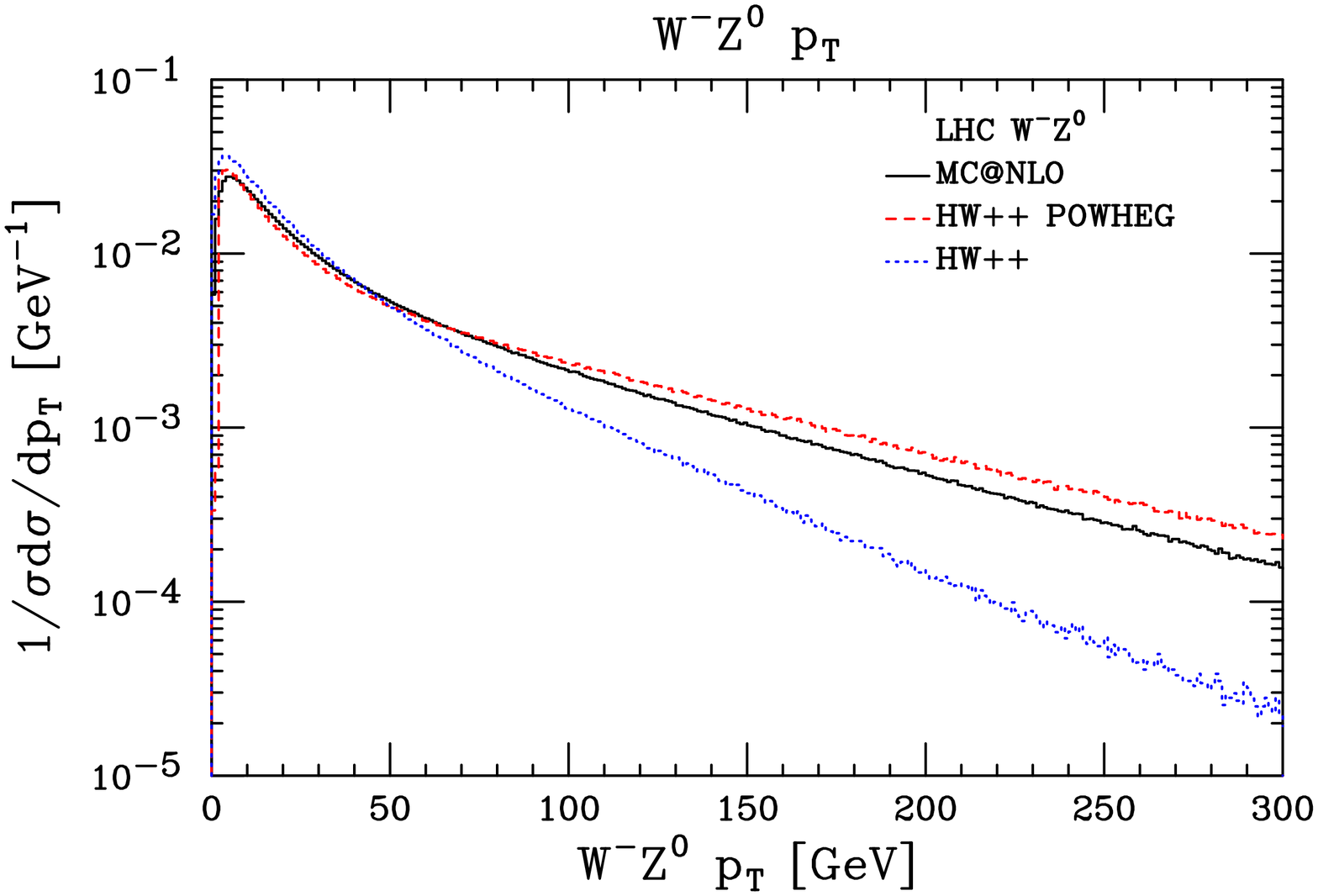} 
\par\end{centering}

\caption{The transverse momentum spectrum of the produced weak boson pair system
at Tevatron (left) and LHC (right) energies. Predictions from the
\noun{Mc}{\footnotesize @}\noun{nlo }and\noun{ Herwig++ Powheg} simulations
are present as black and red dashed lines respectively. Results from
the leading order \noun{Herwig++} parton shower simulation are also
shown as blue dotted lines. For the case of a single emission this
quantity is equivalent to the radiative variable $p_{T}$ introduced
in Section~\ref{sub:Kinematics-and-phase}. }

\label{fig:VV_system_pT_spectra} 
\end{figure}
\begin{figure}[t]
\noindent \begin{centering}
\includegraphics[scale=0.3,angle=90]{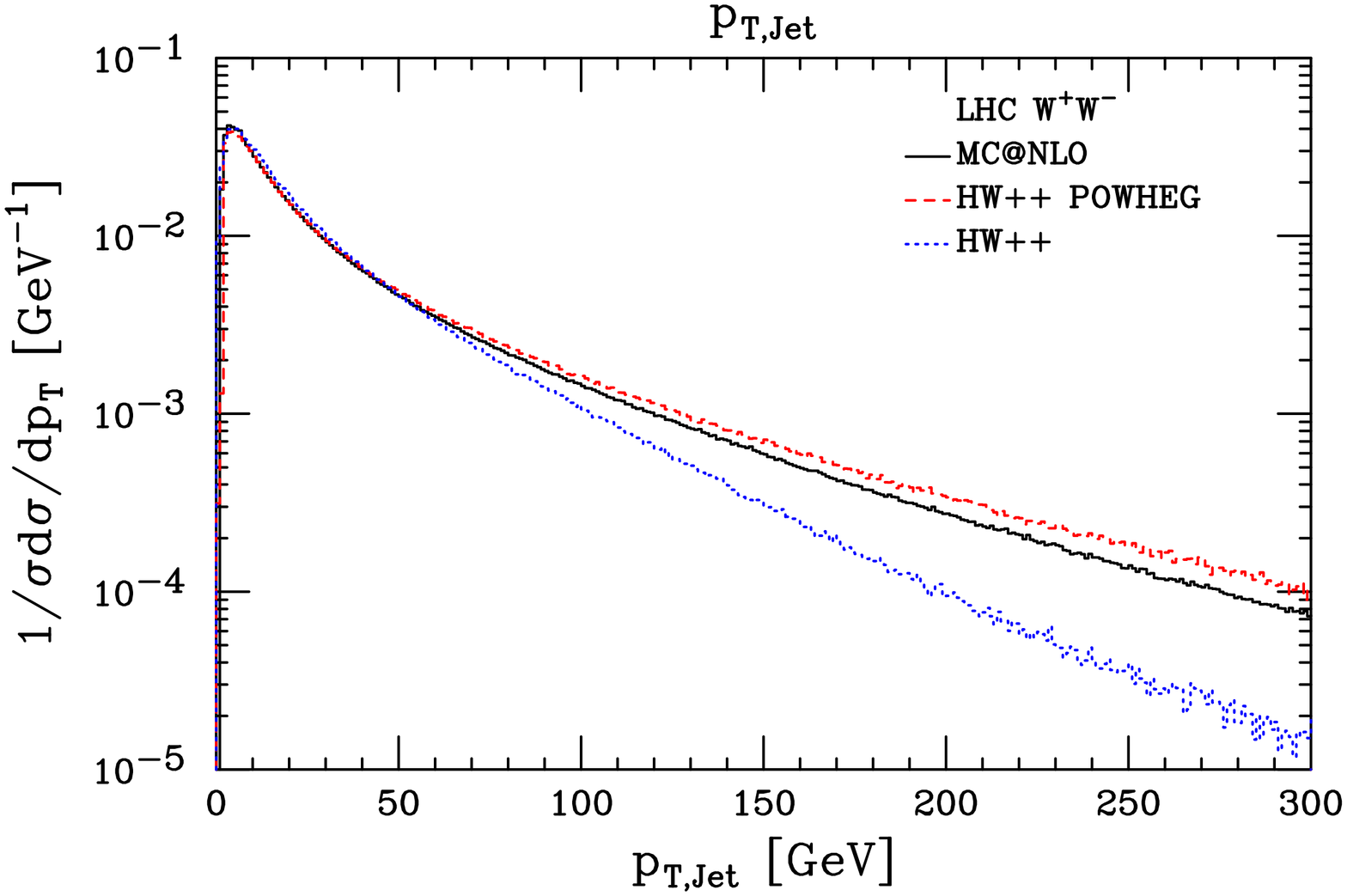}\hfill{}\includegraphics[scale=0.3,angle=90]{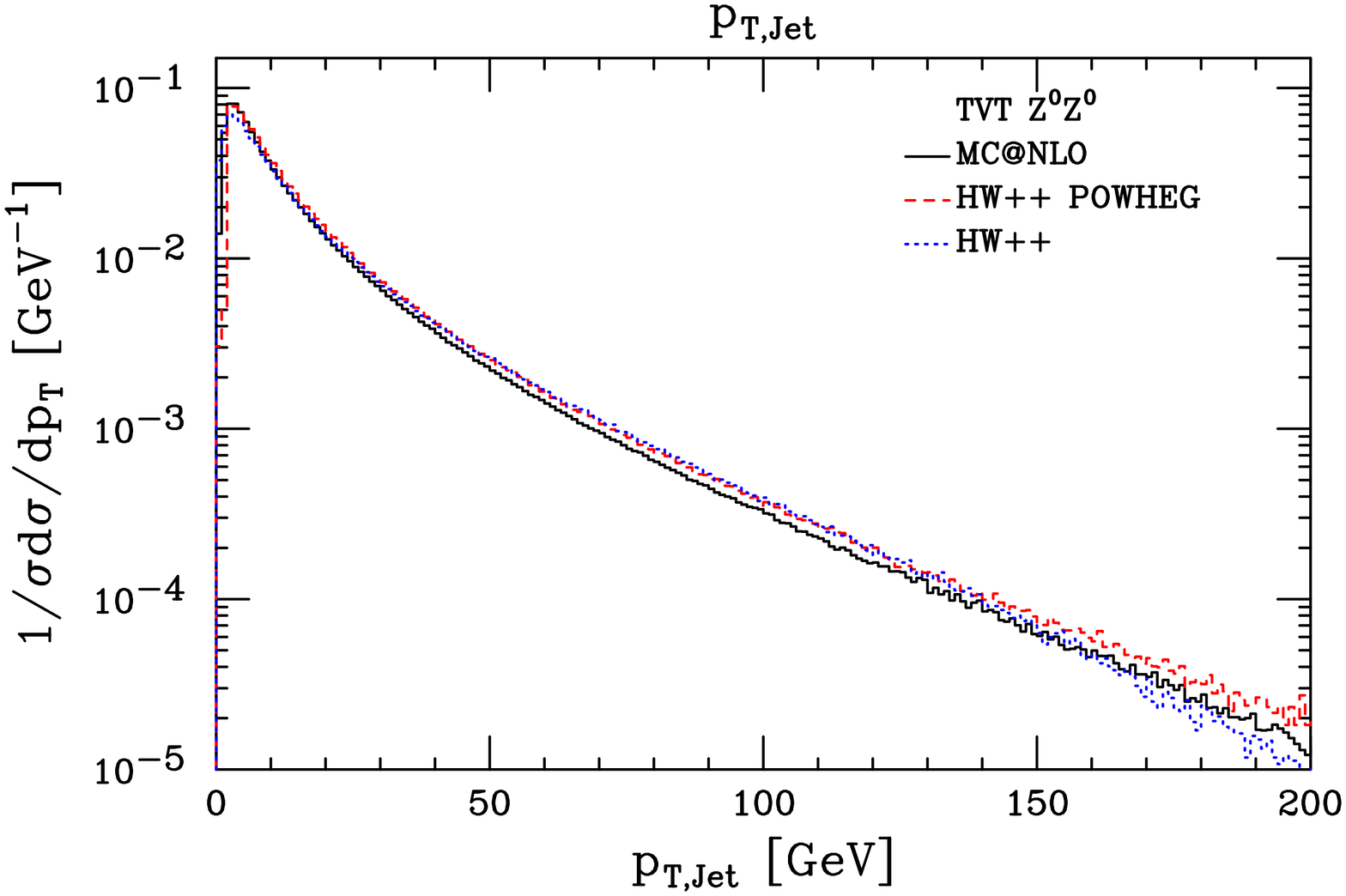}
\par\end{centering}

\begin{centering}
\vspace{7mm}

\par\end{centering}

\begin{centering}
\includegraphics[scale=0.3,angle=90]{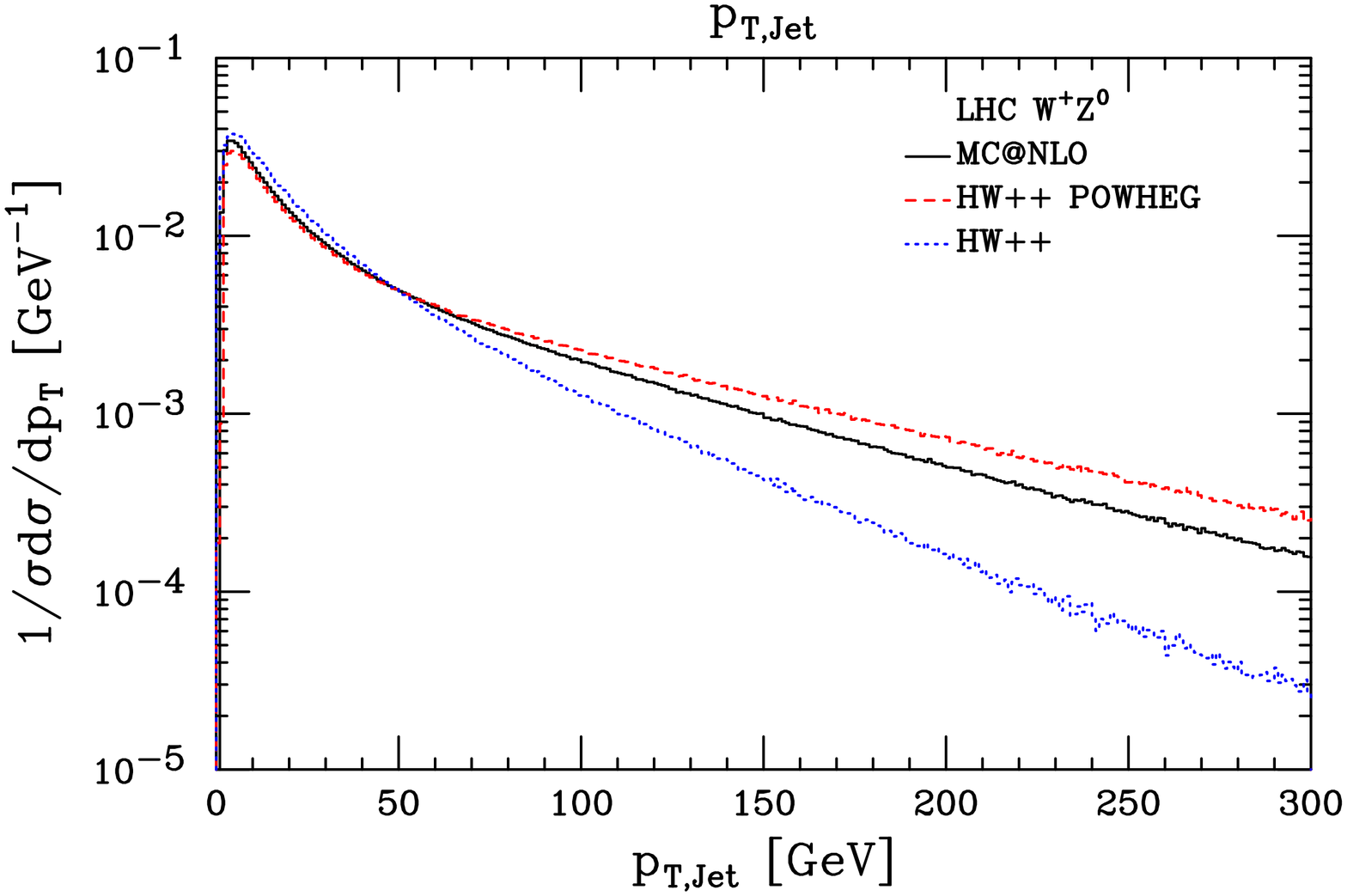}\hfill{}\includegraphics[scale=0.3,angle=90]{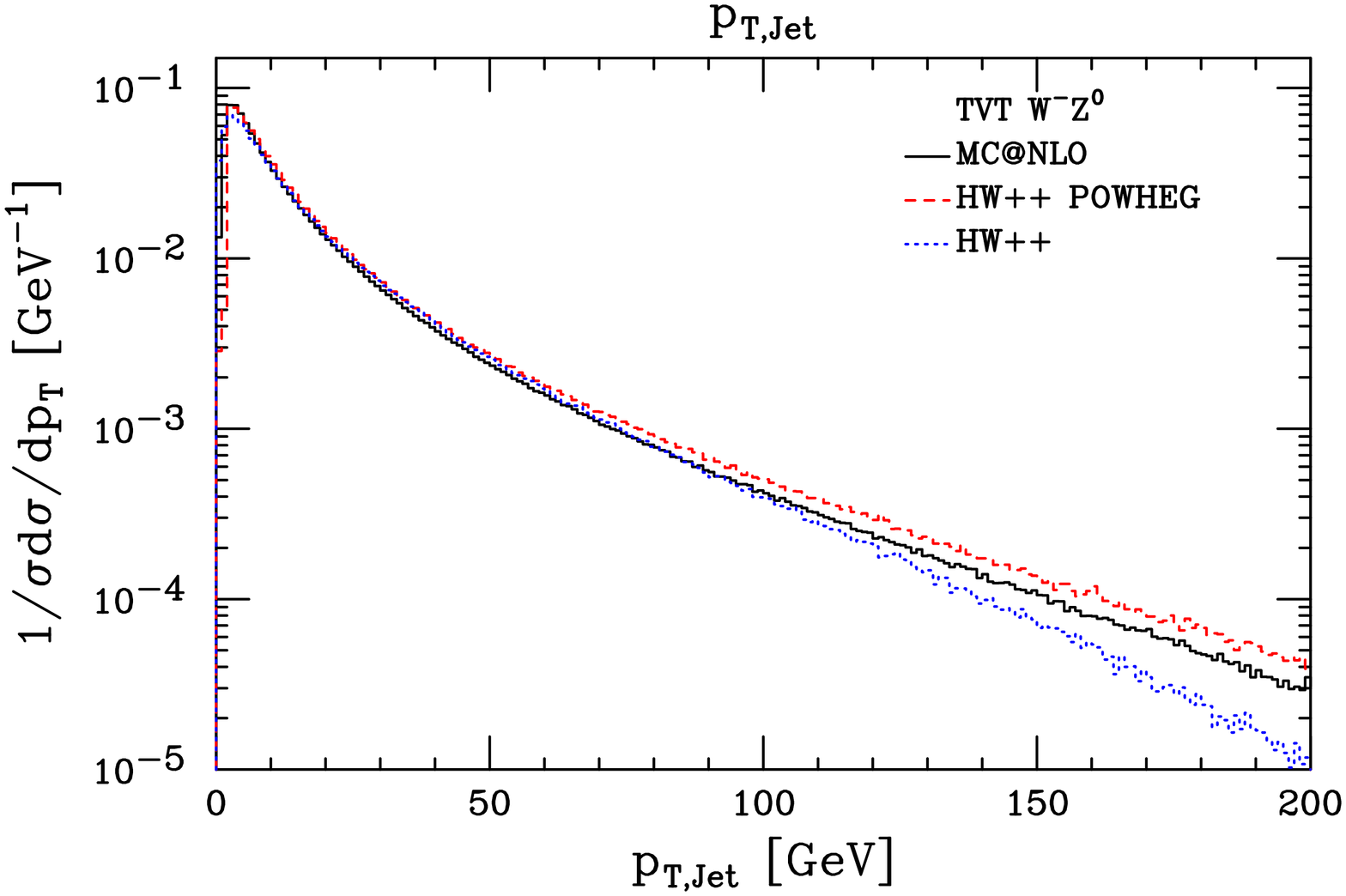} 
\par\end{centering}

\caption{The transverse momentum of the hardest jet in weak boson pair production
at the LHC (left) and Tevatron (right), assuming a nominal LHC centre-of-mass
energy, $\sqrt{s}=14\,\mathrm{TeV}$. The colouring of the histograms
is the same as in figure \ref{fig:VV_system_pT_spectra}. As for the
case of weak boson pair system, for a single emission this quantity
is equal to the radiative variable $p_{T}$ in Section~\ref{sub:Kinematics-and-phase}. }

\label{fig:Hardest_jet_pT_spectrum} 
\end{figure}
Figures~\ref{fig:VV_system_pT_spectra} and \ref{fig:Hardest_jet_pT_spectrum}
exhibit the transverse momentum spectra of the di-vector boson system
and the hardest emitted jet. These distributions directly reflect
the $p_{T}$ dependence of the hardest emission cross section Eq.\,\ref{eq:powheg_1},
modulo small smearing effects due to the truncated and vetoed parton
showers. All three approaches are seen to agree well in the low $p_{T}$
regions, where the parton shower approximation is expected to be reliable.
In the high $p_{T}$ regions one can see that in all cases the parton
shower by itself underestimates the production rate with respect to
\noun{Mc}{\footnotesize @}\noun{nlo }and \noun{Powheg}, moreover,
we note that the degree to which it is underestimated is worse at
LHC energies, than at the Tevatron, where the available phase space
is more constrained (\emph{cf.} Fig.~\ref{fig:VV_p2_and_y}). 

In general the degree of overlap between the \noun{Mc}{\footnotesize @}\noun{nlo
}and \noun{Powheg }results is quite good, with differences between
the two sets of predictions only being noticeable in the high $p_{T}$
tails for LHC energies. Such behavior has already been observed, and
its nature well documented, in publications concerning other \noun{Powheg}
simulations \cite{Alioli:2008tz,Hamilton:2009za}; it is basically
a further manifestation of the effect discussed at the end of Section~\ref{sub:Inclusive_observables},
namely, that the distribution of the hardest emission in the \noun{Powheg}
method (Eq.\,\ref{eq:powheg_1}) is given by $\overline{B}\left(\Phi_{B}\right)/B\left(\Phi_{B}\right)$
multiplied by the NLO real emission cross section, whereas in \noun{Mc}{\footnotesize @}\noun{nlo}
it is given by the real emission cross section alone. It follows that
the \noun{Powheg} predictions tend to exceed those of \noun{Mc}{\footnotesize @}\noun{nlo}
and fixed order calculations when $p_{T}$ is large, by a factor of
the order of the\noun{ NLO }total cross section\noun{ }\emph{K}\emph{\noun{-}}factor.
In keeping with this one sees that the relative differences seen in the
tails of the transverse momentum spectra reflect the size of the relevant
\emph{K-}factors (\emph{cf.} Fig.~\ref{fig:VV_p2_and_y}) and, accordingly,
they are somewhat larger at the LHC than at the Tevatron. 

Note that within the \noun{Powheg} formalism  it is possible to introduce
so-called \emph{damping factors} \cite{Alioli:2008tz}, which act in such a way
as to reduce the effects of the multiplicative
$\overline{B}\left(\Phi_{B}\right)/B\left(\Phi_{B}\right)$ term, leading
to cross sections in the high $p_{T}$ domain closer to those of fixed order
calculations. However, there is no theoretical motivation to implement, or not
to implement such damping, since the contribution from this factor, in the
regions of phase space corresponding to high $p_{T}$ emission, is formally
subleading -- the differences with respect to \noun{Mc}{\footnotesize @}\noun{nlo}
and fixed order predictions being indicative of the associated theoretical
uncertainties.

\begin{figure}[t]
\noindent \begin{centering}
\includegraphics[scale=0.31,angle=90]{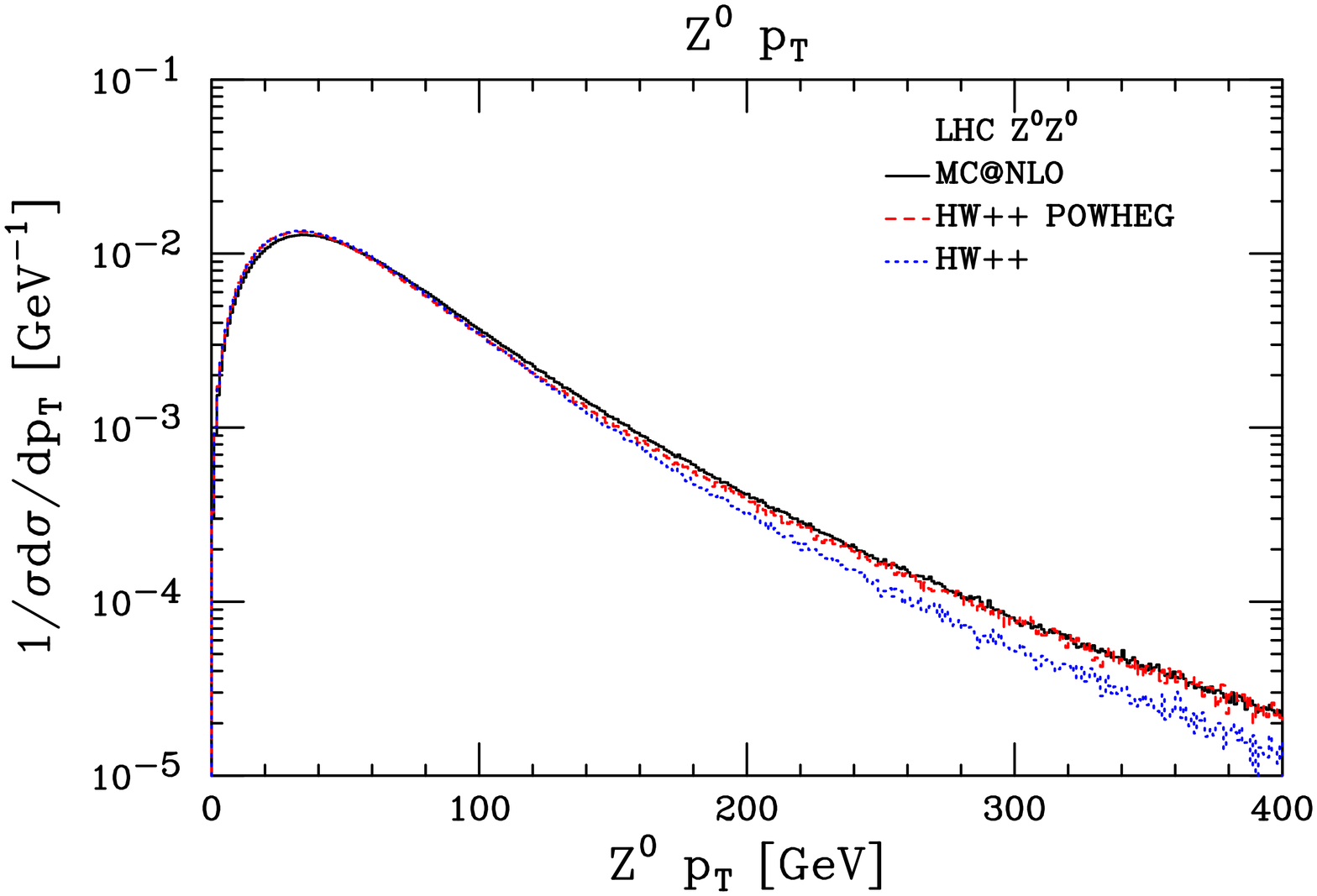}\hfill{}\includegraphics[scale=0.31,angle=90]{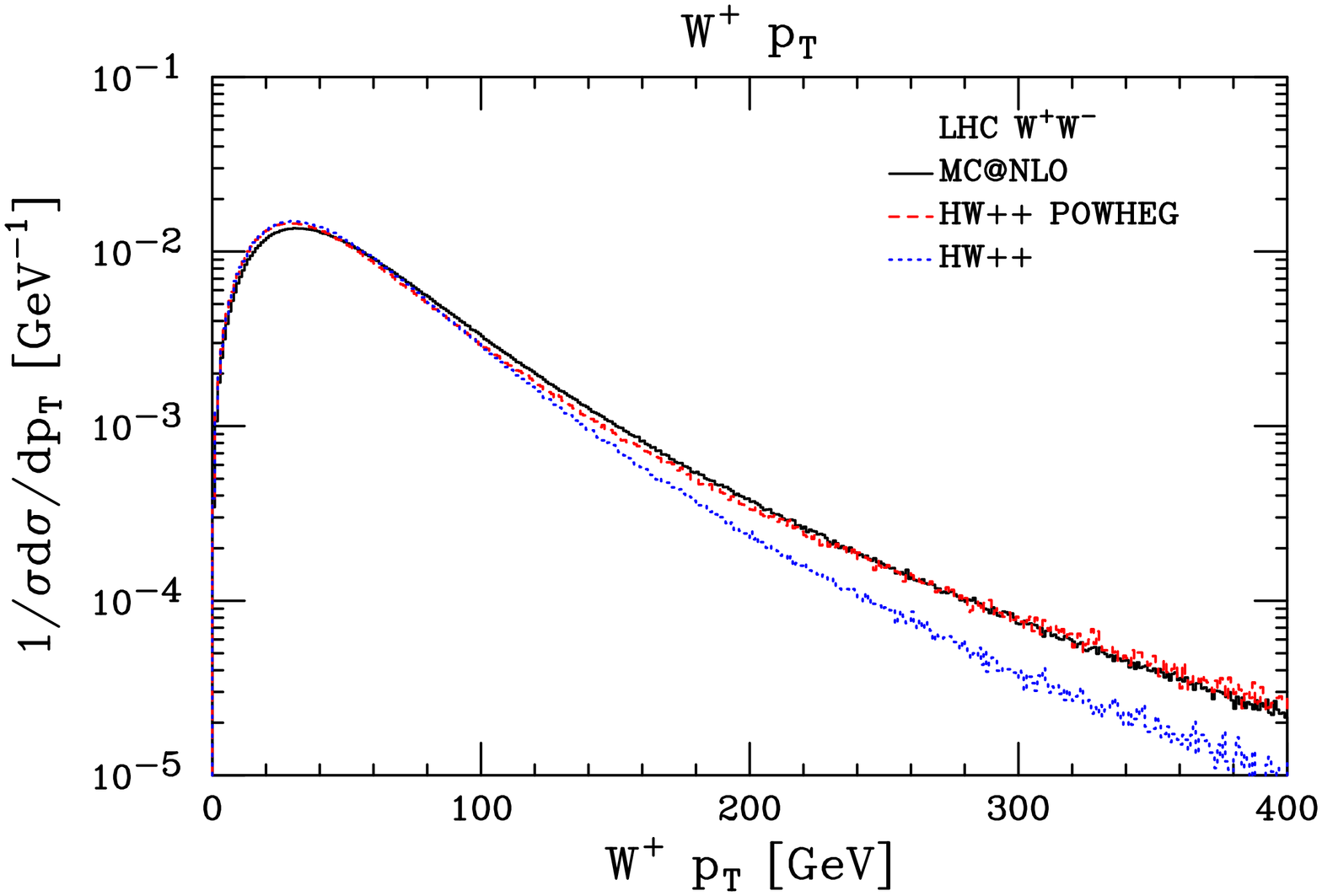}
\par\end{centering}

\begin{centering}
\vspace{8mm}

\par\end{centering}

\begin{centering}
\includegraphics[scale=0.31,angle=90]{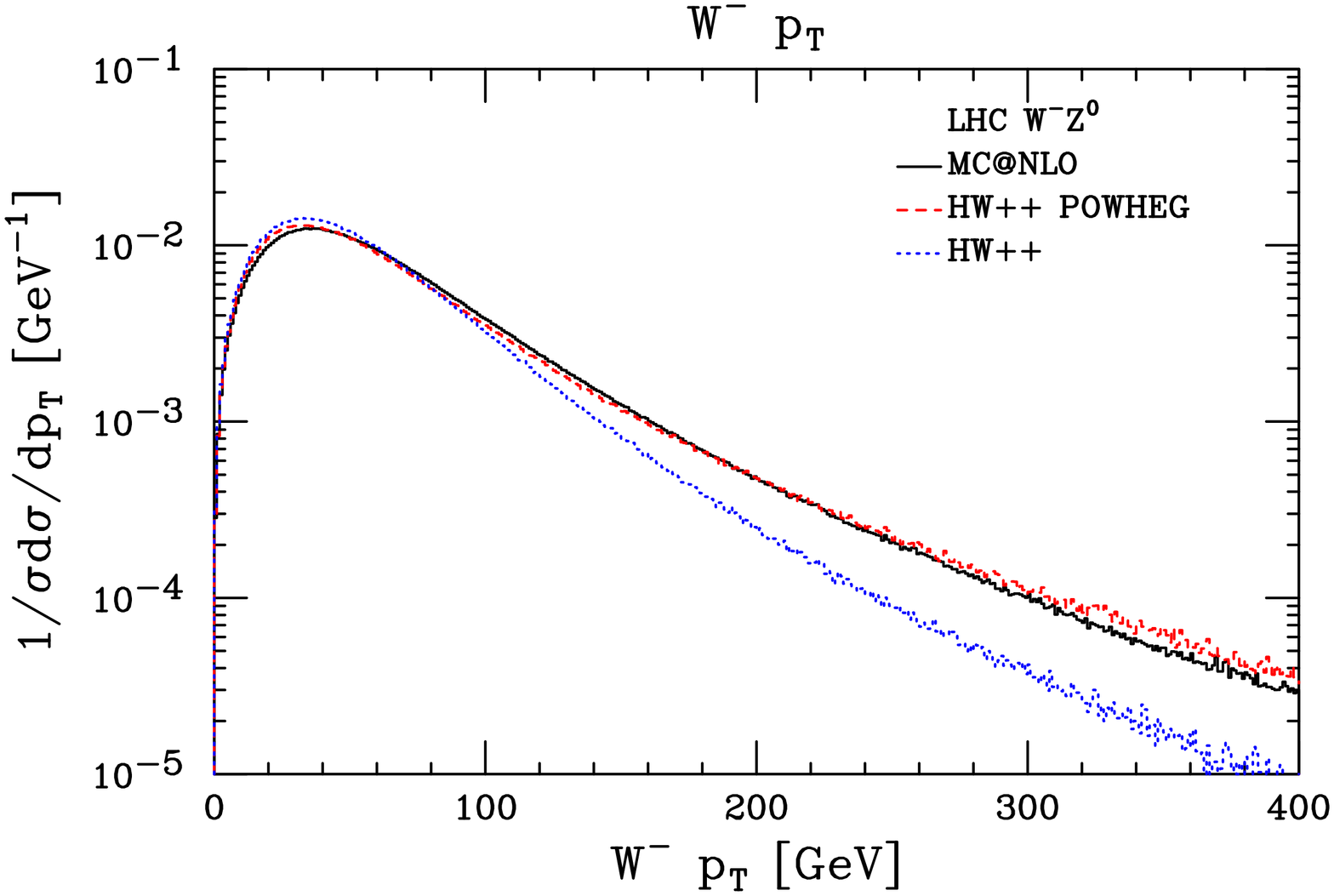}\hfill{}\includegraphics[scale=0.31,angle=90]{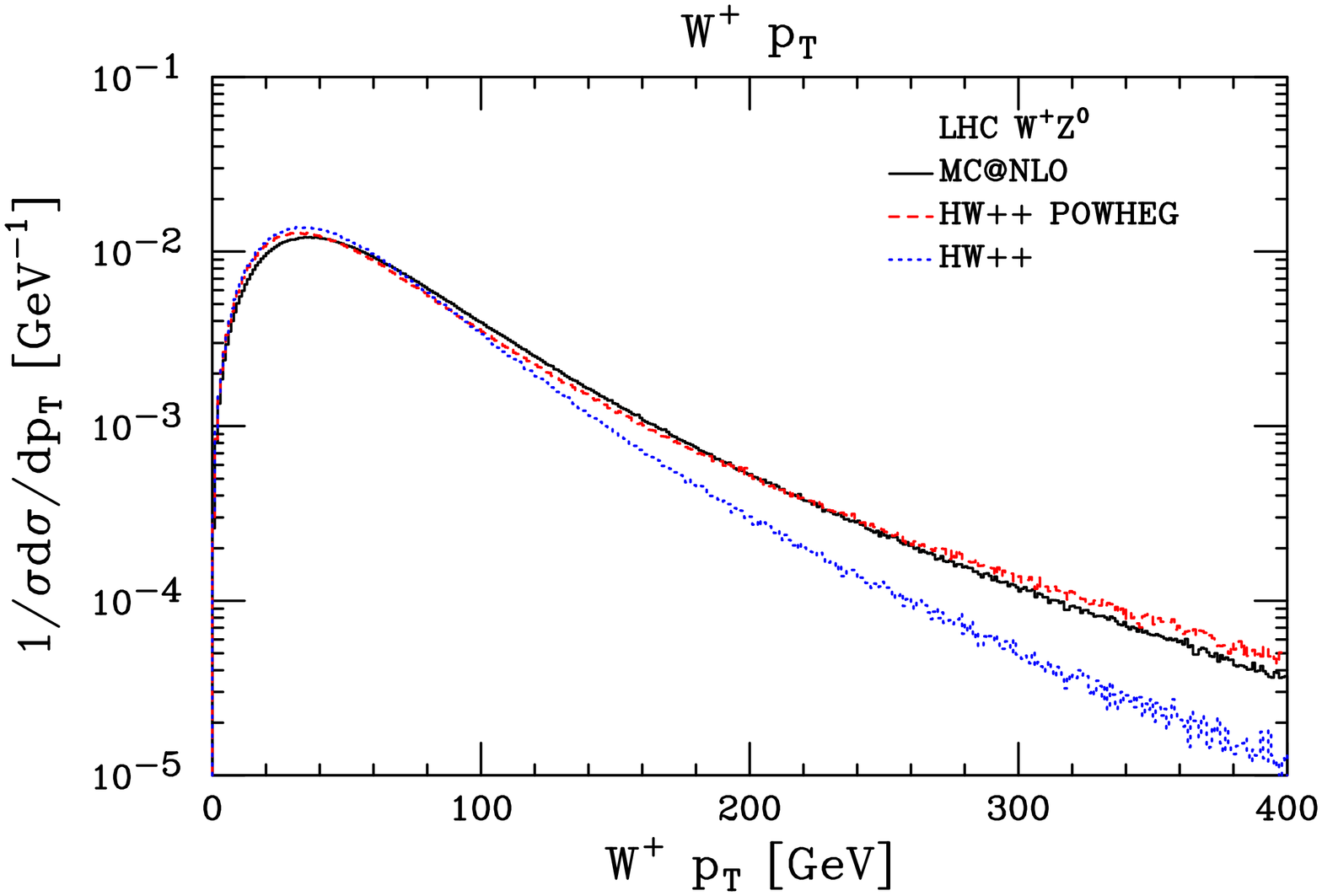} 
\par\end{centering}

\caption{The transverse momentum spectra of individual weak gauge bosons in
di-vector boson production at the LHC, assuming a hadronic centre-of-mass
energy of 14 TeV. Predictions from the \noun{Mc}{\footnotesize @}\noun{nlo
}and\noun{ Herwig++ Powheg} simulations are present as black and red
dashed lines respectively. Predictions from the default \noun{Herwig++}
simulation, with the \noun{Powheg} feature disabled, are represented
by blue dotted lines. }

\label{fig:Single_V_boson_pT_spectra} 
\end{figure}

In Figure~\ref{fig:Single_V_boson_pT_spectra} we plot the transverse
momentum spectra of individual vector bosons in each of the weak boson
pair production channels, at nominal LHC energies only. In contrast
to the transverse momentum spectrum of the diboson system, in fixed
order perturbation theory this observable receives contributions at
leading order for all values of $p_{T}$. Thus one expects that any
differences between the leading and next-to-leading parton shower
predictions should be small, of order $\alpha_{\mathrm{S}}$. It is
then remarkable that these distributions show the next-to-leading
order parton shower predictions, in good agreement with one another,
exceeding the leading order parton shower prediction by up to a factor
of four at high $p_{T}$. 

This curious result has already been noted and investigated in the
original calculation of the fixed order NLO corrections to $ $$W^{\pm}Z$
production in Ref.~\cite{Frixione:1992pj}. The same effect was subsequently
observed in the case of $W^{+}W^{-}$ production and subject to the
same analysis in Ref.~\cite{Frixione:1993yp}. The detailed studies
carried out in these publications conclude that the enhancement seen
at high $p_{T}$ is greatly dominated by contributions arising from
$qg$ initiated real emission corrections. The reason for this large
$qg$ contribution was considered to be twofold. Firstly, it was noted
that the luminosity for the $qg$ channel was more than one order
of magnitude greater than that of the $q\bar{q}$ channel at the LHC;
secondly, in $qg$ reactions, when the radiated quark and one of the
weak bosons are produced with sufficiently high transverse momenta,
the other weak boson may be produced as a `soft' emission from the
recoiling quark -- a process which carries a large logarithmic enhancement
factor, $\log^{2}\left(p_{T,Z}^{2}/m_{W}^{2}\right)$. It is further
noted in Ref.~\cite{Frixione:1993yp} that fewer partonic subprocesses
can participate in this enhancement mechanism in the case of $W^{+}W^{-}$
production than in $W^{\pm}Z$ production, accordingly, we observe
that the magnitude of the effect is somewhat smaller in the latter
case. 

Having studied several $p_{T}$ spectra we now turn to examine other
distributions sensitive to the generation of the radiative variables
and subsequent parton showering. As noted in earlier works concerning
Higgs boson production, Ref.~\cite{Hamilton:2009za}, the rapidity
correlation between the leading jet and the recoiling colourless system
is an interesting observable to examine from this point of view: for
the case of a single hard emission the rapidity correlation $\mathrm{y}_{k}-\mathrm{y}$
can be expressed purely in terms of $\Phi_{R}$. In order to provide
some additional physical insight regarding the nature of this quantity,
we note that in the limit where the radiated parton is produced in
the region perpendicular to the colliding beam partons, in the partonic
centre-of-mass frame\begin{equation}
\lim_{\theta_{k}\rightarrow\frac{\pi}{2}}\,\,\mathrm{y}_{k}-\mathrm{y}=-\frac{2}{1+x}\,\left(\theta_{k}-\frac{\pi}{2}\right)\,,\label{eq:yJ-yVV_approx_formula}\end{equation}
where $\theta_{k}$ denotes the polar angle of the emitted parton
in that frame. Furthermore, when the radiated parton is emitted along
the $\pm z$ directions, $\mathrm{y}_{k}-\mathrm{y}$ tends to $\pm\infty$. 

\begin{figure}[t]
\noindent \begin{centering}
\includegraphics[scale=0.31,angle=90]{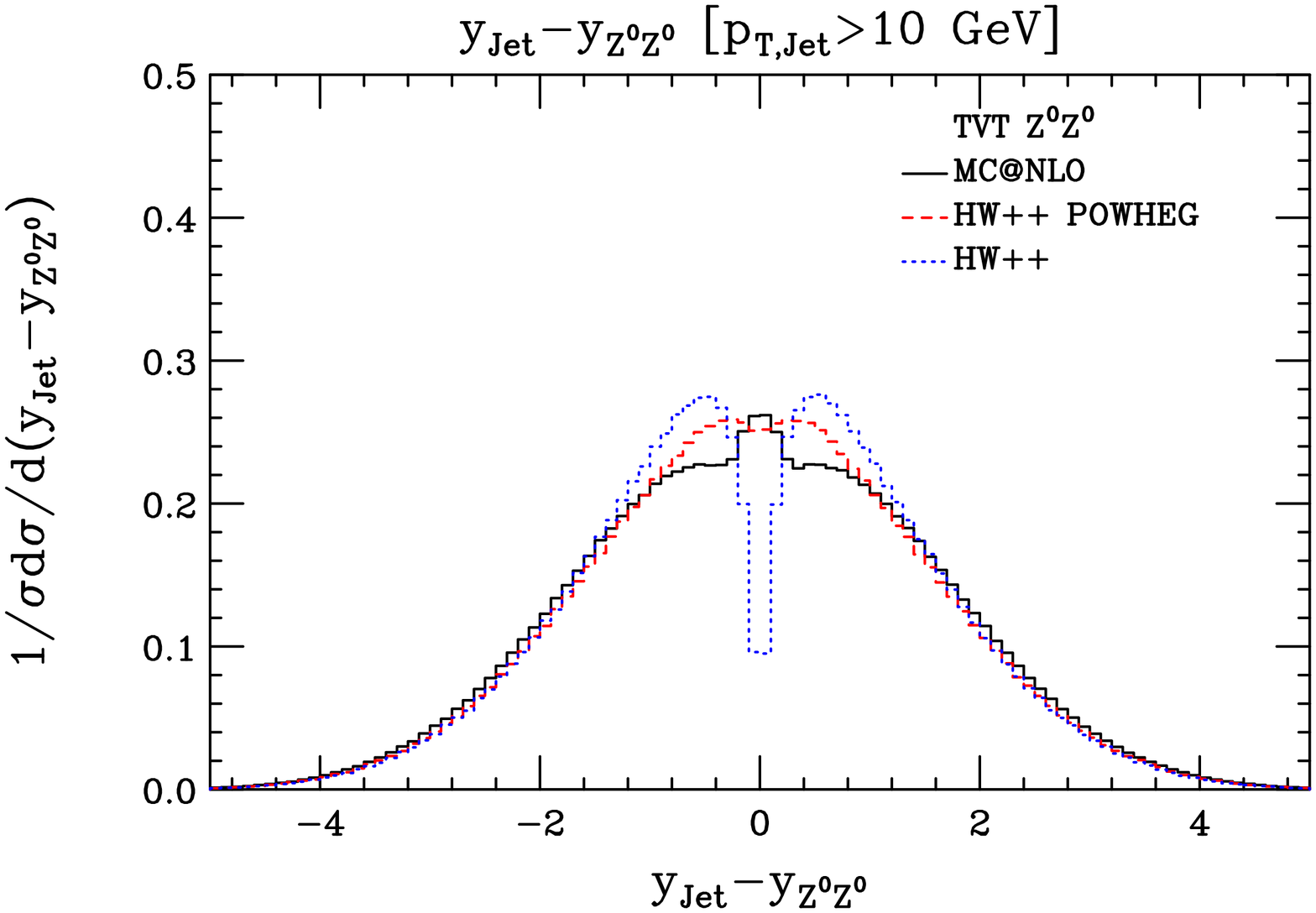}\hfill{}\includegraphics[scale=0.31,angle=90]{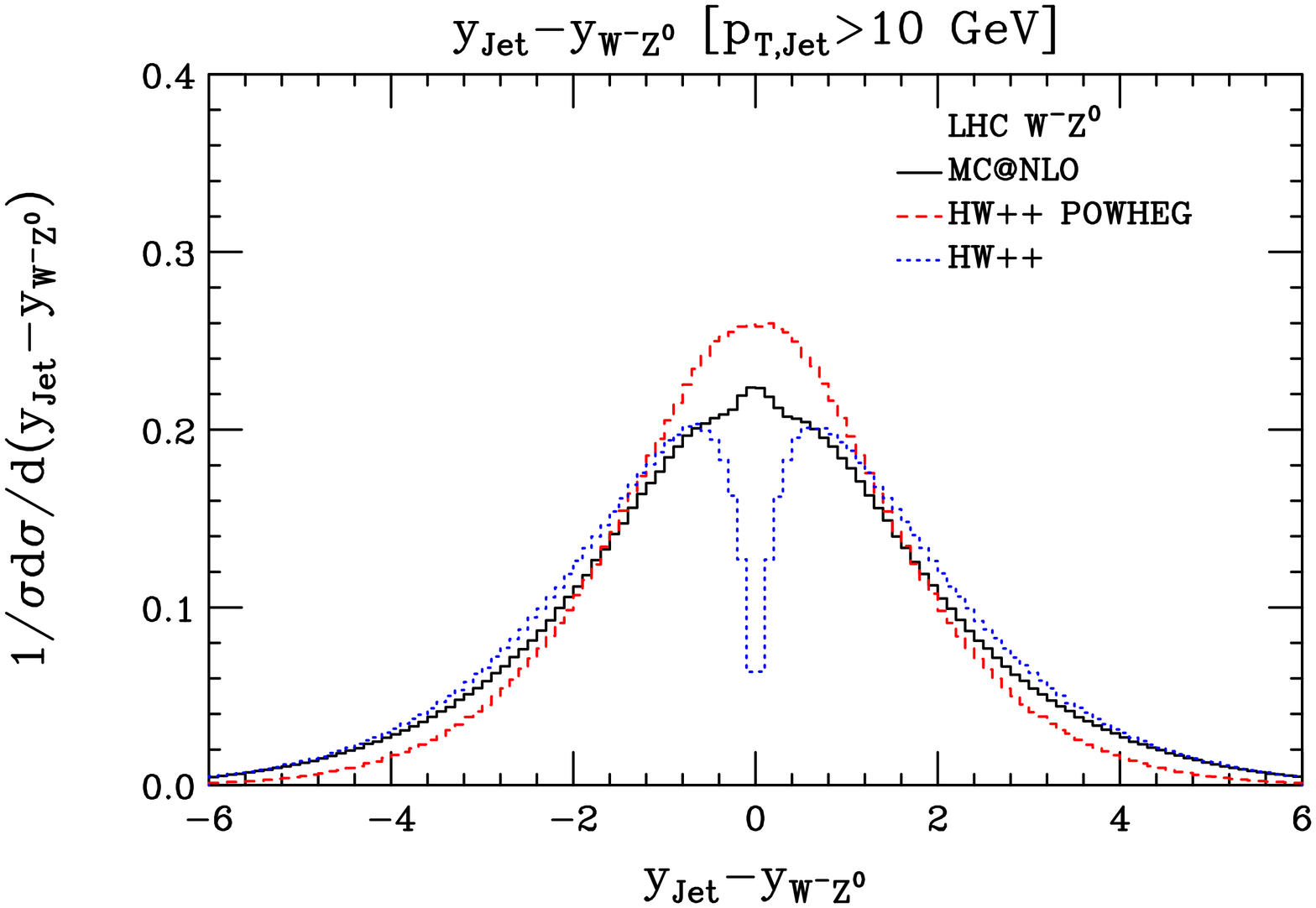}\vspace{8mm}

\par\end{centering}

\noindent \begin{centering}
\includegraphics[scale=0.31,angle=90]{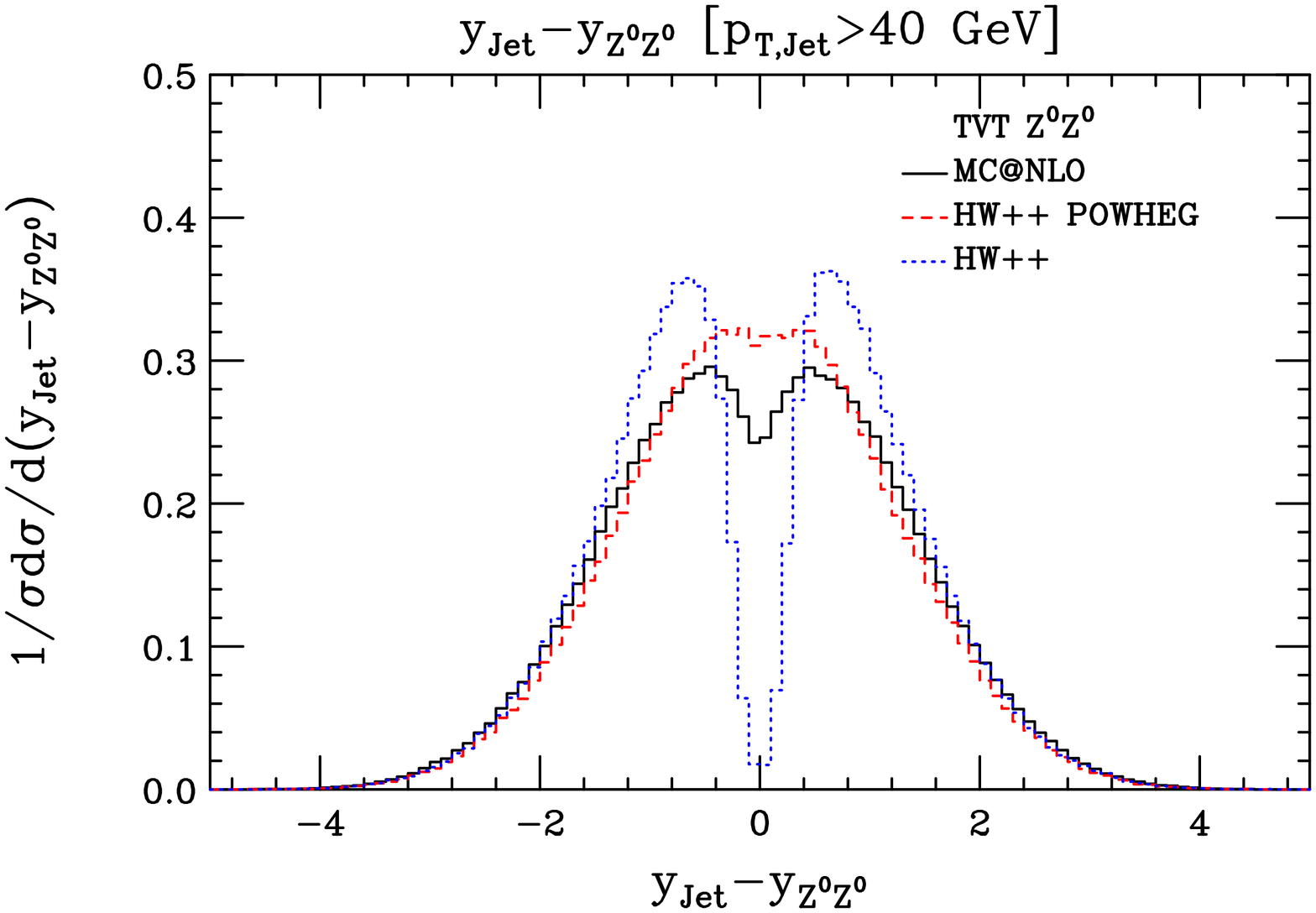}\hfill{}\includegraphics[scale=0.31,angle=90]{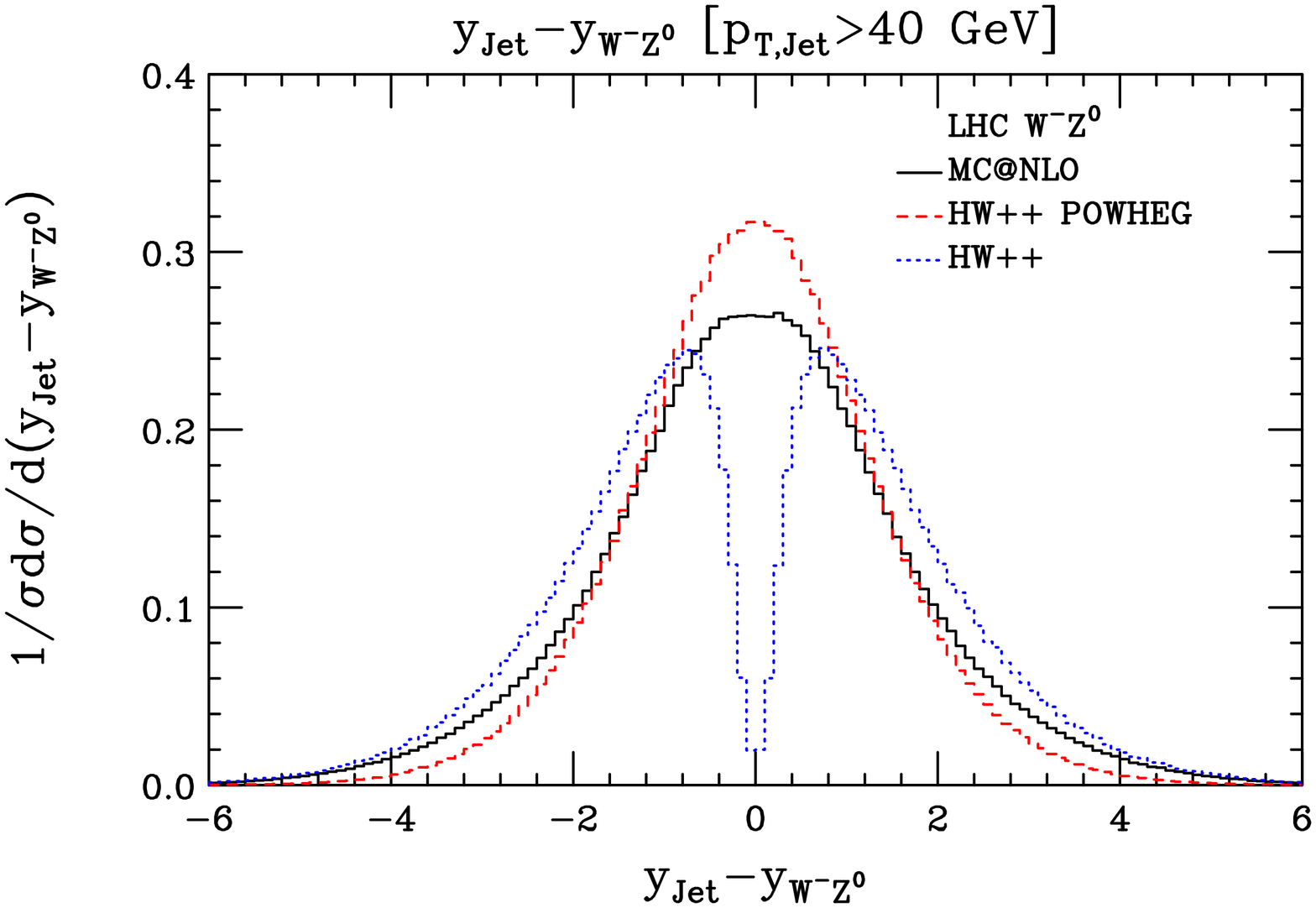}\vspace{8mm}

\par\end{centering}

\noindent \begin{centering}
\includegraphics[scale=0.31,angle=90]{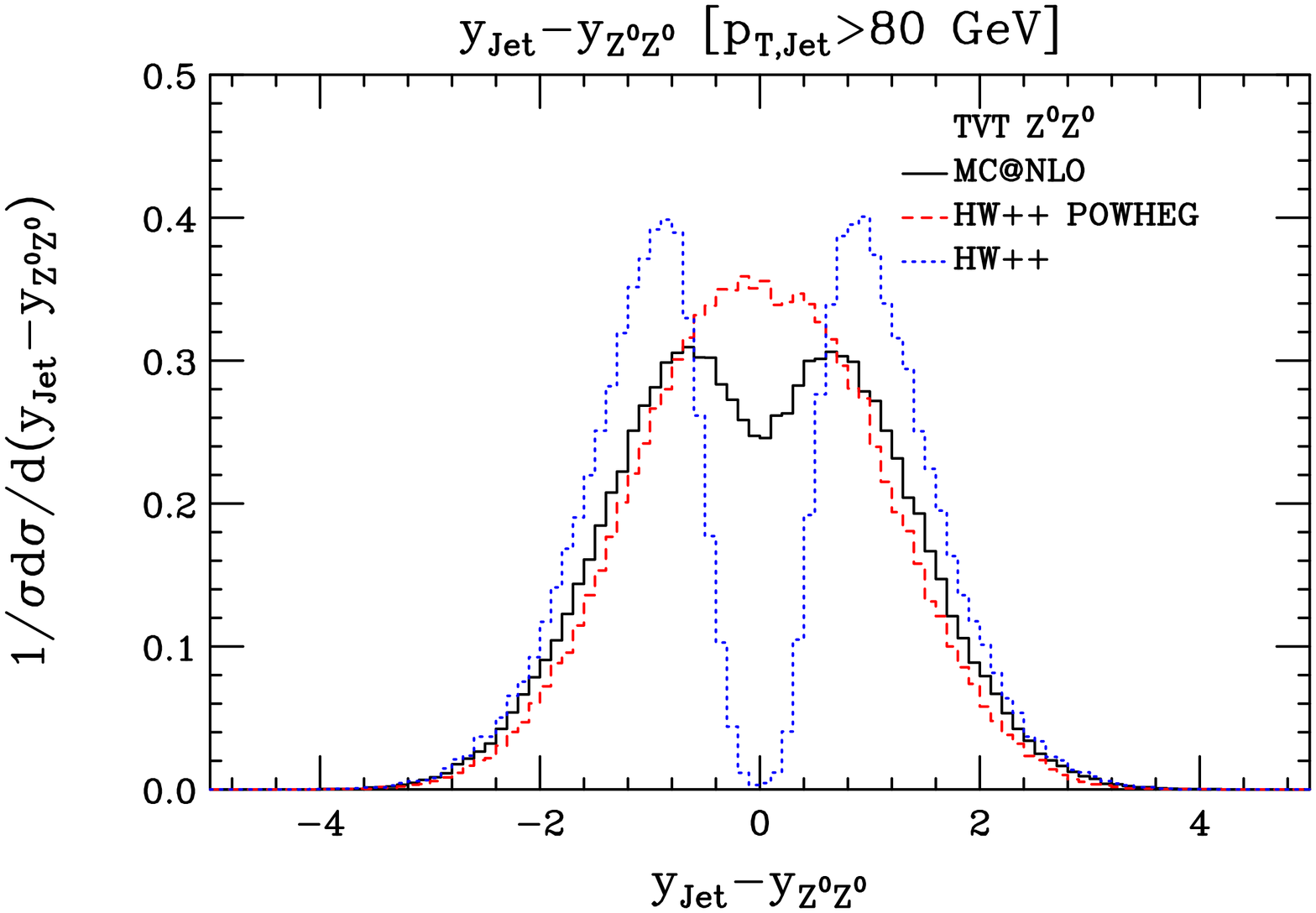}\hfill{}\includegraphics[scale=0.31,angle=90]{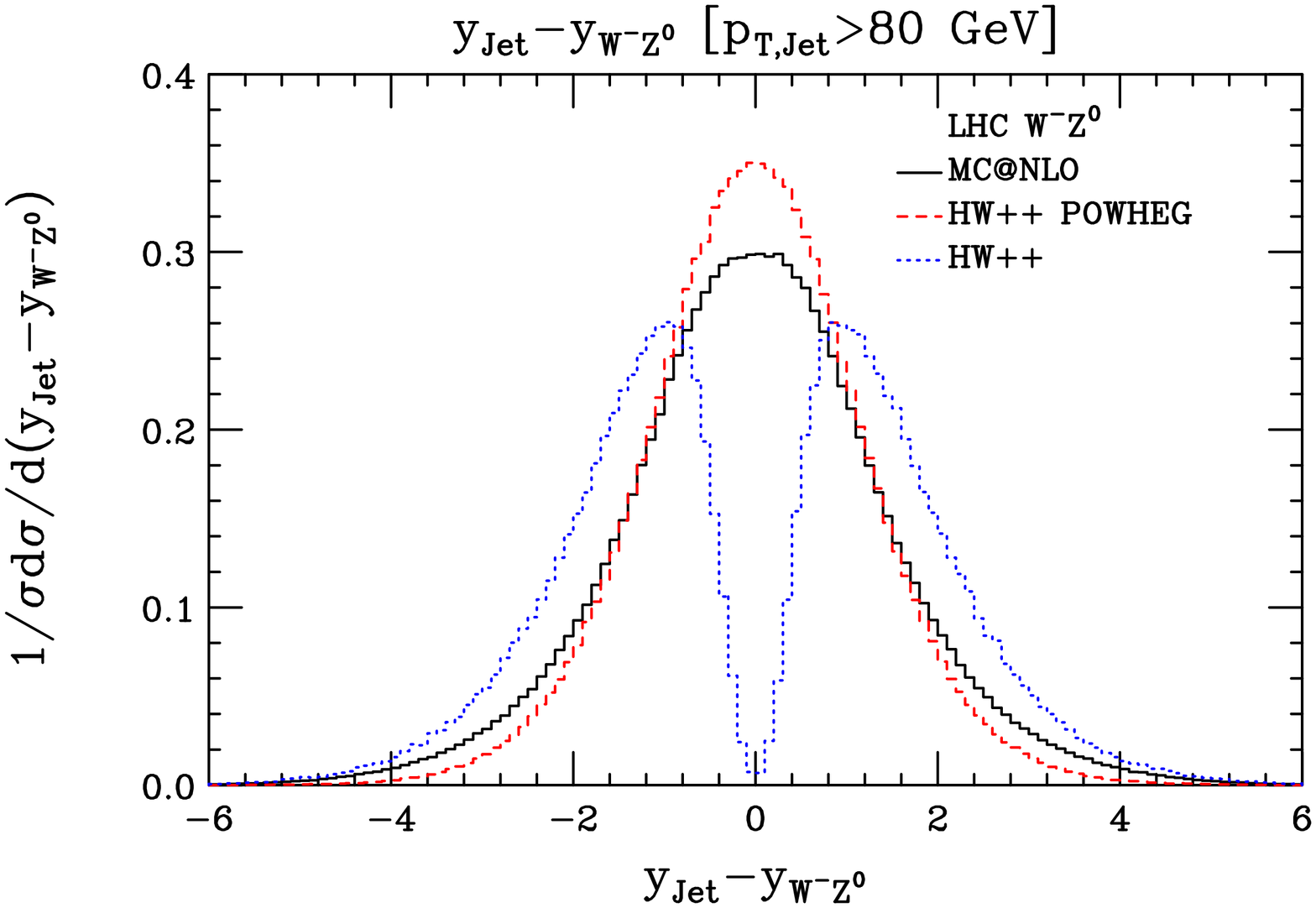} 
\par\end{centering}

\caption{In this figure we plot the difference in rapidity between the hardest
jet and the di-vector boson system, with different cuts imposed on
the transverse momentum of the leading jet. On the left hand side
we show predictions for these observables in $ZZ$ production at the
Tevatron, while on the right hand side they are shown for the case
of $W^{-}Z$ production at the LHC. }

\label{fig:yJet-yVV_rapidity_correlation} 
\end{figure}

In Figure~\ref{fig:yJet-yVV_rapidity_correlation} we show predictions
for $\mathrm{y}_{k}-\mathrm{y}$ distributions in $ZZ$ production
and $W^{-}Z$ production at the Tevatron and LHC respectively. For
each process we have considered how the results are affected by varying
the $p_{T}$ cut on the leading jet. The general trends seen in these
plots are qualitatively the same as those obtained in Higgs boson
production and Higgs boson production in association with a $W^{\pm}/Z$
boson \cite{Hamilton:2009za}, so too are our conclusions relating
to them.

We remind the reader that, in general, parton shower Monte Carlo programs
may not populate the full real emission phase space, resulting in
so-called \emph{dead-zones}. This is certainly true of the \emph{\noun{Herwig
}}and\emph{ }\emph{\noun{Herwig++}} simulations. The presence of dead
zones in the real emission phase space follows directly from the scale
choice used to initiate the parton shower evolution%
\footnote{For explicit phase space computations and maps concerning the origin
of dead zones and their connection to the choice of the initial evolution
scales see Refs.~\cite{Gieseke:2003rz,Hamilton:2009za,Torrielli:2010aw}.%
}, moreover, they are typically located in regions of phase space associated
with high $p_{T}$ emissions. In \noun{Herwig }and \noun{Herwig++}
the dead zone for the first emission in processes such as this, comprised
of a single initial-state colour dipole, is centred on $\theta_{k}=\frac{\pi}{2}$
$\left(\mathrm{y_{k}-}\mathrm{y}=0\right)$, moreover, the angular
breadth of the dead zone increases symmetrically and monotonically
about this point with the energy of the emitted radiation (see \emph{e.g.}
Fig.~7 of Ref.~\cite{Hamilton:2009za}). The \emph{dip} feature
seen in the ordinary parton shower results (blue dots) directly reflects
the angular characteristics of this unpopulated region of phase space:
in all cases, as the $p_{T}$ cut on the leading jet increases the
dip becomes broader and deeper. 

Since \noun{Mc}{\footnotesize @}\noun{nlo }and \noun{Powheg }aim to
fully include NLO corrections within the parton shower simulation,
they naturally populate all of the real emission phase space. In keeping
with this, we observe that the predictions from these two approaches
do not show the same significant dip in the central region of the
distributions. Whereas the \noun{Powheg }simulation fills this phase
space independently of the detailed workings of the parton shower
to which it is subsequently attached, the \noun{Mc}{\footnotesize @}\noun{nlo
}approach involves carefully augmenting the parton shower simulation
by the difference between its own approximate real emission cross
section and the true real emission cross section in the NLO calculation.
In particular, this means that the distribution of radiation from
\noun{Mc}{\footnotesize @}\noun{nlo} in the dead zone follows \emph{exactly}
the fixed order NLO calculation, while either side of it the distribution
differs from this by $\mathcal{O}\left(\alpha_{\mathrm{S}}^{2}\right)$
terms.

Having noted this NNLO discontinuity in the radiation pattern, it
is then understandable that the \noun{Mc}{\footnotesize @}\noun{nlo}
predictions (black) can exhibit some minor irregularities and differences
with respect to \noun{Powheg} (red) in the central region and that
these should reflect, somewhat, the trends seen in the results obtained
using the parton shower alone. On the contrary, the response of the
\noun{Powheg} predictions to the increasing $p_{T}$ cut on the leading
jet lends itself to a more straightforward interpretation based on
simple kinematic reasoning, namely, that the $\mathrm{y}_{k}-\mathrm{y}$
distribution should become more central, as the phase space available
for small angle emissions -- which populate the tails -- becomes reduced
relative to that available for large angle emissions. Furthermore,
we point out that for observables employing cuts which exclude the
softer regions of phase space, such as $\mathrm{y}_{k}-\mathrm{y}$
with a $p_{T}$ cut of 80 GeV on the leading jet, one expects that
the \noun{Powheg} predictions exceed those of conventional NLO calculations
due to the $\overline{B}\left(\Phi_{B}\right)/B\left(\Phi_{B}\right)$
factor multiplying the real emission cross section in Eq.\,\ref{eq:powheg_1}.
This behaviour is apparent in Figure~\ref{fig:yJet-yVV_rapidity_correlation}.
Finally, we remark that we have reproduced all of these distributions
using our \noun{Powheg }simulation with the truncated shower feature
disabled, with no observable consequences.

\begin{figure}[t]
\noindent \begin{centering}
\includegraphics[scale=0.3,angle=90]{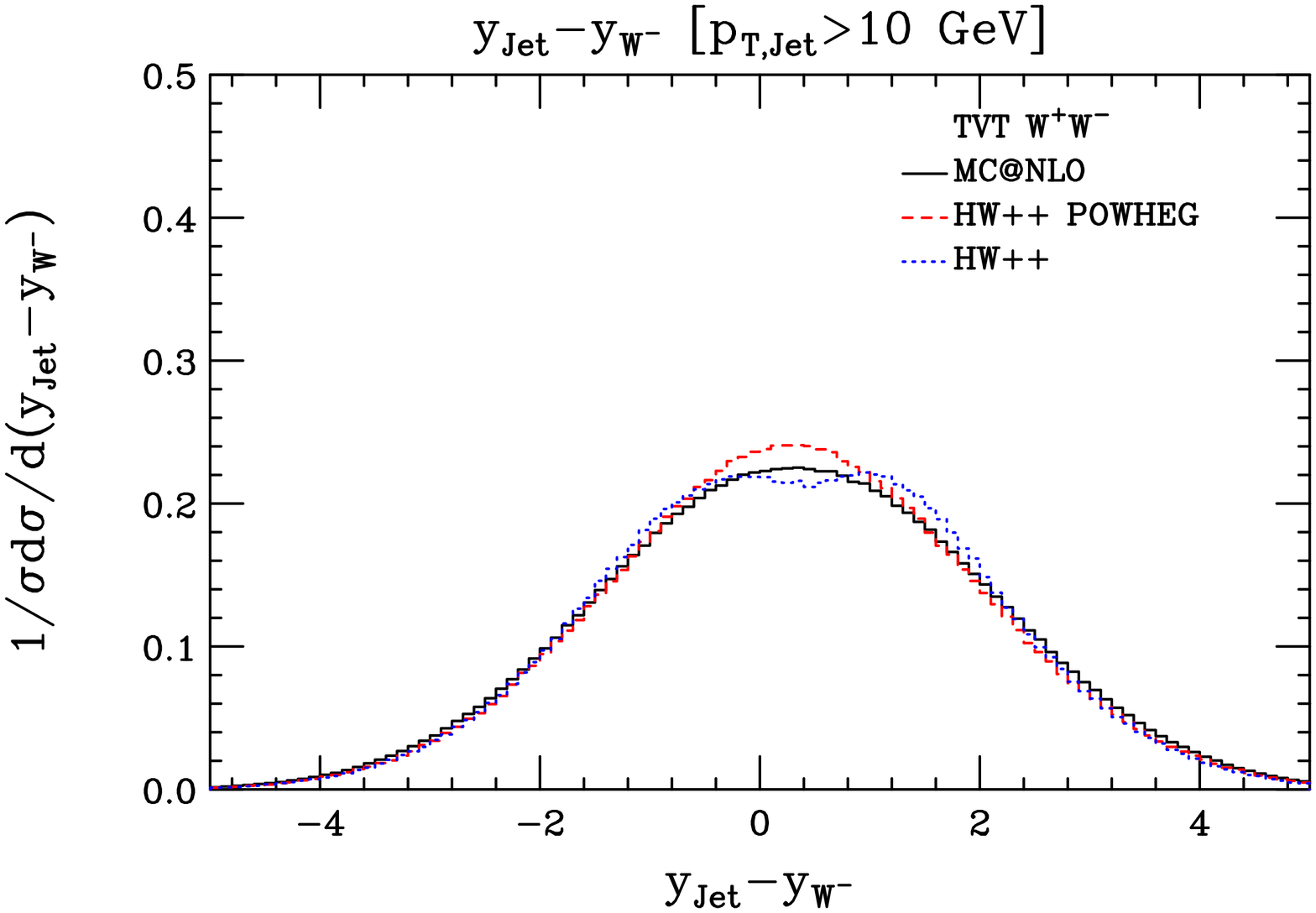}\hfill{}\includegraphics[scale=0.3,angle=90]{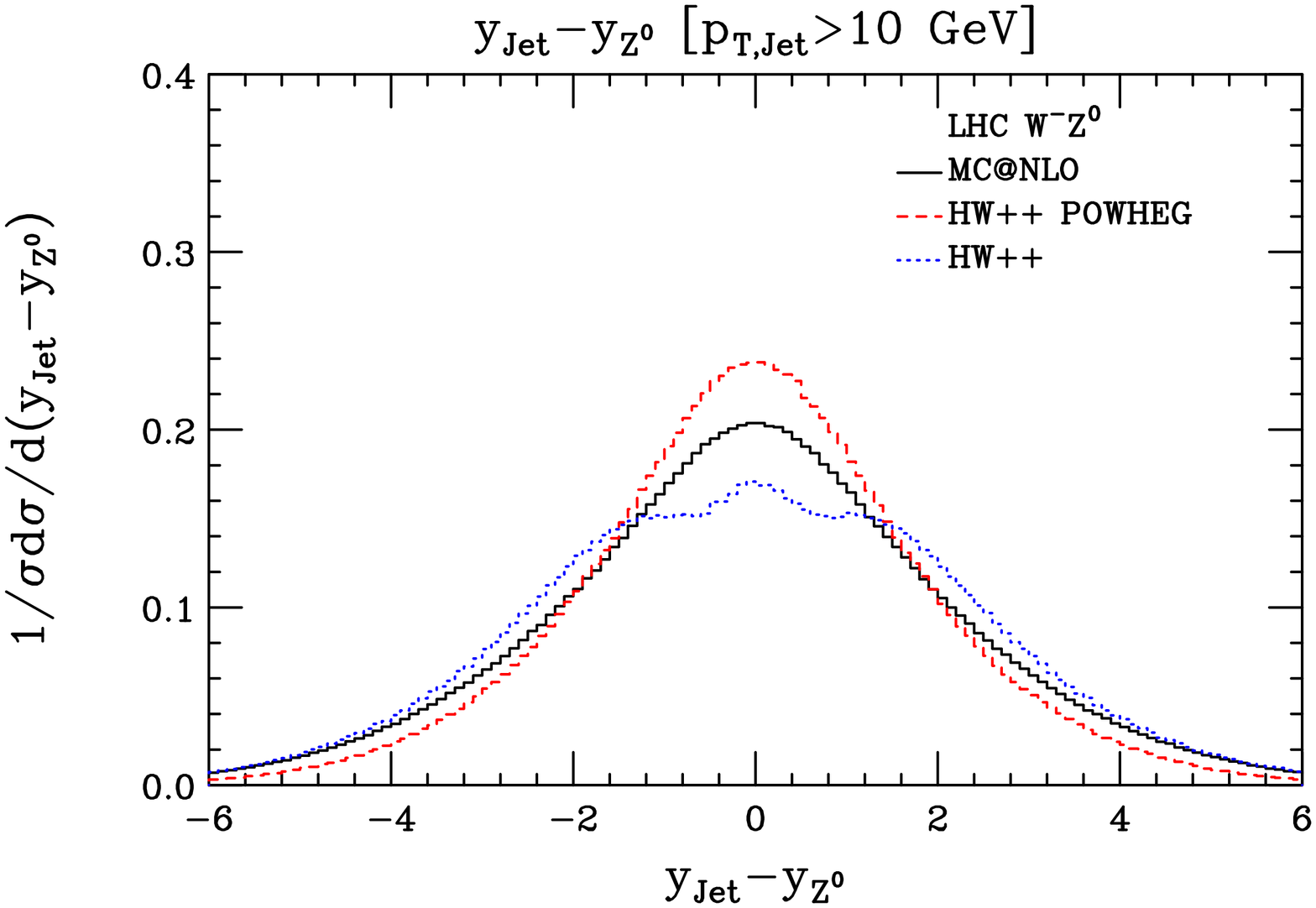}\vspace{7mm}

\par\end{centering}

\noindent \begin{centering}
\includegraphics[scale=0.3,angle=90]{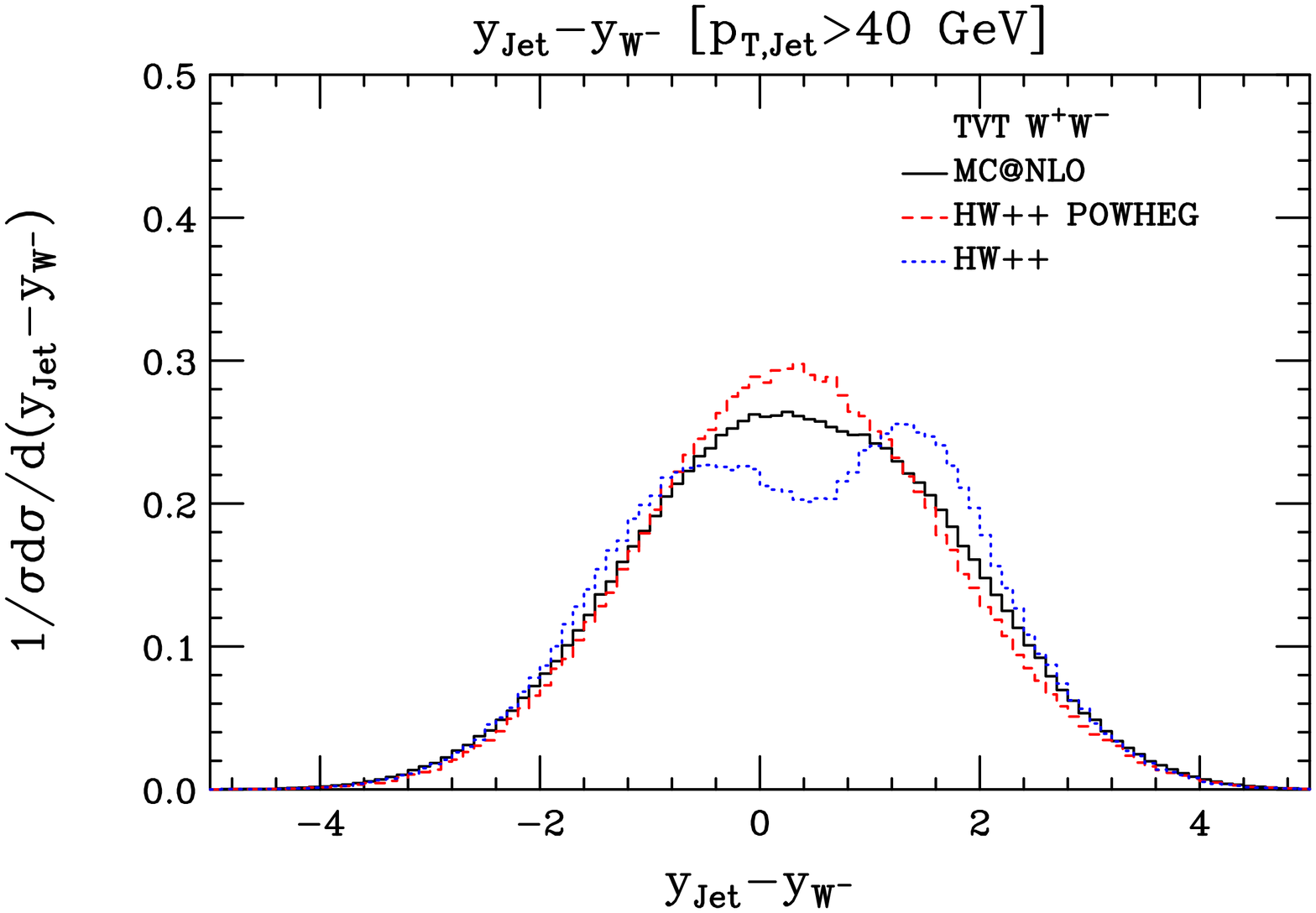}\hfill{}\includegraphics[scale=0.3,angle=90]{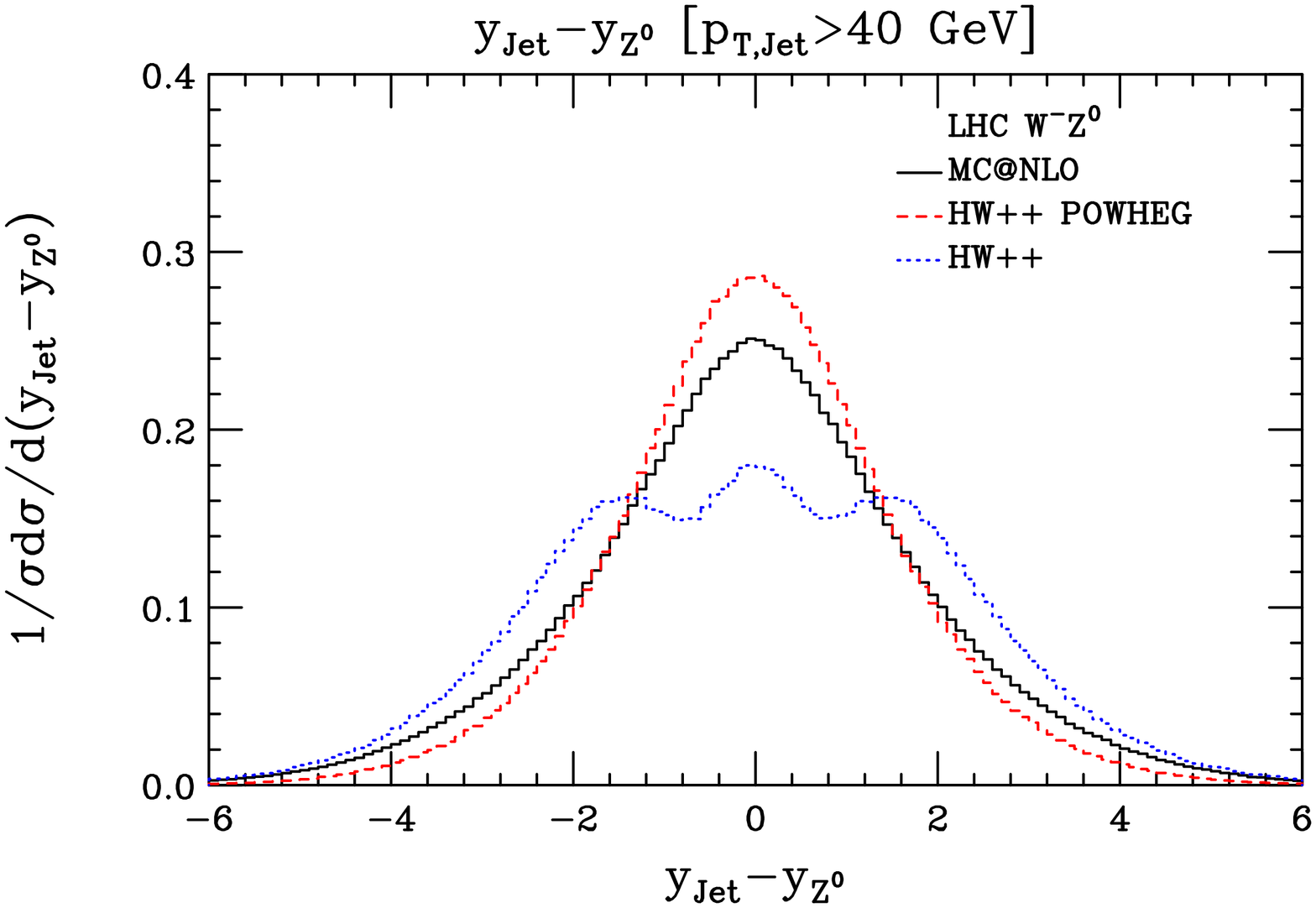}\vspace{7mm}
 \includegraphics[scale=0.3,angle=90]{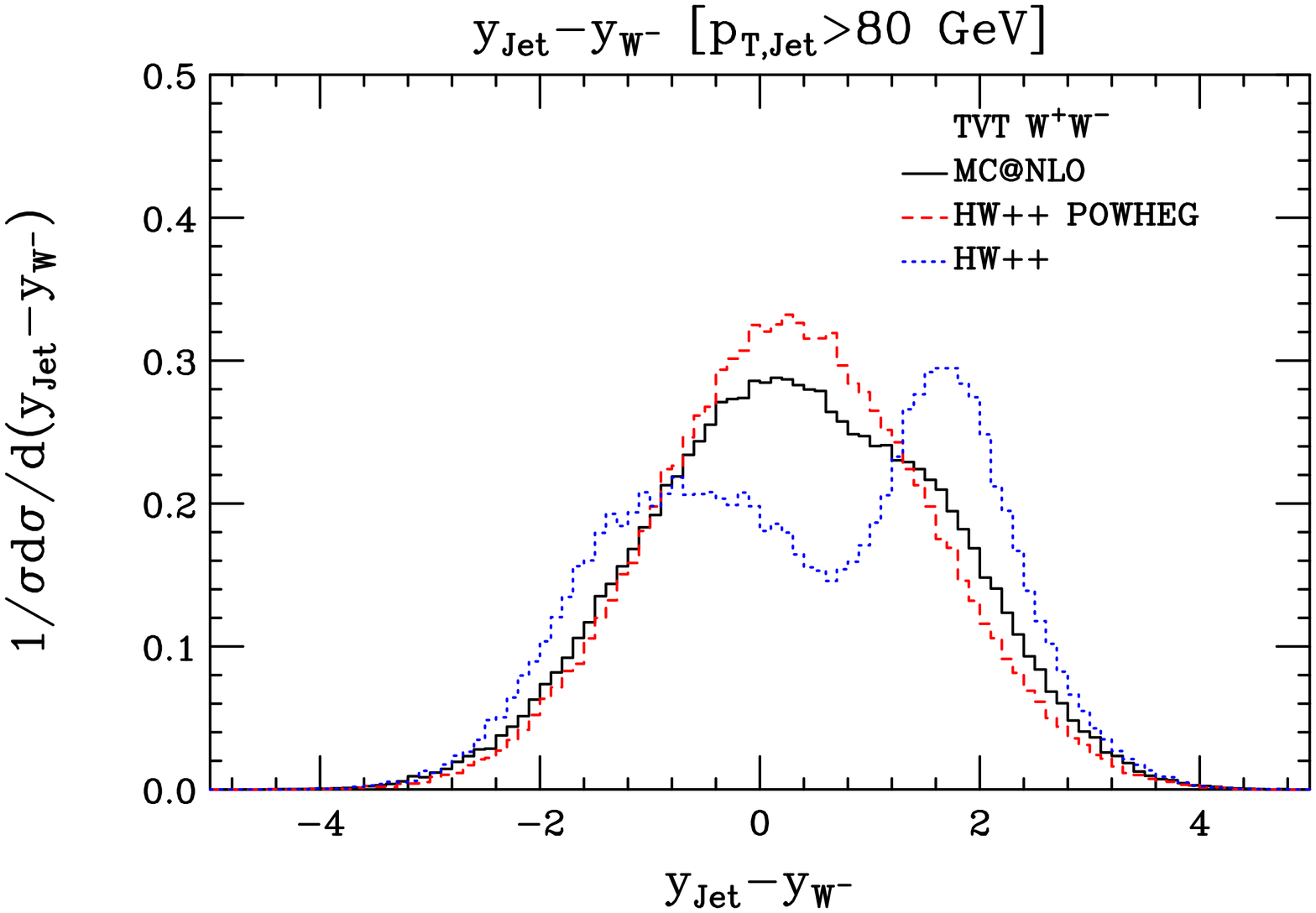}\hfill{}\includegraphics[scale=0.3,angle=90]{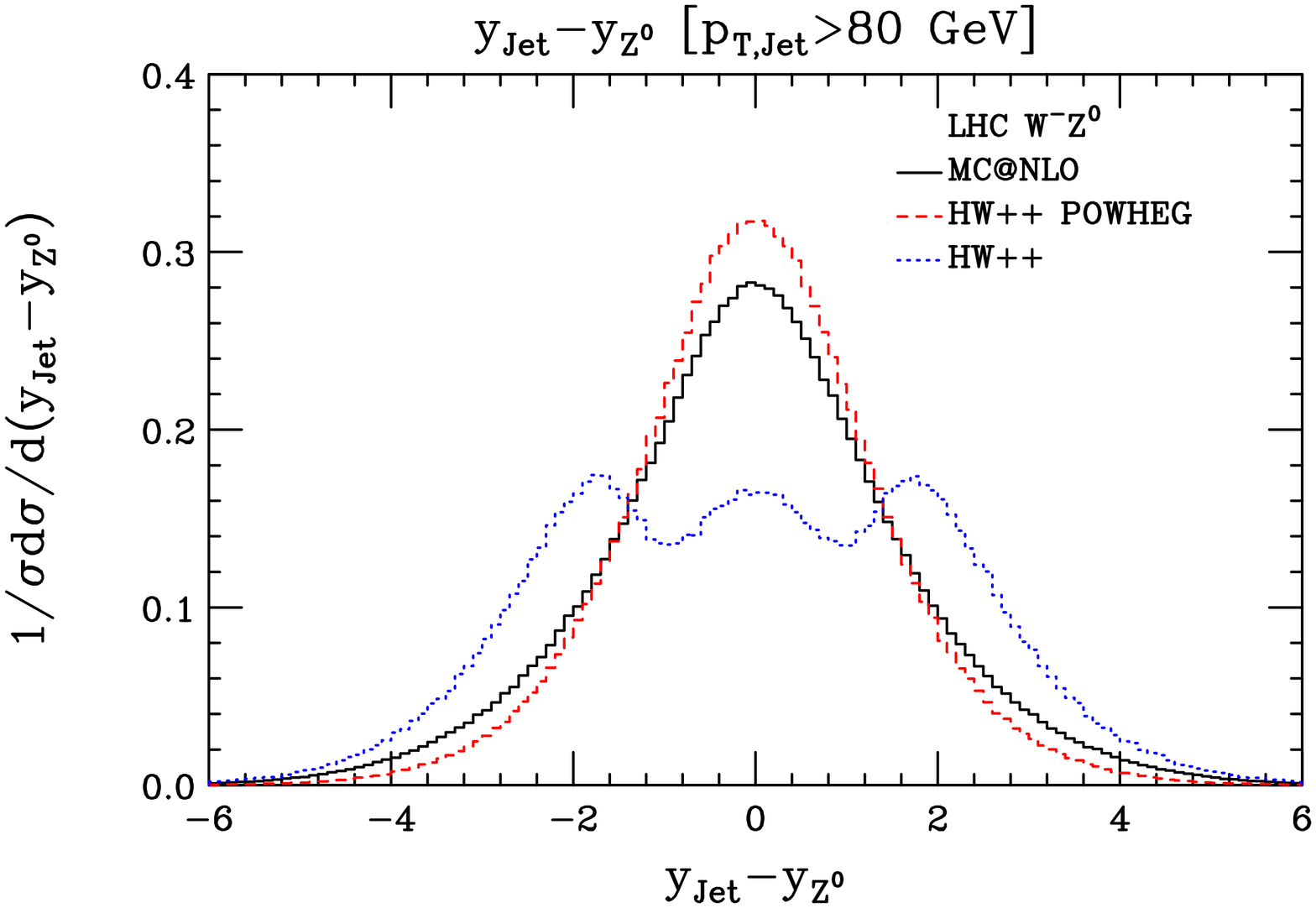} 
\par\end{centering}

\caption{On the left of this figure we show the difference in rapidity between
the hardest jet and the $W^{-}$ boson in $W^{+}W^{-}$ production
at the Tevatron, with $p_{T}$ cuts of 10, 40 and 80 GeV imposed on
the transverse momentum of the leading jet. Analogously, on the right
hand side, we show the rapidity correlation between the hardest jet
and the $Z$ boson in $W^{-}Z$ production events.}

\label{fig:yJet-yV_rapidity_correlation} 
\end{figure}
Figure~\ref{fig:yJet-yV_rapidity_correlation} shows the differences
in rapidity between the leading jet and \emph{one} of the vector bosons
in $W^{+}W^{-}$ production at the Tevatron and $W^{-}Z$ production
at the LHC. Since these distributions are closely related to those
in Fig.~\ref{fig:yJet-yVV_rapidity_correlation}, we argue that the
differences seen in the leading order parton shower predictions with
respect to \noun{Mc}{\footnotesize @}\noun{nlo} and \noun{Powheg}
are again attributable to the dead zone in the former; \emph{a fortiori
}considering the variation of the uncorrected parton shower results
with respect to the changing $p_{T}$ cut on the leading jet. We contend
that the peculiar shape of the pure parton shower results reflect
the impression left by the dead zone in $\mathrm{y}_{k}-\mathrm{y}$
convoluted with the rapidity distribution of the vector bosons with
respect to one another. In all of the distributions the agreement
between the \noun{Mc}{\footnotesize @}\noun{nlo }and \noun{Powheg
}results is quite satisfactory: note that both approaches formally
only offer a leading order description of this quantity. Some small
distortion can be seen on the right of the \noun{Mc}{\footnotesize @}\noun{nlo
}distribution at the Tevatron, for a $p_{T}$ cut of $80$ GeV on
the leading jet, which we tentatively suggest is indicative of the
asymmetric parton shower prediction. 

\begin{figure}[t]
\noindent \begin{centering}
\includegraphics[scale=0.31,angle=90]{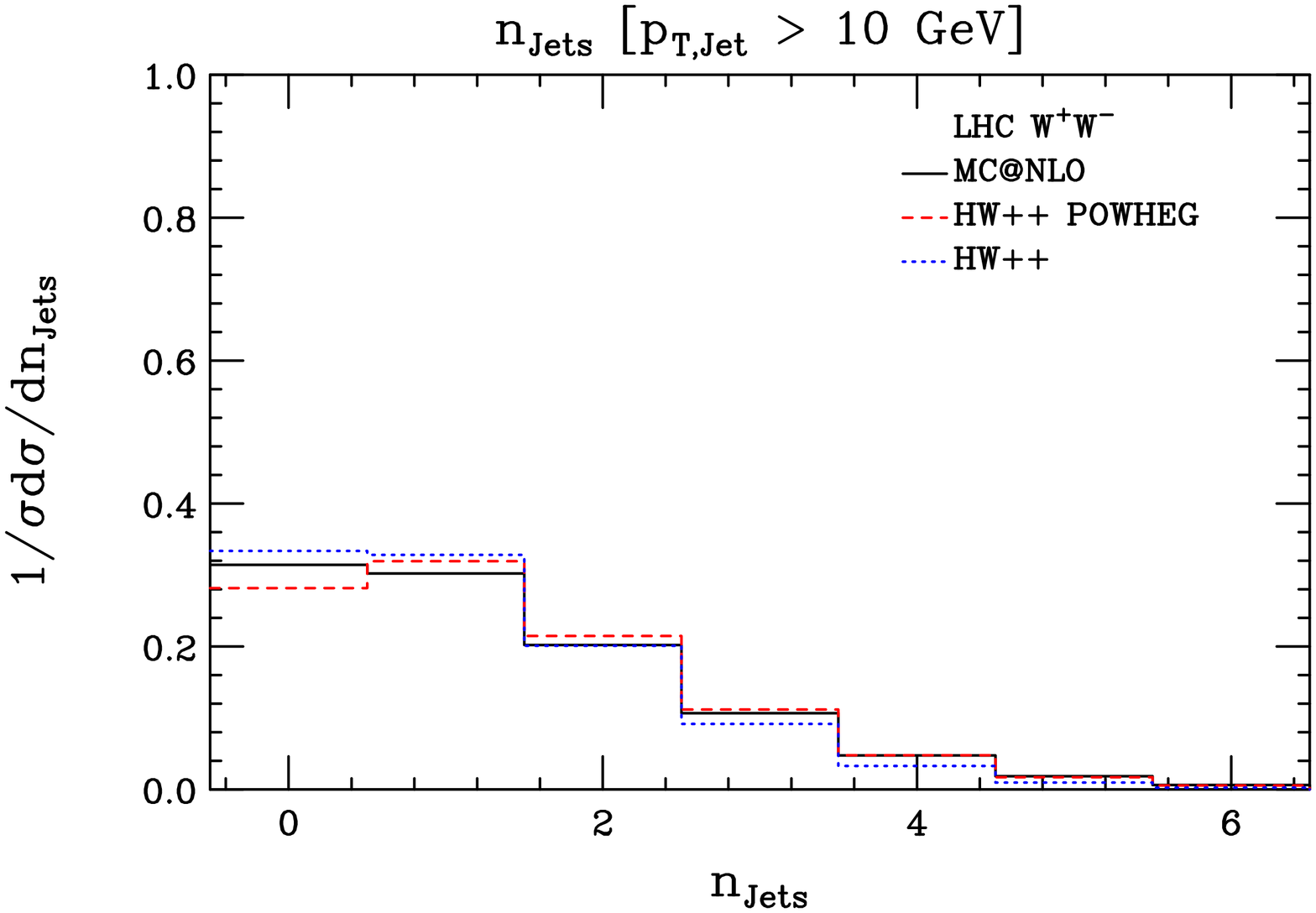}\hfill{}\includegraphics[scale=0.31,angle=90]{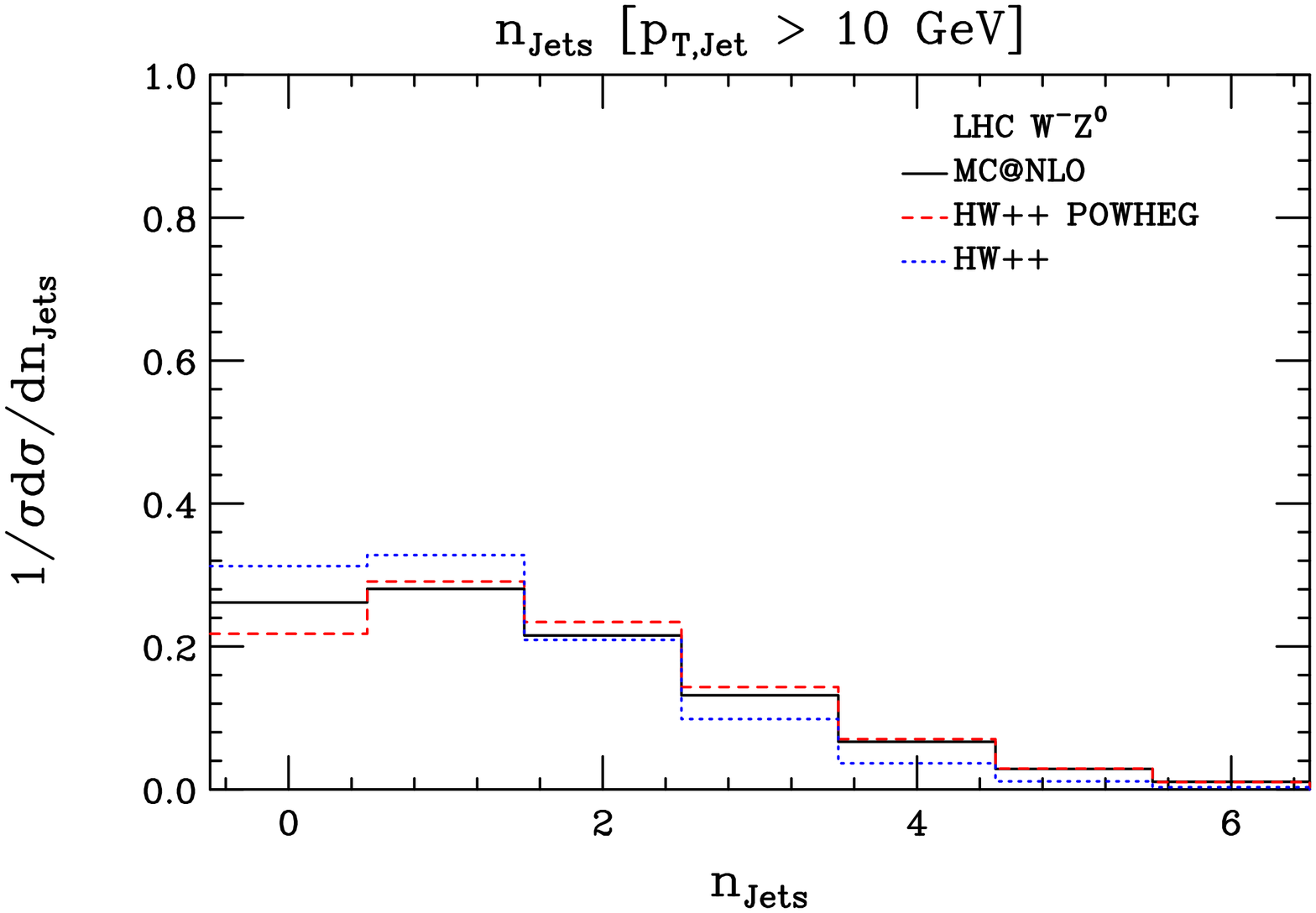}\vspace{8mm}
 \includegraphics[scale=0.31,angle=90]{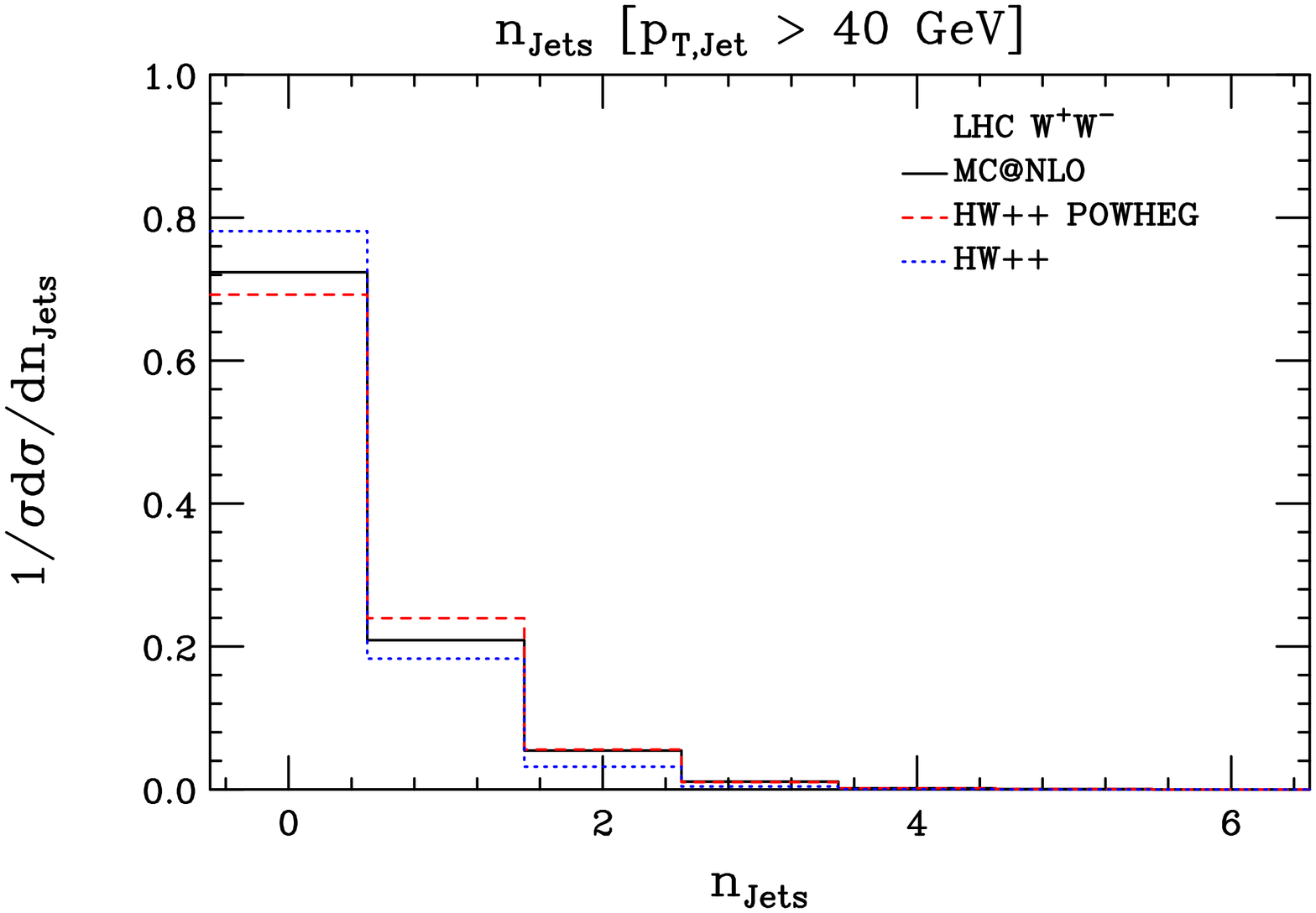}\hfill{}\includegraphics[scale=0.31,angle=90]{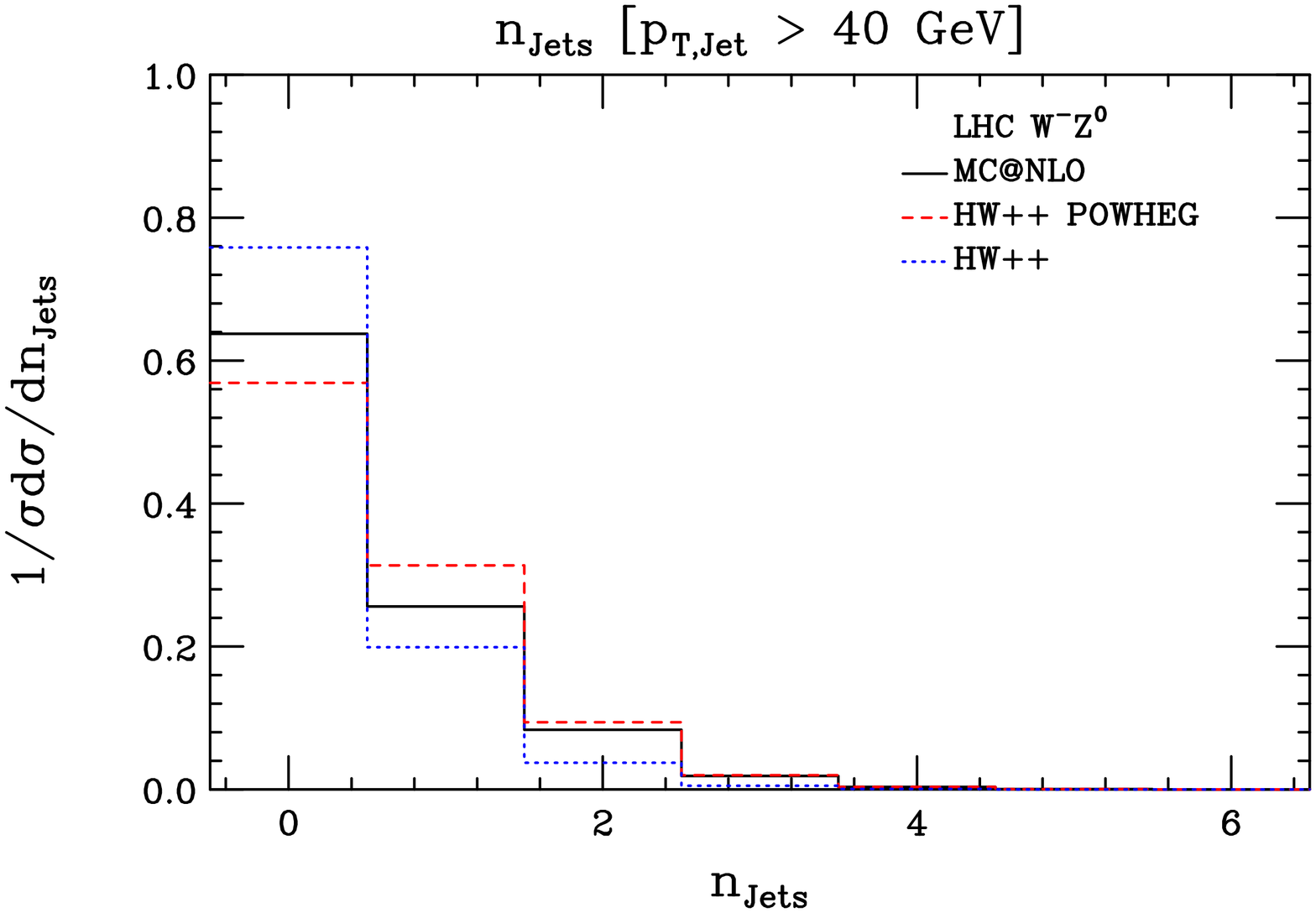}\vspace{8mm}
 \includegraphics[scale=0.31,angle=90]{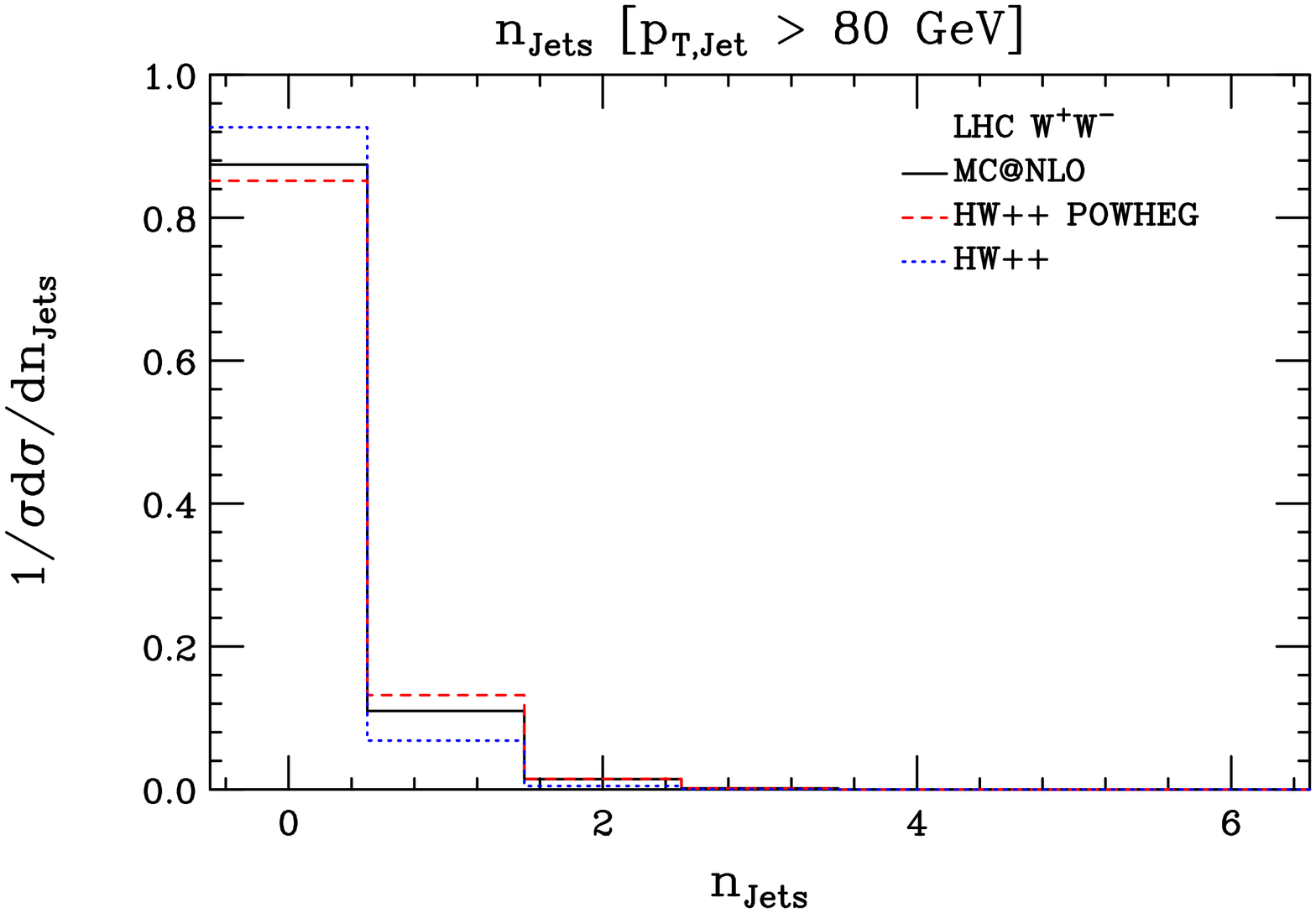}\hfill{}\includegraphics[scale=0.31,angle=90]{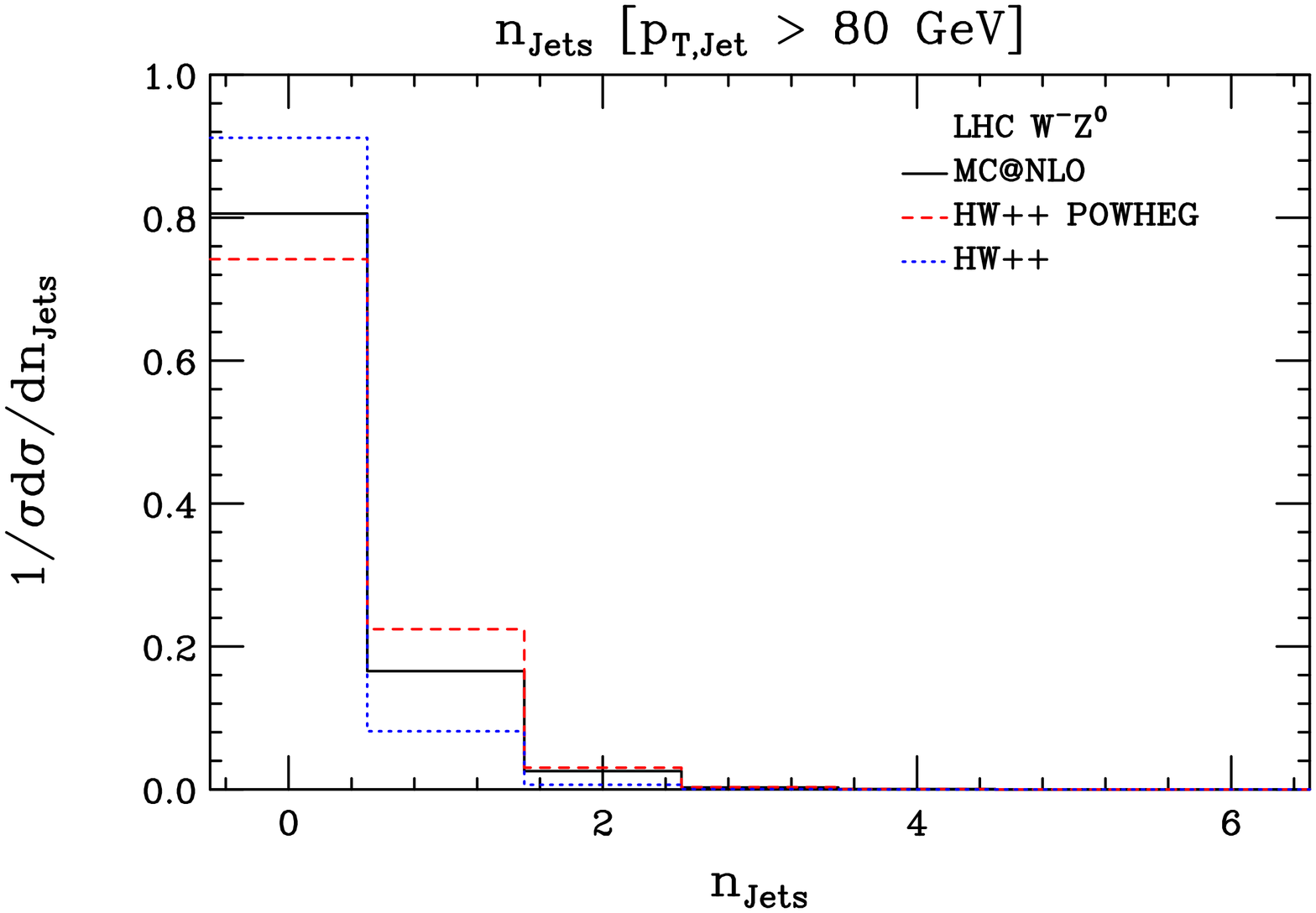} 
\par\end{centering}

\caption{Here we show jet multiplicity distributions in $W^{+}W^{-}$ and $W^{-}Z$
production at the LHC, given a centre-of-mass energy of $14\,\mathrm{TeV}$.
The three pairs of distributions, running from the top to the bottom
of the figure, result from applying three different sets of $p_{T}$
cuts (10, 40 and 80 GeV) to all of the jets in each event. }

\label{fig:n_jets} 
\end{figure}
LHC predictions for jet multiplicity distributions in $W^{+}W^{-}$
and $W^{-}Z$ production can be found in Figure~\ref{fig:n_jets},
assuming a hadronic centre-of-mass energy of 14 TeV. As one expects,
in all cases jet multiplicities decrease rapidly as the $p_{T}$ cut
on the leading jet is increased. One can also see that the results
obtained using the parton shower alone show a tendency to overestimate
the number of events without any jets in comparison to the other methods.
This is also expected given our earlier discussions concerning the
dead zone in the parton shower phase space. Finally we note that the
\noun{Powheg} predictions for the number of events with one jet are
above those of \noun{Mc}{\footnotesize @}\noun{nlo} by an amount characteristic
of the NLO \emph{K-}factor. As noted previously, this systematic effect
can be directly attributed to the presence of the $\overline{B}\left(\Phi_{B}\right)/B\left(\Phi_{B}\right)$
factor multiplying the real emission part of the \noun{Powheg }hardest
emission cross section (Eq.\,\ref{eq:powheg_1}). Once again we note
that the presence of this term modifies the distribution of hard radiation
by terms of NNLO significance only.

\section{Conclusion\label{sec:Conclusion}}

In this article we have presented an implementation of the \noun{Powheg}
NLO matching formalism for simulations of weak boson pair production
and decay, in the double pole approximation. These simulations have
been integrated within the \noun{Herwig++} Monte Carlo event generator,
including truncated shower effects to account for colour coherence
phenomena. 

In constructing this NLO event generator we have employed novel relations
between the $W^{\pm}Z$ cross sections and those of $W^{+}W^{-}$
and $ZZ$ production. Total cross sections and parton level NLO distributions
were found to be in excellent agreement with predictions obtained
from the the \noun{Mcfm} NLO Monte Carlo program. 

The shapes of the emission spectra from the full simulation, including
parton shower effects, are seen to generally compare well with those
of \noun{Mc}{\footnotesize @}\noun{nlo }in a wide variety of kinematic
distributions\noun{ --} both of which exhibit large corrections with
respect to the default parton shower predictions. Where minor differences
have arisen between our results and \noun{Mc}{\footnotesize @}\noun{nlo}
they have been studied in detail.

As noted in previous works comparing \noun{Mc}{\footnotesize @}\noun{nlo}
and \noun{Powheg }\cite{Alioli:2008tz,Hamilton:2009za} we observe
a tendency for \noun{Powheg }to produce slightly more hard radiation
than \noun{Mc}{\footnotesize @}\noun{nlo}. The explanation given for
this effect in those publications is seen to hold well here, specifically,
that the $\overline{B}\left(\Phi_{B}\right)/B\left(\Phi_{B}\right)$
factor which multiplies the real part of the \noun{Powheg }hardest
emission cross section leads to an enhancement of high $p_{T}$ radiation
with respect to the corresponding NLO prediction; the differences
being formally of order $\alpha_{\mathrm{S}}^{2}$. As in Refs.~\cite{Alioli:2008tz,Hamilton:2009za}
we also find that the \noun{Mc}{\footnotesize @}\noun{nlo} program
exhibits a sensitivity to the phase space partitioning in the underlying
parton shower simulation \noun{(}\emph{cf}.\emph{ }Figs.~\ref{fig:yJet-yVV_rapidity_correlation}
and \ref{fig:yJet-yV_rapidity_correlation}), however, as with the
enhancement of hard radiation in \noun{Powheg,} this is formally representative
of NNLO effects. 

All weak boson boson pair production simulations presented here are
due for inclusion in the next public release of the \noun{Herwig++
}event generator. 

\acknowledgments

I wish to thank my colleagues in the \noun{Herwig++} and \noun{Powheg-Box}
collaborations for valuable input in the course of this work. I am
also very grateful to Pavel Demin, Fabio Maltoni and the rest of CP3
Louvain for providing access to their high performance computing cluster. 

\appendix

\section{Regularized and unregularized splitting functions\label{sec:Splitting-functions}}

We write the `customary' regularized Altarelli-Parisi functions in
terms of $\rho$-distributions as\[
P_{i,\widetilde{ic}}\left(x\right)=P_{i,\widetilde{ic}}^{\rho}\left(x\right)+C_{i,\widetilde{ic}}\left(p_{i,\widetilde{ic}}+4\ln\eta\right)\delta\left(1-x\right)\,,\]
 where\begin{align*}
P_{gg}^{\rho}\left(x\right) & =2C_{A}\left[\frac{x}{\left(1-x\right)_{\rho}}+\frac{1-x}{x}+x\left(1-x\right)\right], & \mbox{ }\mbox{ }\mbox{ }\mbox{ }\mbox{ }\mbox{ }\mbox{ }\mbox{ }\mbox{ }C_{gg} & =C_{A}, & \mbox{ }\mbox{ }\mbox{ }\mbox{ }\mbox{ }\mbox{ }\mbox{ }\mbox{ }\mbox{ }p_{gg} & =\frac{2\pi b_{0}}{C_{A}},\\
P_{qq}^{\rho}\left(x\right) & =C_{F}\left[\frac{1+x^{2}}{\left(1-x\right)_{\rho}}\right], & \mbox{ }\mbox{ }\mbox{ }\mbox{ }\mbox{ }\mbox{ }\mbox{ }\mbox{ }\mbox{ }C_{qq} & =C_{F}, & \mbox{ }\mbox{ }\mbox{ }\mbox{ }\mbox{ }\mbox{ }\mbox{ }\mbox{ }\mbox{ }p_{qq} & =\frac{3}{2},\\
P_{qg}^{\rho}\left(x\right) & =C_{F}\left[\frac{1+\left(1-x\right)^{2}}{x}\right],\\
P_{gq}^{\rho}\left(x\right) & =T_{R}\left[x^{2}+\left(1-x\right)^{2}\right],\end{align*}
 and \[
b_{0}=\frac{1}{4\pi}\left(\frac{11}{3}C_{A}-\frac{4}{3}T_{R}n_{\mathrm{lf}}\right)\,,\]
 with all other $p_{i,\widetilde{ic}}$ and $C_{i,\widetilde{ic}}$
being equal to zero.

\bibliographystyle{jhep}
\bibliography{Herwig++}

\end{document}